\pdfoutput=1
\documentclass[10pt,oneside,final]{ucdthesis}

\usepackage[letterpaper,left=0.75in,right=0.75in,top=1.25in,bottom=0.5in]{geometry}
\usepackage{tikz}
\usepackage{algorithm}
\usepackage{algpseudocode}
\usepackage{aas_macros}
\usepackage{colonequals}

\usepackage{fancyhdr}

\usepackage{amsmath}
\usepackage{amsfonts}

\usepackage{graphicx}

\usepackage{hyperref}

\usepackage[round]{natbib}

\usepackage{amsmath}
\usepackage{bm}

\usepackage{color}
\definecolor{MyDarkBlue}{rgb}{0,0.08,0.45}

\DeclareMathOperator{\tr}{tr}
\DeclareMathOperator{\arctanh}{arctanh}
\newcommand{\D}[2]{\ensuremath{\frac{d{#1}}{d{#2}}}}
\newcommand{\Dp}[2]{\ensuremath{\frac{\partial{#1}}{\partial{#2}}}}
\newcommand{\Ht}{\ensuremath{\mathcal{H}}}
\newcommand{\Lt}{\ensuremath{\mathcal{L}}}
\newcommand{\Pt}{\ensuremath{\mathcal{P}}}

\newcommand{\eqnref}[1]{equation~\ref{#1}}
\newcommand{\Eqnref}[1]{Equation~\ref{#1}}
\newcommand{\eqnrefand}[2]{equations~\ref{#1} and \ref{#2}}

\begin{document}

\isdissertation{1}

\iselectronic{1}

\title{\Large{Modeling Techniques for Measuring Galaxy Properties in Multi-Epoch Surveys}}
\author{James Francis Bosch}

\degreemonth{September}
\degreeyear{2011}
\degree{Doctor of Philosophy}

\chair{J. Anthony Tyson}
\othermembers{David Wittman, Christopher Fassnacht}
\numberofmembers{3}

\prevdegrees{B.A. (University of California, Berkeley) 2004}

\field{Physics}
\campus{Davis}

%
%

\pagestyle{fancyplain} 
\setlength{\topmargin}{-0.2in}
\fancyheadoffset[R]{0.4in}
\fancyhead[R]{\thepage}
\fancyfoot[C]{}
\renewcommand{\headrulewidth}{0pt}
\renewcommand{\sectionmark}[1]{}

\begin{frontmatter}
\maketitle

\begin{acknowledgments}
\ssp
It's hard to imagine how different I must have been when I started
graduate school 6 years ago.  I can remember being intimidated by the
idea of having to come up with a research project to work on, and I
had barely heard of gravitational lensing and knew nothing of LSST.
And I thought of myself as an \emph{IDL} programmer, a fact that would
no doubt amuse my current officemates.

What I did know back then was that I would be working with Tony
Tyson.  Tony's persuasive efforts were 90\% of the reason I decided to
come to UC Davis, and I have never regretted that decision.  I have so
much to thank Tony for -- he has been an awesome advisor, and I've been
continually amazed both by how he seems to be an expert at everything
and how hard he fights to make time for his students in what
seems an unimaginably busy schedule full of Very Important People.
Even more importantly, he was a 
bottomless source of interesting ideas when I needed them, but he
encouraged me to follow my own ideas once I started to come up with
them myself.  Of course, he was also instrumental in steering me away
from the ideas that wouldn't have been very interesting, and helping
me to discover that while I may not be a conventional astronomer, I'm
definitely not interested in just being a programmer either. I also
have Tony to thank for a lot of financial support; while I like 
teaching, it's a definite luxury not to have to do that too much in
graduate school.  I should acknowledge in particular his DOE grant 
DE-FG02-07ER41505 and the NSF Graduate Fellowship that he and Pat Boeshaar
were instrumental in helping me win.

Martin Dubcovsky has played a huge part in implementing the algorithms
I describe in this paper, and he took care of a disproportionate part
of the boring and frustrating parts of the code while I did the fun
stuff.  It has been a great experience working with the rest of the
LSST DM team, and I look forward to continuing that collaboration in
the future. 
I'd particularly like to thank Jeff Kantor and Tim Axelrod for always
indulging the grad student with ambitious ideas that didn't fit into
their plans, and Robert Lupton for providing a lot of experience-based
wisdom.  Robert has been a tireless
advocate for astronomers who value their programming skills, and I
thank him for both figuratively and literally helping to carve out a
place for me in the field.

I may not have contributed much to the survey, but there's no doubt my
graduate student experience was greatly 
enhanced by being a part of the DLS group.  David Wittman has been like a
second advisor, and I'd like to thank him in particular for helping me
to focus on practical problems rather than just fun ones.  Perry Gee
did a tremendous amount of work behind the scenes keeping our
computers happy (and perhaps more keeping me happy with our
computers).  I never had a chance to work directly with Chris Roat
much, but a lot of my research is based on what he started.  The
DLS postdocs of the last few years -- Sam Schmidt, James Jee, Russell
Ryan, Paul Thorman, Bego\~{n}a Ascaso -- brought a huge amount of new
life into the group, and I had many interesting discussions with all of
them.  That has also been true for Chris Morrison and Will Dawson, who
kept me on my toes trying to stay ahead of the impressive new
graduate students, and especially my good friend Ami Choi, who did so
many of the same things I did at Davis one step ahead of me, making it
easy to follow her trail.

I probably should have interacted with the rest of the cosmology
group more than I did, but I'd like to thank all the faculty for
making it a great environment for graduate students, and one I'm convinced
is still getting better.  In particular, Chris Fassnacht managed being a
teacher, mentor, 
colleague and friend simultaneously better than anyone I've met.  And
I'd like to thank Andy Albrecht for for his infectious and motivational
love of cosmology -- I always wanted to be a theorist after talking
with him, and I can't think of many other times when I've had that
thought.  In my first few years, I learned a tremendous amount from
the senior graduate students, especially Matt Auger and Michael
Schneider, who introduced me to the Church of Bayes and showed
incredible patience in answering stupid questions.

I should also mention a few faculty members outside the cosmology
group, notably Michael Gertz, Dan Cebra, and Ethan Anderes, who
helped me as mentors, friends, and sounding boards.

Finally, a big thank-you to my parents, Dan and Jane Bosch, for making
science such a big part of our family.  My mom has always had a love
for mathematical puzzles and teaching, and my dad has a ravenous
appetite for discovering and discussing new ideas, and both of
those have played a huge role in helping me find my way in the world.

\end{acknowledgments}

\tableofcontents
\newpage
\begin{inlineabstract}
  \addcontentsline{toc}{chapter}{\abstractname}
  Data analysis methods have always been of critical importance for
  quantitative sciences.  In astronomy, the increasing scale of
  current and future surveys is driving a trend towards a separation
  of the processes of low-level data reduction and higher-level
  scientific analysis.  Algorithms and software responsible for
  the former are becoming increasingly complex, and at the same time
  more general -- measurements will be used for a wide variety of
  scientific studies, and many of these cannot be anticipated in
  advance.  On the other hand, increased sample sizes and the
  corresponding decrease in stochastic uncertainty puts greater
  importance on controlling systematic errors, which must
  happen for the most part at the lowest levels of data analysis.
  Astronomical measurement algorithms must improve in their handling of
  uncertainties as well, and hence must be designed with detailed
  knowledge of the requirements of different science goals.  In this
  thesis, we advocate a Bayesian approach to survey data reduction as
  a whole, and focus specifically on the problem of modeling
  individual galaxies and stars.  We present a Monte Carlo algorithm
  that can efficiently sample from the posterior probability for a
  flexible class of galaxy models, and propose a method for
  constructing and convolving these models using Gauss-Hermite
  (``shapelet'') functions.  These methods are designed to be
  efficient in a multi-epoch modeling (``multifit'') sense, in which
  we compare a generative model to each exposure rather than combining
  the data from multiple exposures in advance.  We also discuss
  how these methods are important for specific higher-level
  analyses -- particularly weak gravitational lensing -- as well as
  their interaction with the many other aspects of a survey reduction
  pipeline.
\end{inlineabstract}

\end{frontmatter}

%

\ssp

\chapter{Introduction and Motivation}
\label{sec:intro}
\section{From Images to Catalogs}
\label{sec:intro:images-to-catalogs}

Vision is the most powerful and well-developed of the human senses,
and our brains excel at difficult pattern matching and image
analysis tasks that still vex our most advanced computer
algorithms.  In many scientific disciplines, however, automated image
processing techniques already perform better than the human eye in
many respects; while the human eye excels at discovery and
classification, computers are far more quantitative and repeatable.
Almost all automated image analysis can be seen as a form of modeling
and/or image reconstruction.  Algorithms generally attempt to divide an image
into parts they can separately understand, and model these
independently in such a way that we can build a theoretical
approximation to the original image.  The success of such approaches
depends crucially on our ability to define reasonable models and
methods to constrain their free parameters with the data.
Astronomical images are vastly simpler than most of the images our
brains process daily, and are usually less complex than most
scientific imaging in other fields, such as medicine or the earth
sciences.  Most of the night sky is dark, and while upon closer inspection
it is full of billions of faint objects, we can represent the
vast majority of these individually using one of two models, one for
stars and one for galaxies.  However,
astronomical images also have a much lower signal-to-noise ratio than
images in other disciplines, so while it is relatively easy to build a
model-based reconstruction of the sky that roughly resembles an image,
it can be much harder to validate that those models reflect reality,
and to solve the inverse problem of constraining those models using
the available data.

Unlike the beautiful images produced for press releases and public
outreach, raw astronomical data is rarely pretty.  In fact, the most
scientifically interesting data is often the ugliest in its raw form,
where detector artifacts and distortions, noise, resolution limits,
and foreground and background contaminants conspire to make the signal
of interest difficult to extract.  In the current era of large,
public surveys, much of the data reduction must be done without a
specific scientific signal in mind; the goal is to generate catalogs
that allow a wide range of studies without recourse to the raw image
data.  While it will always be important for astronomers to fully
understand the processes that generate high-level data products like
catalogs, the technical skills involved in developing survey data
reduction pipelines have diverged somewhat from the more traditional
astronomical skills involved in analyzing catalog data and obtaining
and reducing smaller, more targeted observations.  We should not
consider the first set of skills the domain of engineers and
statisticians, however, as many of the problems are 
unique to astronomy, and astronomers have a long history of developing
algorithms that are useful even beyond their field.  Indeed, many
subfields of astronomy, such as weak gravitational lensing, clearly
require both large surveys and the close attention of practitioners to
the pixel-level data reduction algorithms, though most may be less
interested in the technical details of how those data reduction
algorithms are implemented.

\subsection{Early Automated Data Reduction}

The first modern wide-area astronomical surveys were carried out in the
1950s at the Palomar and Lick Observatories in California,
covering essentially the entire northern sky.  Copies of the Palomar
Observatory 
Sky Survey (POSS) were made available for purchase, giving
birth to a public survey in which outside researchers could make
use of the data in ways unforeseen by the original researchers.
Data access was limited to the plates; there were no official
catalogs, and most analysis was done by eye.  George Abell's original
catalog of 2712 galaxy clusters, published in 1958 and based on a visual
inspection of the POSS plates, is still in wide use today as a
definitive list of massive, nearby clusters.  A large catalog of
galaxies counts from the Lick survey, published by
\citet{ShaneWirtanen1967},  was notable in marking a transition to a
more statistical approach to astronomy; unlike the catalogs of mainly
bright and well-resolved galaxies published earlier, the Lick survey
enabled studies of the spatial distribution of faint galaxies,
revolutionizing our perspective on the structure of the universe.

Producing catalogs from these surveys and their successors by eye was
an incredibly labor-intensive process, however, and the early 1970s
saw the first computer-based methods for automated astronomical data
analysis.  These depended on plate scanning machines, which were
almost immediately capable of much more precise measurements than the
human eye.  More importantly, automated measurements were far more
repeatable and consistent, eliminating a major source of human bias in
the process of measurement. Software for detecting, segmenting, and
measuring the properties of stars and galaxies continued to be developed
throughout the 1970s, and was ultimately packaged into well-defined,
widely-used software systems such as FOCAS \citep{FOCAS} and IRAF in
the early 1980s.  Throughout the 1980s,
NASA took the lead in advocating digital public data, releasing public images
and catalogs from its space-based telescopes and digitizing
photographic plate surveys.  The latter culminated in the Digitized Sky
Survey, a scanned version of POSS and its successors in both the
northern and southern hemispheres.

\subsection{Modern Surveys and ``User Friendly'' Data}

The primary technological limitation throughout the 1970s and 1980s was
detector technology; photographic plates suffered from nonlinearities,
distortions, and low throughput, and early arrays of photodiodes or
photomultiplier tubes suffered from poor resolution and photometric
and astrometric instability.
This changed as larger and less-noisy CCD arrays became available
throughout the 1980s and 1990s, and while CCD-based surveys have yet
to match the 
all-sky coverage of photographic plates at comparable resolutions,
even early generation CCDs produced much higher-quality data with a
much larger dynamic range over
smaller areas.  A key feature of CCD observations was that systematic
effects due to the detector could be calibrated out through repeated
observations; QE variations across the chip did not change between
observations, and thus
could be separated from variations in the sky background.  This
enabled observations to ultimately go much deeper, as multiple
exposures could be combined without being limited by a irreducible
systematic noise floor\citep{Tyson1986}.  Deeper, higher-quality data
demanded improved data reduction 
algorithms and software, and enabled qualitatively new types of
measurements (such as weak gravitational lensing) that were not
possible with photographic plates.

Telescopes with CCD cameras and similar infrared detectors started to
survey large areas of the sky in the middle and late 1990s.  
Surveys such as the 2-Micron All Sky Survey (2MASS, \citealt{2MASS})
and the Sloan Digital Sky Survey (SDSS, \citealt{SDSS}) necessitated a
more ``user-friendly'' approach to public 
data, in which more and more of the low-level data reduction was done
by the survey collaboration and facility, with less done by public ``users''
of the data.  This allowed many reduction steps to be done more 
infrequently, with the results published to the community in the form of
regular data releases.  It also allowed the
algorithms to be tuned by specialists who understood the data best,
freeing other astronomers to focus more on higher-level science questions.
Meanwhile, easy-to-use software packages -- most notably 
SExtractor \citep{SExtractor}, a successor to FOCAS -- became
standards that defined algorithmic choices for many individual
data reduction tasks.

The SDSS SkyServer database \citep{SkyServer} takes this user-friendly
public data concept one step further, providing a queryable database
full of measurements that are more carefully tuned and calibrated than
could be achieved independently using available software like
SExtractor.  While downloading and using
the SDSS image data directly is no longer prohibitively difficult from
a technological standpoint, the quality of the catalogs and the ease
with which users can access them have made image access less
important.  Many papers based entirely on SDSS data use only the
public 
catalogs, and others that do use the SDSS images or data from
other telescopes rely heavily on the database for sample selection,
high-quality measurements, and calibration.  We cannot attribute the
success of the SDSS entirely to its data-reduction and public
interface advantages; it is, after all, the largest optical survey to
date in many respects, and it makes use of world-class instruments.  But
it is fair to state that the usefulness of the survey would have been
greatly diminished without the quality or the ease-of-use of its public
catalogs.

\subsection{Future Challenges}

The data reduction challenge will become
even more acute in the near future, with the advent of even larger
surveys such as that of the Large Synoptic Survey Telescope (LSST,
\citealt{LSST}).\footnote{The public interface also becomes
  considerably more difficult, but we will not address that question
  substantively here.}
The challenge here is both technical and scientific.  The sheer data
volume means the catalogs \emph{must} be more generically useful,
because it is simply impractical to provide the computing power or
network bandwidth for
most users to operate on the pixels directly, and the official nightly
and yearly processing must be highly optimized and highly parallelized
to keep up with the torrent.  Meanwhile, the vastly increased survey
volume makes control of systematic errors much more important;
systematic effects that could previously be ignored because they were
smaller than the measurement uncertainties must now be
addressed.  Algorithms must also be improved because future surveys
will go to much lower surface brightness levels, forcing us to model
faint features that were previously in the noise -- while we have gone
deeper in small areas of the sky, the LSST will cover half the sky
at a depth comparable to all but the deepest space-based observations,
and many of the methods used for low surface brightness studies on
narrow surveys (especially space-based surveys) will not be usable at
those scales.
Furthermore, most of the next generation of surveys are
purely photometric, while the current state of the art, SDSS, is also
a spectroscopic survey.  To bring 
much of the SDSS science to LSST scale, we must improve methods (such
as photometric redshift techniques) that can make up for the lack of
spectroscopic data.  Increased reliance on these algorithms may put
additional requirements on our basic data reduction algorithms as
well.  Finally, for the first time it will become completely
impractical for data quality analysis to rely heavily on human
inspection; we must have automated ways to flag both expected and
unexpected errors in the data.

One of the most fundamental changes from SDSS to the surveys that are
now coming online is that almost all current and future surveys are
multi-epoch surveys.  In
some cases, this means each piece of sky is observed perhaps 5-10 times
in each bandpass.  For LSST, the average patch of sky may be observed
200 times in each band, and some smaller areas may have thousands
of overlapping exposures.  There are some indications from current
medium-size surveys that this large
difference between the single-exposure depth and the full survey depth
requires a qualitatively different approach to object measurement.
Rather than combine images and perform measurements on a coadd, we
may need to model all exposures simultaneously.  This
procedure, called ``multifit'' within the LSST collaboration, is not
new to astronomy -- but in the past it has been mostly limited to small
areas, and generally used only on relatively easy-to-model point
sources.  Early work in multifit shear estimation by
\citet{RoatMultifit} and \citet{ADASSMultifit}, in addition to coining the term, demonstrated
notable improvements under certain conditions.  The multifit approach is
formally superior when speed is not a concern, but it remains to be
seen whether a more careful coadd-based approach can also meet the
requirements of future surveys.

\section{Inverse Problems and Generative Models}
\label{sec:intro:inverse-problems}
\subsection{Astronomy as an Inverse Problem}

While we can improve detectors and build our
telescopes at better sites to minimize the contamination of the scientific
signals we are interested in, astronomy is inherently an
observational science.  We usually do not have the luxury of designing
an experiment to isolate an interesting effect or control 
systematic errors by (for instance) modulating the source; we are
always limited at some level by what nature gives us, and we must
\emph{infer} the signal of interest from that data.

In this situation, the only recourse is to model these contaminants,
from purely observational details like out-of-focus optics, to
inconveniently-positioned celestial bodies that may obscure more
fascinating ones.  Sometimes these models are based on a physical
understanding of the contamination process; often they are derived
from our knowledge of how the process affects other aspects of our
data -- we can use stars to constrain a point spread function (PSF) model,
for instance.  With an appropriate model, we can attempt to remove
these annoyances from our data, by subtracting off the sky background or
deconvolving the PSF.  Astronomical data is inevitably noisy, however,
and while that noise is often well-understood, it still makes this sort of
direct inversion approach nonrobust.

We will advocate a generative model (i.e. \emph{Bayesian}, a term we
will explain more fully in the next chapter) approach to this
problem.  We formally describe the whole system, from the
astrophysical processes to the observatory, as a single model with many
parameters that can reproduce the observed data.  Some parameters
of the model control how photons are 
emitted from astrophysical sources, while others determine how they
make their way through the telescope and all the noise processes that
affect them along the way.  For a given set of parameters
$\bm{\omega}$, then, we can compute the probability of those
parameters given the data $\bm{z}$; we will call this the posterior
probability $P(\bm{\omega}|\bm{z})$.
We can then integrate this distribution over the 
parameters that only characterize the observational system, and
possibly some others that characterize astrophysical objects we consider
unimportant.  The resulting marginalized
distribution is then in essence a quantitative scientific result: it
tells us the probability of some set of physical parameters, given
the data.

Unfortunately, we cannot generally
produce parameterized astrophysical models of the full sky, partly because
we do not understand some processes well enough, and partly because
we expect the initial conditions to be at some level inherently
random, so some astrophysical predictions are always statistical in nature.
Even if we can construct a fully generative astrophysical model of one
patch of the sky, experience tells us that we may be able to obtain
the same scientific result much more efficiently if we compare our
physical predictions to a catalog rather than raw pixels.  In this
context, a catalog is essentially a characterization of the
probability of the parameters of an intermediate model of the sky.
The catalog ``model'' may not be physical, but it is nevertheless
sufficient to represent the sky with enough flexibility to closely
resemble the observed data.  Most importantly, this model should still be
generative; we should still have parameters that represent
the idealized, intrinsic sky and others that represent our observing
conditions and procedures, and marginalize the catalog over the
latter.  This approach has also recently been advocated by
\citet{Hogg2011}, who use a slightly different nomenclature; they
use ``catalog'' to refer to a single set of generative model
parameters, rather than a characterization of the distribution.  This
difference in terminology should not obscure the fact that we
essentially agree.

Most survey reduction pipelines are not consciously designed from the
perspective of a grand generative model, however; they are built by piecing together
well-tested algorithms that have proven their usefulness in producing
robust scientific results in the past.  One such procedure is known as
``detection'' or ``segmentation''; we have mature algorithms that with little
ambiguity 
can divide an image into many small regions, where each is
populated by at most a few distinct astronomical objects.  These
regions can then be modeled almost entirely independently.  From the
generative modeling standpoint, this is the feature that
makes a catalog representation useful: the probability of a model of
the sky can be written as the product of the independent probabilities
of a large number of distinct objects.  In a multi-epoch survey,
the data that correspond to a single astronomical object may be found
on multiple exposures, but it is still easy to identify and separate
from data that ``belongs'' to another distinct object or group of
objects, by detecting and segmenting on a coadd and transforming these
regions.  From a more traditional standpoint, building a catalog can
be viewed as a ``dimensionality reduction'' operation on the image
data; from our perspective, it is perhaps more analogous to noting that
our grand full-survey covariance matrix is almost block-diagonal,
with each block corresponding to the objects that inhabit one
small region of the sky.

\subsection{Modeling Individual Objects}

There are two principal problems in modeling these segmented pixel
regions.  The first is how to parameterize the intrinsic model of an 
astronomical object.  If we have multiple objects in a region, we can
simply fit multiple objects simultaneously, so this is conceptually
no more difficult than fitting a single object.
A model for stars, quasars, and other point-sources is relatively easy
to build.  These objects are almost always
completely unresolved, so their intrinsic model is a delta function,
and they can be parameterized simply by a flux and a centroid.  We can
easily go one step further and add motion parameters, or allow for
variable objects by allowing the flux to be different on different
exposures.  Galaxies are significantly harder.  While they can safely be
considered motionless and nonvariable, their complicated morphologies
are extremely difficult to parameterize.  The huge range in
resolution and signal-to-noise ratio (S/N) between bright, nearby galaxies, and the distant
galaxies that make up the bulk of a typical survey virtually
guarantees that any model flexible enough to fit the former will be
under-constrained by the available data for the latter.  Moreover,
complex models with more parameters are more expensive to fit to the
data, and the computational challenges associated with future surveys
are already considerable.  On the other hand, basing measurements on
idealized models that do not have the flexibility to fit the data will
result in systematic errors due to underfitting biases.  We will not
deal with extended objects other than galaxies in this paper; while
such objects certainly exist, they are rare enough (and different
enough) to require very different approaches.

The second problem in modeling individual astronomical objects is how to
characterize the probability of the intrinsic model parameters in an
efficient way, especially when the data are split across multiple
exposures.  There is no guarantee that these probability distributions
will be well approximated by analytic functions, and they may be highly
asymmetric in some dimensions and possibly multimodal.  We certainly
cannot ignore correlations between different parameters of the model,
or consider it to be safely represented by a covariance term without
justification.  Finally, we also need to marginalize the intrinsic
probability over the observational or ``calibration'' parameters that
determine the PSF, background, astrometric, and photometric models.
These are not well-constrained by the small region of data we use in
fitting an individual object, but they are part of the model as well,
and we cannot in general consider them to be perfectly known.

The goal of this thesis is to address both of these questions: what
models to use when fitting galaxies (and to a lesser extent, stars),
and how to efficiently characterize the marginal posterior of the parameters
that describe those models.  To that end, we motivate
and describe a specific Monte Carlo algorithm and an associated 
class of galaxy models that will
allow multi-exposure modeling of galaxies at LSST scales.  Some aspects,
notably those described in Chapter~\ref{sec:shapelets},
have been fully implemented 
and tested on both real and simulated data.  Most, however, are merely
proposed pieces in a much larger data reduction pipeline that will be
built by a large collaboration over the next decade.  Many of the most
interesting ideas are thus as yet untested, and our goal here is to
provide a detailed formal motivation for them, and sketch out a rough
plan for practical implementation.  It should be emphasized that this
is a description of a particular algorithm and a set of galaxy models
that take advantage of the multifit approach and address some of
its challenges.  We do not suggest that these algorithms or models are
necessary for multi-exposure modeling, but they do help to illustrate
its advantages and disadvantages.

\subsection{Science Goals}

We will try to remain agnostic about the particular science goals --
and even the specific properties of galaxies that we wish to measure
-- for most of the paper, and discuss some of these applications in
more detail in Chapter~\ref{sec:applications}.  Our focus is on the
intermediate, catalog-level models -- after all, the premise that
these models can be considered a fair representation of the sky is
(from the Bayesian perspective) at
the heart of the notion that we can use catalogs rather than raw
pixels to constrain physical parameters.

We will pay particular
attention to the problem of ellipticity measurement for weak
lensing, however.  Understanding dark matter and dark energy using
cosmological weak lensing is one of the most important science
goals for most upcoming wide-area photometric surveys, and it also
puts some of the strictest demands on systematic error control \citep{GREAT08}.  Weak
lensing is also an important tool for understanding galaxy formation
and evolution, as well as the properties of groups and clusters of
galaxies.  The essential inputs to weak lensing analyses are unbiased
ellipticity estimates for faint, distant, and often barely-resolved
galaxies.  The spatial correlation of these ellipticities allows us to
infer the mass in front of the source galaxies, but the effect of
lensing on the ellipticity is usually tiny compared to a galaxy's
intrinsic ellipticity and the modification of the observed ellipticity
due to the PSF.  Carefully making use of a PSF model -- and the
uncertainties in the PSF model -- will be an important part of our efforts.

The choice of galaxy model is also important for ellipticity
estimation, because a model that is not flexible enough to fit the
data can result in an underfitting bias.  Over a limited S/N and
resolution range, simple elliptically-symmetric S\'{e}rsic models
\citep{deVaucouleurs1948,Sersic1963} have 
been very successful at matching galaxy profiles, but they suffer from
serious parameter degeneracies that make them problematic for faint
and poorly-resolved galaxies, and they cannot capture the more
complicated morphologies of better-resolved galaxies.  Basis expansion
techniques, particularly shapelets \cite{BJ02,R03}, have 
likewise done well in capturing more complicated morphologies over a
limited S/N and resolution range, but fail to reproduce the cores and
wings of galaxies well.  Recent work has shown that the systematic
errors associated with using S\'{e}rsic or shapelet models in
weak-lensing analysis are too high for future experiments to achieve
their lensing science goals \citep{VB09,Melchior2010}, simply because
the morphologies of galaxies are systematically different from the
best-fit S\'{e}rsic or shapelet models used to estimate the
ellipticity.

Improved galaxy models will lead to improved
photometry, morphological classifiers, and deblending for galaxies,
which will also play a role in studies of the formation and evolution
of galaxies and larger
structures.  The limiting factor in many of these areas is the quality
of the photometric redshift estimates, which are based on a
higher-level modeling problem than the ones we will describe, but
depend strongly on the quality of the photometry and its
calibration.  And while we will focus more on galaxy modeling, many
aspects of our algorithm also apply to point source models.
In particular, astrometric measurements of very faint stars are only
possible using multi-exposure modeling techniques, and are extremely
important for mapping the structure of the Milky Way \citep{Lang2009}.  Variable point
sources such as supernovae and strongly-lensed quasars are also
important cosmological and astrophysical tools (see,
e.g. \citealt{supernova-cosmology-review,strong-lensing-review}),
and multi-exposure modeling may also have an
role to play in measuring these blended light curves accurately.

Overall, we should view improving survey data reduction techniques as a worthy
goal even without specific science implications.  There are always new
discoveries that we cannot anticipate, and these are most often made
at the very limits of the data where these improvements will have the
most impact.  And even scientific results that do not push these
boundaries should make use of the best possible accounting of
measurement errors and calibration uncertainties.

\section{Measurements on Multi-Exposure Data}

\label{sec:intro:motivation}

The advantage of a generative modeling approach becomes even more pronounced when
the data consist of multiple observations taken under different
conditions.  With a modeling approach, information from multiple
observations can naturally be combined by constructing the probability for
each observation and computing their product.  Without a model,
combining these measurements is significantly more problematic.

The most common procedure for measuring the properties of galaxies in
multi-epoch data is to perform the
measurements on a coadd.  Exposures are resampled to a common pixel grid and
combined to form a single coadd image, which is then treated as a fair
representation of the data of all of the individual exposures.  However,
coadding exposures taken under different observing conditions is
inevitably suboptimal for a number of reasons.  The primary reason is
the fact that there are two different measures of the quality of an
exposure: the signal-to-noise ratio (which is set by the exposure
length and photometric conditions) and the resolution (set by the size
of the PSF).  For realistic data, it is impossible to assign weights to
the exposures in a way that optimizes both of these
quantities.\footnote{A formally optimal method proposed by
  \citet{Kaiser2004} assumes stationary noise, no masked pixels, and a 
  spatially invariant PSF.  This method has (to the author's
  knowledge) not been tested in practice, and the performance of this
  method when these conditions are not met is unknown.}  In fact,
unless each exposure is convolved to match the PSF of the exposure
with the lowest image quality, the coadd will contain discontinuities
in the PSF that can make sources on exposure borders unusable for
PSF-sensitive measurements like those required in gravitational
lensing.  This marks a clear choice on the sacrifices to be made when
making a coadd.  If we choose to match the PSFs when constructing the
coadd, we can sacrifice depth, by removing images with large PSFs, or
sacrifice resolution by using all images.  If we do not choose to
match the PSFs, we can sacrifice area by masking out objects that lie
on exposure borders, or we can accept the systematic errors that may
arise due to a discontinuous PSF.  It is also important to build a PSF
model on the coadd using the PSF model from individual exposures,
rather than attempting to rebuild a PSF model from scratch using the
images of stars on the coadd; because the coadd PSF is not continuous,
its spatial variation cannot be modeled as well by smooth spatial
functions.

Building a coadd also ``locks-in'' various calibrations, making it
difficult to propagate uncertainties in the calibrations into
measurements.  Exposures must be background-subtracted and
resampled based on their astrometric solutions before being combined,
and their weights depend on the photometric calibration for each
exposure.  When exposures are PSF-matched, knowledge of the PSF model
is necessary; if not, it still remains an ingredient
in the PSF model for the coadd.  The uncertainty in these calibration
models -- the background model, PSF model, and astrometric and photometric
solutions -- 
cannot be represented as uncertainties in the coadd pixel values,
despite the fact that these pixel values depend on the calibration
models.  While uncertainties in the per-exposure background,
photometric solutions, and to a lesser extent PSFs can be partially
represented as uncertainties on these same quantities on the coadd,
this propagation of uncertainty is not complete, and uncertainty in
the relative astrometry cannot be represented in this way
at all.  While ignored in virtually all astronomical modeling
approaches to date, uncertainties in calibration can easily dominate
the error budget.  Sometimes these uncertainties can be approximately
propagated and added in quadrature to the final measurement errors,
but this is no substitute for a correct accounting of these important
sources of error.  Estimating uncertainties through simulations is
also an important approach that must be used in concert with Bayesian
methods, but when we discover through simulations that our
methods do not correctly estimate the uncertainty, we should always prefer
to explain and correct those problems directly by fixing our method,
rather than by adding opaque error terms based on the simulation results.

As we have discussed, with a generative modeling approach, calibrations
enter as additional 
parameters, and we can marginalize our model probabilities to account for
uncertainties in these parameters.  In many cases, such as background
or PSF modeling, the effect of a particular calibration parameter will
be limited to the model's realization on a single exposure, and it
will be possible to marginalize over these parameters on the
per-exposure probabilities before forming the full multi-exposure
probability product.

While it should be clear at this point that a multi-epoch modeling
approach has the potential to address many of the downsides of
operating on an image coadd, it presents one big hurdle of its own:
evaluating the model on many exposures is much more expensive than
evaluating it on the coadd.  While building the coadd is also an
expensive process, it is still necessary for detection, even if
we fit simultaneously to all exposures for the final measurements.
Furthermore, more flexible models generally have more parameters,
and the difficulty of fitting a model increases dramatically
as the dimensionality of the problem is increased.  Many of
the parameters used in common galaxy models -- particularly radius,
ellipticity, and profile slope -- are nonlinear and are often highly
degenerate, making the use of traditional least-squares fitting with a
``greedy'' optimizer dubious at best, especially in estimating the
uncertainty of the parameters.  A more robust approach is to use Monte
Carlo methods to sample from the probability of the model, but this
represents a serious computational challenge when each Monte Carlo
point requires an evaluation (and convolution) of the galaxy model on
each of tens or hundreds of exposures.  The best we can do is to
mitigate some of these problems -- in particular, we will make
use of an approximate coadd-based modeling result as much as possible,
and our approach attempts to reduce the number of model evaluations
needed to produce a viable Monte Carlo sample.

\section{Ellipse-Transformed Basis Function Models}

Throughout the rest of this paper, we will focus on galaxy models with
a particular form: our models will consist of a linear combination
of basis functions that are parametrized by an \emph{ellipse transform}.
This geometric transform is a combination of scaling, rotation, and
translation that maps the unit circle at the origin to a particular
ellipse.  For an ellipse with semimajor and semiminor 
axes $a$ and $b$ and position angle $\varphi$, the transform matrix
$\bm{T}$ has the form
\begin{equation}
  \bm{T}[\bm{\phi}] =
  \left[
    \begin{array}{ c c }
      \cos{\varphi} & -\sin{\varphi} \\
      \sin{\varphi} & \cos{\varphi}
    \end{array}
    \right]
  \left[
    \begin{array}{ c c }
      a & 0 \\
      0 & b
    \end{array}
    \right]
  \label{eqn:ellipse-transform-matrix}
\end{equation}
$\bm{T}[\bm{\phi}]$ and the translation vector
$\hat{\bm{\theta}}[\bm{\phi}]$ are parameterized by the 5-element
  vector $\bm{\phi}$, which represents an arbitrary
parameterization of the ellipse (we make no assumption about using
$\{a,b,\varphi\}$, for instance; see
Appendix~\ref{sec:appendix:ellipse-parameterizations} for alternatives).
By transforming the arguments
to the basis functions by $\bm{T}^{-1}$ and $-\hat{\bm{\theta}}$, we
effectively transform the basis 
functions themselves by $\bm{T}$ and $\hat{\bm{\theta}}$, and align
the basis functions with the ellipse defined by $\bm{\phi}$. The full
``above the atmosphere'' model thus has the form 
\begin{equation}
  g[\bm{\theta},\bm{\phi},\bm{\alpha}] = \bigl|\bm{T}[\bm{\phi}]\bigr|
  \sum_{i=1}^N
  \Psi_i\!\left[\bm{T}[\bm{\phi}]^{-1}(\bm{\theta} -
    \hat{\bm{\theta}}[\bm{\phi}])\right]\alpha_i 
  = \sum_{i=1}^N \tilde{\Psi}_i\left(\bm{\theta},\bm{\phi}\right) \alpha_i
  \label{eqn:model-ideal}
\end{equation}
\begin{equation}
  \tilde{\Psi}_i\left[\bm{\theta},\bm{\phi}\right] \equiv
  \bigl|\bm{T}[\bm{\phi}]\bigr|\cdot
  \Psi_i\!\left[\bm{T}[\bm{\phi}]^{-1}(\bm{\theta} -
    \hat{\bm{\theta}}[\bm{\phi}])\right]
\end{equation}
where $\bm{\theta}$ is a pair of angular coordinates, $\Psi_i$
is a set of arbitrary 2-d basis functions, and $\bm{\alpha}$ are the
basis function coefficients (which are also parameters of the model).

To obtain a model image we can compare with data, we must convolve the
model with the PSF and transform from angular coordinates
$\bm{\theta}$ to image coordinates $\bm{x}$:
\begin{align}
  f[\bm{x},\bm{\phi},\bm{\alpha},\bm{\tau},\bm{\beta},\bm{\rho}]
  &= b[\bm{x},\bm{\beta}] + \int\!d\bm{x}^\prime\,k[\bm{x}-\bm{x}^\prime\!,\bm{\rho}]\,
  g\left[\bm{W}[\bm{\tau}](\bm{x}^\prime - \hat{\bm{x}}[\bm{\tau}]),\bm{\phi},\bm{\alpha}\right]
   \\
  &= b[\bm{x},\bm{\beta}]
   + \sum_{i=1}^N 
   \alpha_i\int\!d\bm{x}^\prime\,
   k[\bm{x}-\bm{x}^\prime\!,\bm{\rho}]\,
   \tilde{\Psi}_i\!\left[\bm{W}[\bm{\tau}](\bm{x}^\prime -
     \hat{\bm{x}}[\bm{\tau}]), \bm{\phi}\right]
   \label{eqn:model-observed}
\end{align}
Here we have introduced $\bm{\tau}$, $\bm{\beta}$, and $\bm{\rho}$ as
the parameters of the most important calibration models: the local
image to sky linear transform $\bm{W}$ and offset $\hat{\bm{x}}$, the
sky background model $b$, and the PSF model $k$.  The photometric
calibration model is not represented explicitly, but if we allow the
PSF model to normalize to some factor other than unity, we can
consider the photometric calibration parameters to be included in the
vector $\bm{\rho}$ with no loss of generality.  Convolution is a
linear operation, so we retain the feature that the model is nonlinear
in the ellipse parameters $\bm{\phi}$ and linear in the coefficients
$\bm{\alpha}$.  In Chapter~\ref{sec:modeling}, we will
develop a Bayesian Monte Carlo method that takes advantage of this
form, and in Chapter~\ref{sec:shapelets} we will discuss
how to build an appropriate set of basis functions for modeling
galaxies as well as how to convolve them efficiently.  As we have
mentioned, Chapter~\ref{sec:applications} will address specific
measurements that make use of these models.
Chapter~\ref{sec:implementation} will discuss some technical
implementation details of the algorithm, and place it in the larger
context of a full survey reduction pipeline.  We will discuss a few
other related modeling methods that represent the current
state-of-the-art in Chapter~\ref{sec:related-work}, and conclude in
Chapter~\ref{sec:conclusion}.

\section{Notation}
\noindent
Through the paper, we will adhere to the following conventions for our
mathematical notation:
\begin{itemize}
\item All probability distributions will be written as
  $P(a|b,c)$, with an uppercase $P$ and standard parenthesis, and
  should be read as ``the probability of $a$ given $b$ and $c$''.  This
  is by definition normalized with respect to $a$, and unlike other
  functions, a change in the order of the variables that does not
  cross the ``$|$'' boundary does not change the definition of the
  distribution; $P(a|b,c) = P(a|c,b)$.  We will consistently use the
  same symbols for most parameters, so this will not cause confusion.
\item We will use square brackets, as in $f[a]$, to denote the
  the functional dependency of a quantity, and reserve parentheses
  for grouping and precedence (except, as previously noted, for
  probability distributions).  We will occasionally ignore the
  dependence of some quantities on certain variables when those
  variables can be considered fixed through a large section of the
  text.
\item A boldface lowercase variable such as $\bm{a}$ indicates a
  vector, while a boldface uppercase variable such as $\bm{A}$
  indicates a matrix or occasionally a higher-order tensor.  The
  notation $a_{i}$ or $A_{i,j}$ is used to refer to the
  elements of tensors.
\item We will use
  $\bm{a}_k$ to refer to the $k$-th block of a vector $\bm{a}$; note
  that here $\bm{a}$ is still boldface, in contrast to when we refer
  to a single element of $\bm{a}$.
  We will use a colon ($:$) in the subscript to indicate all blocks
  in one dimension, as in $\bm{A}_{:,k}$, which refers to the
  $k$-th block column of a matrix $\bm{A}$.  We will sometimes use a
  single index to refer to one block of a block-diagonal matrix; if
  $\bm{F}$ is block-diagonal, then $\bm{F}_i\equiv\bm{F}_{i.i}$.
\item We will use $\bm{a}_{[i]}$ to refer to the $i$-th vector in a
  sample of random vectors.
\item We will generally reserve lowercase Greek letters for model
  parameters, and use accents such as $\hat{\alpha}$ and
  $\bar{\alpha}$ to refer to specific instances or estimators of these
  parameters, especially in Chapter~\ref{sec:modeling}.  One notable
  exception is $\bm{\theta}$, which will generally refer to a pair of
  angular coordinates.
\item We will occasionally need to make use of 3-tensor
  products in a context where traditional sum notation would be
  confusing, because we are already using subscript indices for other
  purposes.  Instead, we will use the following notation:
  \begin{align}
    \bm{D} = \bm{C}\!\times\!
    \left\{\begin{smallmatrix} \bm{a} & \circ & \circ \end{smallmatrix}\right\} 
    &\longleftrightarrow D_{j,k} = \sum_{i}C_{i,j,k}\,a_i \\
    \bm{d} = \bm{C}\!\times\!
    \left\{\begin{smallmatrix} \bm{a} & \circ & \bm{b} \end{smallmatrix}\right\} 
    &\longleftrightarrow d_j = \sum_{i,k}C_{i,j,k}\,a_i\,b_k
  \end{align}
  That is, each of the three positions inside the braces indicates one
  dimension of the tensor, and an open circle in one of these
  positions indicates that dimension will not be modified, while a
  variable name indicates an inner product.
  We have no need to form the product of a 3-tensor with anything other
  than a vector.
\end{itemize}
\chapter{Bayesian Modeling of Astronomical Images}

\label{sec:modeling}

\section{Prior and Posterior Probabilities}

\label{sec:modeling:probability-math}

\subsection{Maximum Likelihood and Bayes' Theorem}

In traditional maximum likelihood modeling, the goal is to maximize
the probability of the data ($\bm{z}$,  here) given a model and its
parameters ($\bm{\phi},\bm{\alpha}$): this is the likelihood
$P(\bm{z}|\bm{\phi},\bm{\alpha})$.  When the noise is Gaussian,
this is equivalent to minimizing the familiar $\chi^2$ statistic:
\begin{equation}
  \chi^2 = -2\ln P(\bm{z}|\bm{\phi},\bm{\alpha})
\end{equation}
For most astronomical observations, the per-pixel uncertainties are
indeed very close to Gaussian (and in the raw data are very nearly
uncorrelated, resulting in a diagonal covariance matrix).  Different
sets of model parameter values may not be 
equally likely \emph{a priori}, however, and in Bayesian modeling our
ultimate goal is the \emph{posterior} probability
$P(\bm{\phi},\bm{\alpha}|\bm{z})$, which is related to the likelihood
via Bayes' Theorem:
\begin{equation}
  P(\bm{\phi},\bm{\alpha}|\bm{z}) = P(\bm{z}|\bm{\phi},\bm{\alpha})\frac{P(\bm{\phi},\bm{\alpha})}{P(\bm{z})}
\end{equation}
The key difference here is the \emph{prior} probability of the model
$P(\bm{\phi},\bm{\alpha})$; the \emph{evidence} $P(\bm{z})$ can be
viewed largely as a normalization constant, though it will play a role
in model selection questions addressed later in the paper.
Mathematically, Bayes' Theorem is simply the result of the
standard laws of probability.  The core of Bayesian statistics is
\emph{interpreting} this equality to
mean that the posterior is the ultimate quantity of interest, and that
the prior is a legitimate input.  For a full discussion of Bayesian
methods and a comparison with classical or ``frequentist'' approaches,
we recommend \citet{Sivia2006}. 

The inclusion of a seemingly arbitrary, user-selected prior is
naturally a common criticism of Bayesian techniques, and it may seem
most appropriate to use a flat (constant) prior.  Even a flat prior
will generally be limited to a finite volume of parameter space,
reflecting the desirability of a normalizable prior and the fact that
some combinations of parameter values may not be physical.  When
we have enough information to constrain all the properties of an object,
the posterior should be approximately proportional to the likelihood;
we will want the prior distribution to be much broader than the
likelihood in these cases.  The preferred role of the prior is to
``fill in'' only those properties of the model that are not constrained by
the data, allowing a flexible model to be used with a much greater
range of S/N and resolution.

While it may be much more difficult to construct, it can be highly
advantageous to use a non-flat empirical prior that reflects the
intrinsic distribution of astronomical objects.  In particular, if we
can train an informative prior on particularly high-resolution or deep
data, this prior can provide additional modeling power for
poorly-resolved and faint objects, using the typical properties of
galaxies as observed from space to help infer the properties of
similar galaxies from the ground.  Parameterizing and training such a
prior is a difficult procedure, however, and is largely a subject for
future research.  This is ultimately the best choice
for a prior distribution, and we consider the ability of our algorithm
to make use of an arbitrary prior to be an important feature, even if
we cannot make full use of it today.

\subsection{Calibration Parameters}

The as-observed model of \eqnref{eqn:model-observed} also depends on the
calibration nuisance parameters $\bm{\tau}$, $\bm{\beta}$, and $\bm{\rho}$.  We
must first marginalize over these to obtain
$P(\bm{z}|\bm{\phi},\bm{\alpha})$:
\begin{equation}
  P(\bm{z}|\bm{\phi},\bm{\alpha}) = \int\!d\bm{\tau}\!\int\!d\bm{\beta}\!\int\!d\bm{\rho}\,
  P(\bm{z}|\bm{\phi},\bm{\alpha},\bm{\tau},\bm{\beta},\bm{\rho})\,
  P(\bm{\tau})\,P(\bm{\beta})\,P(\bm{\rho})
  \label{eqn:calibration-marginalization}
\end{equation}
Unlike the weak priors we expect to use for $\bm{\phi}$ and
$\bm{\alpha}$, the nuisance parameter priors will be fairly strong, as
these encode the uncertainty in our calibrations; the common procedure
of holding these parameters fixed is equivalent to a using
delta-function prior.  In truth, these are not priors at all -- they
are more properly posterior probabilities that are conditioned on a
different set of data than what we are using to fit the object at
hand.  For instance, the PSF model is typically built from the images
of many isolated, high S/N stars.  Calling these ``priors'' is really
just a notational device; while we could carry around an extra
symbol in all of our probabilities to mark our dependence on this
external data, it will simplify the notation considerably if we
consider this dependence implicit. To be formally correct we should
also consider the possible overlap of the calibration dataset and the
modeling data -- we may, for instance, want to apply our modeling
algorithm to one of the stars used to construct the PSF model -- but
in all realistic cases the modeling data is a small fraction of the
calibration data, and thus any overlap should be safe to ignore.

\subsection{Monte Carlo Sampling}

Another difference from maximum likelihood techniques is that we want
to draw Monte Carlo samples from the posterior, rather than just find
its maximum.  With a large enough sample size, this allows for a much more
rigorous accounting of uncertainties and model degeneracies.  We will
refer to the random error in a Monte Carlo estimate as the Monte Carlo
variance; it is always inversely proportional to the sample size, but can
affected by other factors as well.  This error is not deterministic --
Monte Carlo errors are intrinsically random.  However, a
Monte Carlo estimator can often be constructed to be unbiased, and
requires no assumption about the form of the distribution.  This makes
Monte Carlo methods more robust, even though they may add additional
stochastic uncertainty to measurements.

If we want to estimate some model property $u[\bm{\phi},\bm{\alpha}]$
($u$ may be the flux, ellipticity, or radius of the model, for instance),
we will want to compute its posterior expectation:
\begin{gather}
  \hat{u} =
  \int\!d\bm{\phi}\!\int\!d\bm{\alpha}\,u[\bm{\phi},\bm{\alpha}]\,P(\bm{\phi},\bm{\alpha}|\bm{z})
  \label{eqn:expectation-integral}
\end{gather}
With a set of Monte Carlo samples
$\{\bm{\phi}_{[n]},\bm{\alpha}_{[n]}\}$ and weights $\{w_{[n]}\}$ drawn
from the posterior $P(\bm{\phi},\bm{\alpha}|\bm{z})$, we can
approximate $\hat{u}$ as
\begin{gather}
  \hat{u} \approx
  \frac{1}{N}\sum_{n=1}^{N}
  w_{[n]} u[\bm{\phi}_{[n]},\bm{\alpha}_{[n]}]
\end{gather}
In fact, the $w_{[n]}$-weighted sample
$\{u[\bm{\phi}_{[n]},\bm{\alpha}_{[n]}]\}$ is effectively drawn from
the posterior $P(u[\bm{\phi},\bm{\alpha}]|\bm{z})$, enabling us to
fully characterize the distribution of any measurement, including its
correlation with other measurements and its confidence limits (with
no assumptions of Gaussianity).

Rather than draw pairs of random vectors
$(\bm{\phi}_{[n]},\bm{\alpha}_{[n]})$, however, we will sample in a
two-stage nested process.  In the outer stage, we draw random vectors
$\bm{\phi}_{[n]}$ with weights $w_{[n]}$.  For each $n$, we will draw
many random vectors $\bm{\alpha}_{[n,m]}$ with weights $w_{[n,m]}$, in the inner
stage.  This is analogous to writing the posterior as a product of
marginal and conditional terms:
\begin{gather}
  P(\bm{\phi},\bm{\alpha}|\bm{z}) =
  P(\bm{\phi}|\bm{z})\,P(\bm{\alpha}|\bm{z},\bm{\phi})
\end{gather}
where the weights are related by
\begin{gather}
  w_{[n]} = \frac{1}{N_m}\sum_{m=1}^{N_m} w_{[n,m]} \approx 
  P(\bm{\phi}_{[n]}|\bm{z}) =
  \int\!d\bm{\alpha}\,P(\bm{\phi}_{[n]},\bm{\alpha}|\bm{z})
  \label{eqn:weights-def-1}
\end{gather}
and the full sum is unity by construction:
\begin{gather}
  \frac{1}{N_n}\sum_{n=1}^{N_n}w_{[n]}=\frac{1}{N_nN_m}\sum_{n=1}^{N_n} \sum_{m=1}^{N_m} w_{[n,m]} =
  \int\!d\bm{\phi}\,P(\bm{\phi}|\bm{z}) = 1
  \label{eqn:weights-def-2}
\end{gather}
Estimators based on this sample will have a higher Monte Carlo variance
than estimators based on a sample that associates a different
$\bm{\phi}$ with each $\bm{\alpha}$, but this nested approach will
allow us to take advantage of the fact that we can draw many
$\bm{\alpha}$ samples at fixed $\bm{\phi}$ much more efficiently. 

In both stages, we will use importance sampling, which uses a set of
random vectors $\bm{x}_{[i]}$ drawn from a known importance distribution
($h$, below) to approximate some target distribution ($p$, below):
\begin{gather}
  \int\!d\bm{x}\,p[\bm{x}] \approx \frac{1}{N}\sum_{i=1}^N w_{[i]} \\
  w_{[i]} = \frac{p[\bm{x}_{[i]}]}{h[\bm{x}_{[i]}]}
\end{gather}
The value of importance sampling lies in the fact that the variance of
the Monte Carlo estimate approaches zero as the
importance distribution approaches proportionality with the target
distribution.  In other words, if we can come up with an analytic
distribution that closely approximates the
distribution we would like to sample from, we can construct
high-quality Monte Carlo estimators, even with relatively small sample
sizes.  The importance sampling estimator is only valid when the
support of the importance distribution contains the support of the
target distribution -- that is, the importance density must be nonzero
everywhere the target density is nonzero.  In practice, it is prudent
to go a bit further, and ensure that the importance distribution has
slightly broader tails than the target distribution.  When the
reverse is true, rare points can have very large weights,
dramatically increasing the variance.

In the next section, we will discuss how to generalize the model to
fit multiple objects simultaneously to data from multiple exposures.
In section~\ref{sec:modeling:marginalizing-calibrations} we will discuss how to
marginalize over the calibration parameters; that is, how to compute the
integral in \eqnref{eqn:calibration-marginalization}.  We will discuss
the inner stage sampling in section~\ref{sec:modeling:inner-sampling},
and the outer stage sampling in 
section~\ref{sec:modeling:outer-sampling}.  We will present the full algorithm in
detail in section~\ref{sec:modeling:algorithm}.

\section{Multiple Objects and Multiple Exposures}

\label{sec:modeling:multifit}

One advantage of writing the model in the form of
\eqnref{eqn:model-ideal} is that it makes it relatively easy to
write compound models that involve fitting multiple nearby objects
simultaneously and/or comparing the model to pixel values from multiple
exposures with different observing conditions.  We begin by pixelizing
\eqnref{eqn:model-observed}, with some abuse of notation for $f$ and $b$:
\begin{align}
  f_i[\bm{\phi},\bm{\alpha},\bm{\tau},\bm{\beta},\bm{\rho}]
  &\equiv
  f[\bm{x}_i,\bm{\phi},\bm{\alpha},\bm{\tau},\bm{\beta},\bm{\rho}]
  = b_i[\bm{\beta}] + \sum_{j=1}^N A_{i,j}[\bm{\phi},\bm{\tau},\bm{\rho}]\,\alpha_j \\
  A_{i,j}[\bm{\phi},\bm{\tau},\bm{\rho}] &\equiv
  \int\!d\bm{x}^\prime\,
  k[\bm{x}_i-\bm{x}^\prime\!,\bm{\rho}]\,
  \tilde{\Psi}_j\!\left[\bm{W}[\bm{\tau}](\bm{x}^\prime - \hat{\bm{x}}[\bm{\tau}]),\bm{\phi}\right] \\
  b_i[\bm{\beta}] &\equiv b[\bm{x}_i,\bm{\beta}]
  \label{eqn:model-pixelized}
\end{align}

We can turn this into a multi-object model simply by forming new
$\bm{\phi}$ and $\bm{\alpha}$ vectors as the direct sum of the
per-object parameter vectors, and applying the same procedure to the
columns of $\bm{A}$:
\begin{align}
  \bm{\phi} &\rightarrow \left[ \begin{array}{ c }
      \bm{\phi}_1 \\ \bm{\phi}_2 \\ \vdots \\ \bm{\phi}_{N_{\mathrm{obj}}}
    \end{array}
    \right] &
  \bm{\alpha} &\rightarrow \left[ \begin{array}{ c }
      \bm{\alpha}_1 \\ \bm{\alpha}_2 \\ \vdots \\ \bm{\alpha}_{N_{\mathrm{obj}}}
    \end{array}
    \right] &
  \bm{A} &\rightarrow \left[ \begin{array}{ c c c c }
      \bm{A}_{:,1} & \bm{A}_{:,2} & \cdots & \bm{A}_{:,N_{\mathrm{obj}}} \\
    \end{array}
    \right]
\end{align}
where $\bm{\phi}_k$ refers to the vector of ellipse parameters for the
$k$-th object. Similarly, we can redefine $\bm{z}$ and $\bm{b}$, along
with the rows of $\bm{A}$ to further extend to multiple exposures:
\begin{align}
  \bm{z} &\rightarrow \left[ \begin{array}{ c }
      \bm{z}_1 \\ \bm{z}_1 \\ \vdots \\ \bm{z}_{N_{\mathrm{exp}}}
    \end{array}
    \right] &
  \bm{b} &\rightarrow \left[ \begin{array}{ c }
      \bm{b}_1 \\ \bm{b}_2 \\ \vdots \\ \bm{b}_{N_{\mathrm{exp}}}
    \end{array}
    \right] &
  \bm{A} &\rightarrow \left[ \begin{array}{ c c c c }
      \bm{A}_{1,1} & \bm{A}_{1,2} & \cdots & \bm{A}_{1,N_{\mathrm{obj}}} \\
      \bm{A}_{2,1} & \bm{A}_{2,2} & \cdots & \bm{A}_{2,N_{\mathrm{obj}}} \\
      \vdots & \vdots & \ddots & \vdots \\
      \bm{A}_{{N_{\mathrm{exp}}},1} &
      \bm{A}_{{N_{\mathrm{exp}}},2} & \cdots &
      \bm{A}_{{N_{\mathrm{exp}}},N_{\mathrm{obj}}} \\
    \end{array}
    \right]
\end{align}

When fitting to multiple exposures, we also need to concatenate
the calibration parameter vectors $\bm{\tau}$, $\bm{\beta}$, and
$\bm{\rho}$ in similar fashion.  Because each per-exposure calibration
vector only affects the model on that exposure, however, we will never
explicitly form the full multi-exposure calibration parameter vectors,
and will develop a procedure to marginalize over them one exposure at
a time in the next section.

Slight complications arise when fitting variable objects or when
fitting to exposures with different filters.  In the first case, the
straightforward solution is to also extend the coefficient vector
$\bm{\alpha}$ and corresponding columns of $\bm{A}$ for each exposure,
so each variable object has a different coefficient vector for each
exposure and one ellipse parameter vector shared across all exposures.
The part of the model matrix $\bm{\alpha}$ that corresponds to a
single variable object would thus be block-diagonal; each set of
coefficients (in columns) only affects a single set of pixels (in
rows).  Typically variable
objects will be point sources, so the size of the per-exposure
coefficient vector is one.  When fitting to multiple filters, we have
the option of having one coefficient vector per object per filter;
this allows the morphology to change slightly as a function of color
while the center, radius and ellipticity of the model remains roughly
constant across filters.  We will discuss these possibilities further
in section~\ref{sec:applications:photometry}.

\section{Handling Calibration Uncertainties}
\label{sec:modeling:marginalizing-calibrations}

The purpose of this section is to derive a method for marginalizing
the likelihood over the calibration parameters $\bm{\tau}$,
$\bm{\beta}$, and $\bm{\rho}$:
\begin{equation}
  P(\bm{z}|\bm{\phi},\bm{\alpha}) = \int\!d\bm{\tau}\!\int\!d\bm{\beta}\!\int\!d\bm{\rho}\,
  P(\bm{z}|\bm{\phi},\bm{\alpha},\bm{\tau},\bm{\beta},\bm{\rho})\,
  P(\bm{\tau})\,P(\bm{\beta})\,P(\bm{\rho})
  \tag{\ref{eqn:calibration-marginalization}}
\end{equation}
A very straightforward solution is to use Monte Carlo importance
sampling, using the prior distribution as the importance
distribution.  This has all the features desired of a good importance
function: because the calibration parameter priors generally provide a
much tighter constraint than the likelihood, the target distribution
(which is the integrand above, the product of the likelihood and the
prior) will be very similar to the importance distribution, up to a
constant factor.  And because the likelihood does provide some small
additional constraint, the importance distribution will generally have
slightly broader tails than the target, ensuring the support
requirements are met.

\subsection{Astrometry Marginalization}

For the astrometric parameters $\bm{\tau}$, this is exactly the
approach we will take.  Because the $\bm{\tau}$ parameters affect the
observed model in almost exactly the same manner as the ellipse parameters
$\bm{\phi}$, it will be most efficient to draw the two in pairs, and
marginalize over $\bm{\tau}$ in the outer stage.  More precisely, for
every random vector $\bm{\phi}_{[n]}$ drawn from the outer stage
importance function on $\bm{\phi}$, we
draw a corresponding random vector $\bm{\tau}_{[n]}$ from
$P(\bm{\tau})$, and both of these are held fixed in the inner stage as
we integrate over the other calibration parameters.
Because the prior $P(\bm{\tau})$ appears in both the numerator and the
denominator of the importance sampling weights, the weights reduce to just
an evaluation of the likelihood:
\begin{gather}
  \int\!d\bm{\tau}\,P(\bm{\tau})\,P(\bm{z}|\bm{\phi},\bm{\alpha},\bm{\tau})
  \approx \frac{1}{N}\sum_{i=1}^N
  \frac{P(\bm{\tau}_{[i]})\,P(\bm{z}|\bm{\phi},\bm{\alpha},\bm{\tau}_{[i]})}{P(\bm{\tau}_{[i]})}
  = \frac{1}{N}\sum_{i=1}^N P(\bm{z}|\bm{\phi},\bm{\alpha},\bm{\tau}_{[i]})
\end{gather}
The integral over $\bm{\tau}$ is thus implicit in the full
outer-stage sample set; we simply ignore the fact that we have a
different random astrometric parameter for each ellipse parameter
vector.  The computation of the integral is not ``free'', however --
we have increased the dimensionality of the space we are sampling, and
to obtain a similar Monte Carlo variance in our estimates we must
increase the outer-stage sample size to compensate.

\subsection{Gaussian Likelihoods and Priors}

We could use a similar procedure to marginalize over the uncertainty in
the background parameters $\bm{\beta}$ and the PSF model parameters
$\bm{\rho}$, but the form of the dependence of the galaxy model on these
parameters allows for an analytic solution when certain reasonable
conditions are met.  One condition is that the priors $P(\bm{\beta})$ and
$P(\bm{\rho})$ be Gaussian; this will rarely be exactly true but will
often be an acceptable approximation.  For the background, we will
also assume a spline, polynomial, or other linear function, so we can
write the background model $\bm{b}$ as
\begin{equation}
  b_i = \sum_{i=1}^{N_{\mathrm{bkg}}} B_{i,j} \,\beta_j
  \longleftrightarrow \bm{b} = \bm{B}\bm{\beta}
\end{equation}
where $\bm{B}$ is the design matrix of the background model.
Similarly, we will require a linear PSF model:
\begin{equation}
  k[\bm{x},\bm{\rho}] = \sum_{i=1}^{N_{\mathrm{psf}}} \Phi_i(\bm{x}) \rho_i
\end{equation}
This still allows for considerable freedom in the PSF model, in that
we do not put any conditions on the basis functions $\Phi$ used to model the
PSF images or the spatial variation of the coefficients (the basis functions are
\emph{not} the same basis functions used to model the galaxies).  The
linearity of the PSF model implies that the galaxy model is also
linear in $\bm{\rho}$:
\begin{align}
  A_{i,j} &= \sum_{k=1}^{N_{\mathrm{psf}}} C_{i,j,k}\,\rho_k
  \longleftrightarrow \bm{A} =
  \bm{C}\!\times\!
  \left\{\begin{smallmatrix}
  \circ & \circ & \bm{\rho}_k
  \end{smallmatrix}\right\}\\
  C_{i,j,k} &\equiv
  \int\!d\bm{x}^\prime\,
  \Phi_k\!\left[\bm{x}_i-\bm{x}^\prime\right]
  \tilde{\Psi}_j\!\left[\bm{W}[\bm{\tau}](\bm{x}^\prime -
    \hat{\bm{x}}[\bm{\tau}]),\bm{\phi}\right]
  \label{eqn:model-convolution-tensor}
\end{align}
Because the model is linear in both $\bm{\beta}$ and $\bm{\rho}$, and
the priors for both are Gaussian, the posterior is also Gaussian in
these parameters, and we will be able to marginalize over them using
analytic Gaussian integrals.

\subsection{Background Marginalization}

We will begin by marginalizing over the background model in the
single-exposure case.  At fixed $\bm{\phi}$ and
$\bm{\tau}$, with pixel values $\bm{z}$
and diagonal pixel covariance matrix $\bm{\Sigma}_z$, the negative log
likelihood is
\begin{align}
  &-\ln P(\bm{z}|\bm{\phi},\bm{\alpha},\bm{\tau},\bm{\beta},\bm{\rho})
  = \frac{1}{2}(\bm{z} - \bm{f}[\bm{\alpha},\bm{\rho},\bm{\beta}])^T\!\bm{\Sigma}_z^{-1}(\bm{z} -
  \bm{f}[\bm{\alpha},\bm{\rho},\bm{\beta}])
   + \frac{1}{2}\ln\left|2\pi\bm{\Sigma}_z\right|
\end{align}
and the marginalization integral is
\begin{equation}
  P(\bm{z}|\bm{\phi},\bm{\alpha},\bm{\tau},\bm{\rho})
  = \int\!
  P(\bm{z}|\bm{\phi},\bm{\alpha},\bm{\tau},\bm{\beta},\bm{\rho})\,
  P(\bm{\beta})\,d\bm{\beta}
\end{equation}
Combining these, the negative log of the integrand is thus
\begin{align}
  L_\beta[\bm{\alpha},\bm{\rho},\bm{\beta}] \equiv&
  -\ln P(\bm{z}|\bm{\phi},\bm{\alpha},\bm{\tau},\bm{\beta},\bm{\rho})
  -\ln P(\bm{\beta})\\
  =& \quad
  \frac{1}{2}(\bm{z}-\bm{f}[\bm{\alpha},\bm{\rho},\bm{\beta}])^T\bm{\Sigma}_z^{-1}
  (\bm{z}-\bm{f}[\bm{\alpha},\bm{\rho},\bm{\beta}])
  +
  \frac{1}{2}(\bm{\beta}-\bar{\bm{\beta}})^T\bm{\Sigma}_\beta^{-1}(\bm{\beta}-\bar{\bm{\beta}}) \nonumber\\
  &\quad+ \frac{1}{2}\ln\left|2\pi\bm{\Sigma}_z\right|
  + \frac{1}{2}\ln\left|2\pi\bm{\Sigma}_\beta\right|
\end{align}
where $\bar{\bm{\beta}}$ and $\bm{\Sigma}_\beta$ are the mean
and covariance of $P(\bm{\beta})$.  We can write $L_\beta$ exactly as
a second-order Taylor series (in $\bm{\beta}$) at $\bar{\bm{\beta}}$:
\begin{align}
  L_\beta[\bm{\alpha},\bm{\rho},\bm{\beta}]
  &= L_\beta[\bm{\alpha},\bm{\rho},\bar{\bm{\beta}}]
  + \bm{c}_\beta[\bm{\alpha},\bm{\rho}]^T(\bm{\beta} - \bar{\bm{\beta}})
  + \frac{1}{2}(\bm{\beta}-\bar{\bm{\beta}})^T
  \bm{H}_\beta(\bm{\beta}-\bar{\bm{\beta}}) \\
  \bm{c}_\beta[\bm{\alpha},\bm{\rho}] &\equiv -\bm{B}^T
  \bm{\Sigma}_z^{-1}(\bm{z}-\bm{f}[\bm{\alpha},\bm{\rho},\bar{\bm{\beta}}])
  \label{eqn:beta-g}
  \\
  \bm{H}_\beta &\equiv \bm{\Sigma}_\beta^{-1} + \bm{B}^T\bm{\Sigma}_z^{-1}\bm{B}
  \label{eqn:beta-h}
\end{align}
The formula for the multidimensional Gaussian integral in this form is:
\begin{equation}
  -\ln\!\int\!e^{-s -
    \bm{c}^T(\bm{x}-\bar{\bm{x}})-\frac{1}{2}(\bm{x}-\bar{\bm{x}})^T\bm{H}(\bm{x}-\bar{\bm{x}})}
  d\bm{x}
     = s + \frac{1}{2}\ln\left|\frac{\bm{H}}{2\pi}\right| - \frac{1}{2}\bm{c}^T\bm{H}^{-1}\bm{c}
  \label{eqn:gaussian-integral}
\end{equation}
so the $\bm{\beta}$-marginalized likelihood is
\begin{align}
  &-\ln P(\bm{z}|\bm{\phi},\bm{\alpha},\bm{\tau},\bm{\rho})
  =
    -\ln\!\int\!d\bm{\beta}\,e^{-L_\beta[\bm{\alpha},\bm{\rho},\bm{\beta}]}
    \\
    &\quad\quad\quad= L_\beta[\bm{\alpha},\bm{\rho},\bar{\bm{\beta}}] 
    + \frac{1}{2}\ln\left|\frac{\bm{H}_\beta}{2\pi}\right|
    -
    \frac{1}{2}\bm{c}_\beta[\bm{\alpha},\bm{\rho}]^T\bm{H}_\beta^{-1}\bm{c}_\beta[\bm{\alpha},\bm{\rho}] \\
    &\quad\quad\quad=
    \frac{1}{2}(\bm{z}-\bm{f}[\bm{\alpha},\bm{\rho},\bar{\bm{\beta}}])^T
    \bm{F}_\beta
    (\bm{z}-\bm{f}[\bm{\alpha},\bm{\rho},\bar{\bm{\beta}}])
    + \frac{1}{2}\ln\left|\bm{H}_\beta\right||\bm{\Sigma}_\beta|
    + \frac{1}{2}\ln\left|2\pi\bm{\Sigma}_z\right| 
    \label{eqn:beta-marginalized}
\end{align}
with
\begin{equation}
    \bm{F}_\beta \equiv
    \bm{\Sigma}_z^{-1} -
    \bm{\Sigma}_z^{-1}\bm{B}\bm{H}_\beta^{-1}\bm{B}^T\bm{\Sigma}_z^{-1}
\end{equation}
As \eqnref{eqn:beta-marginalized} shows, the effect of marginalization
over uncertainty in the background model is fairly intuitive: the
effective pixel noise is increased and correlated.  This adds some
subtlety to our statement in section~\ref{sec:intro:motivation} that the
uncertainty in the calibration models cannot be represented as pixel
uncertainties.  In the case of the background, it can -- but this
correlates the noise in a way that cannot be easily propagated through
the interpolation steps required to build a coadd.

\subsection{PSF Marginalization}

With marginalization over $\bm{\beta}$ complete, the marginalization
integral over $\bm{\rho}$ is
\begin{align}
  P(\bm{z}|\bm{\phi},\bm{\alpha},\bm{\tau}) =
  \int\!d\bm{\rho}\,P(\bm{z}|\bm{\phi},\bm{\alpha},\bm{\tau},\bm{\rho})\,P(\bm{\rho})
  = \int\!d\bm{\rho}\,e^{-L_\rho[\bm{\alpha},\bm{\rho}]}
\end{align}
We will follow the same procedure we used in marginalizing the
background and write $L_\rho$
as a second-order Taylor expansion in $\bm{\rho}$:
\begin{align}
  L_\rho[\bm{\alpha},\bm{\rho}] &\equiv
  -\ln P(\bm{z}|\bm{\phi},\bm{\alpha},\bm{\tau},\bm{\rho})
  -\ln P(\bm{\rho}) \\
  &=\frac{1}{2}(\bm{z}-\bm{f}[\bm{\alpha},\bm{\rho},\bar{\bm{\beta}}])^T
  \bm{F}_\beta
  (\bm{z}-\bm{f}[\bm{\alpha},\bm{\rho},\bar{\bm{\beta}}])
  +
  \frac{1}{2}(\bm{\rho}-\bar{\bm{\rho}})^T\bm{\Sigma}_\rho^{-1}(\bm{\rho}-\bar{\bm{\rho}})
  \nonumber\\
  & \quad\quad\quad + \frac{1}{2}\ln\left|\bm{H}_\beta\right||\bm{\Sigma}_\beta|
  + \frac{1}{2}\ln\left|2\pi\bm{\Sigma}_z\right|
  + \frac{1}{2}\ln\left|2\pi\bm{\Sigma}_{\rho}\right| \\
  &= L_\rho[\bm{\alpha},\bar{\bm{\rho}}] +
  \bm{c}_\rho[\bm{\alpha}]^T(\bm{\rho}-\bar{\bm{\rho}})
  +
  \frac{1}{2}(\bm{\rho}-\bar{\bm{\rho}})^T\bm{H}_\rho[\bm{\alpha}](\bm{\rho}-\bar{\bm{\rho}})
  \intertext{with}
  \bm{c}_\rho[\bm{\alpha}] &\equiv
  \bm{J}[\bm{\alpha}]^T\bm{F}_\beta(\bm{z}-\bm{f}[\bm{\alpha},\bar{\bm{\rho}},\bar{\bm{\beta}}])
  \\
  \bm{H}_\rho[\bm{\alpha}] &\equiv
  \bm{\Sigma}_\rho^{-1} + \bm{J}[\bm{\alpha}]^T\bm{J}[\bm{\alpha}] \\
  \bm{J}[\bm{\alpha}] &\equiv \bm{C}\!\times\!
  \left\{\begin{smallmatrix}
  \circ & \bm{\alpha} & \circ
  \end{smallmatrix}\right\}
\end{align}
We can then insert this result into \eqnref{eqn:gaussian-integral}
to find
\begin{align}
  & -\ln P(\bm{z}|\bm{\phi},\bm{\alpha},\bm{\tau})
  =
  -\ln\!\int\!d\bm{\rho}\,e^{-L_\rho[\bm{\alpha},\bm{\rho}]}
  \\
  &\quad\quad\quad= L_\rho[\bm{\alpha},\bar{\bm{\rho}}] 
  + \frac{1}{2}\ln\left|\frac{\bm{H}_\rho[\bm{\alpha}]}{2\pi}\right|
  -
  \frac{1}{2}\bm{c}_\rho[\bm{\alpha}]^T\bm{H}_\rho[\bm{\alpha}]^{-1}\bm{c}_\rho[\bm{\alpha}]
  \\
  &\quad\quad\quad=
  \frac{1}{2}(\bm{z}-\bm{f}[\bm{\alpha},\bar{\bm{\rho}},\bar{\bm{\beta}}])^T
  \bm{F}_\rho[\bm{\alpha}](\bm{z}-\bm{f}[\bm{\alpha},\bar{\bm{\rho}},\bar{\bm{\beta}}])
  +
  \frac{1}{2}\ln\left|2\pi\bm{\Sigma}_z\right| \label{eqn:rho-marginalized} 
  \nonumber\\
  &\quad\quad\quad\quad\quad\quad\quad\quad\quad
  + \frac{1}{2}\ln\left|\bm{H}_\rho[\alpha]\right|\left|\bm{\Sigma}_\rho\right|
  + \frac{1}{2}\ln\left|\bm{H}_\beta\right|\left|\bm{\Sigma}_\beta\right|
\end{align}
with
\begin{equation}
  \bm{F}_\rho[\bm{\alpha}] \equiv \bm{F}_\beta
  - \bm{F}_\beta\bm{J}[\bm{\alpha}]\bm{H}_\beta[\bm{\alpha}]^{-1}\bm{J}[\bm{\alpha}]^T\bm{F}_\beta
\end{equation}

In this case, $\bm{F}_\rho$ and $\bm{H}_\rho$ are not constant with
respect to $\bm{\alpha}$.  This makes the log likelihood not quite
quadratic in $\bm{\alpha}$, which means marginalization over
$\bm{\rho}$ destroys the Gaussianity of the likelihood in
$\bm{\alpha}$ (a quadratic log likelihood implies a Gaussian
likelihood, and vice-versa).  This non-Gaussianity
is small in the $\bm{\Sigma}_\rho \rightarrow 0$ limit, so it will
generally be an acceptable approximation to simply ignore the
$\bm{\alpha}$-dependency of $\bm{F}_\rho$ and $\bm{H}_\rho$ when the PSF model is
well-constrained (this is analogous to the Fisher matrix
approximation common in cosmological forecasting).  We can also
correct for the non-Gaussianity using the Monte Carlo techniques
developed in the next section, but as we will show, this is
considerably more computationally expensive. 

With that in mind, we will separate \eqnref{eqn:rho-marginalized} into
a quadratic term $L_{\mathrm{g}}$ and a small higher-order correction
term $L_{\mathrm{c}}$.  To simplify the notation, we introduce
\begin{align}
  \bar{\bm{z}} &\equiv \bm{z} - \bm{B}\bar{\bm{\beta}} &
  \bar{\bm{A}} &\equiv \bm{A}[\bar{\bm{\rho}}] &
  \tilde{\bm{F}}_\rho &\equiv \bm{F}_\rho[\tilde{\bm{\alpha}}] &
  \tilde{\bm{H}}_\rho &\equiv \bm{H}_\rho[\tilde{\bm{\alpha}}]
\end{align}
where the fiducial point $\tilde{\bm{\alpha}}$ is the maximum
of $P(\bm{z}|\bm{\phi},\bm{\alpha},\bm{\tau},\bar{\bm{\rho}})$, which
we can solve for by differentiating \eqnref{eqn:beta-marginalized}:
\begin{gather}
  \Dp{}{\bm{\alpha}}\ln
  P(\bm{z}|\bm{\phi},\bm{\alpha},\bm{\tau},\bar{\bm{\rho}})
  = \bar{\bm{A}}^T\!\bm{F}_\beta(\bar{\bm{z}}-\bar{\bm{A}}\bm{\alpha}) = 0 \\
  \left(\bar{\bm{A}}^T\!\bm{F}_\beta\bar{\bm{A}}\right)\tilde{\bm{\alpha}} = \!
  \bar{\bm{A}}^T\!\bm{F}_\beta\bar{\bm{z}}
  \label{eqn:solve-tilde-alpha}
\end{gather}
\Eqnref{eqn:rho-marginalized} can then be written as
\begin{equation}
  -\ln P(\bm{z}|\bm{\phi},\bm{\alpha},\bm{\tau}) =
  L_{\mathrm{g}}[\bm{\alpha}] + L_{\mathrm{c}}[\bm{\alpha}]
\end{equation}
\begin{align}
  L_{\mathrm{g}}[\bm{\alpha}] &\equiv
  \frac{1}{2}(\bar{\bm{z}}-\bar{\bm{A}}\bm{\alpha})^T
  \tilde{\bm{F}}_\rho
  (\bar{\bm{z}}-\bar{\bm{A}}\bm{\alpha})
  + \frac{1}{2}\ln|\bm{H}_\beta||\bm{\Sigma}_\beta|
  + \frac{1}{2}\ln|\tilde{\bm{H}}_\rho||\bm{\Sigma}_\rho|
  + \frac{1}{2}\ln|2\pi\bm{\Sigma}_z|
  \\
  L_{\mathrm{c}}[\bm{\alpha}] &\equiv
  \frac{1}{2}(\bar{\bm{z}}-\bar{\bm{A}}\bm{\alpha})^T
  \left(\bm{F}_\rho[\bm{\alpha}]-\tilde{\bm{F}}_\rho\right)
  (\bar{\bm{z}}-\bar{\bm{A}}\bm{\alpha})
  +
  \frac{1}{2}\ln\frac{|\bm{H}_\rho[\bm{\alpha}]|}{|\tilde{\bm{H}}_\rho|}
  \label{eqn:Lc}
\end{align}
Finally, we expand $L_{\mathrm{g}}$ in a Taylor series in
$\bm{\alpha}$, centered at a new point $\bar{\bm{\alpha}}$ that
eliminates the first-order term:
\begin{gather}
  L_{\mathrm{g}}[\bm{\alpha}] = L_{\mathrm{g}}[\bar{\bm{\alpha}}]
  +
  \frac{1}{2}(\bm{\alpha}-\bar{\bm{\alpha}})^T\!
  \left(\bar{\bm{A}}\tilde{\bm{F}}_\rho\bar{\bm{A}}\right)
  (\bm{\alpha}-\bar{\bm{\alpha}})
  \label{eqn:Lg}
  \\
  \left(\bar{\bm{A}}\tilde{\bm{F}}_\rho\bar{\bm{A}}\right)
  \bar{\bm{\alpha}} =
  \bar{\bm{A}}^T \tilde{\bm{F}}_\rho \bar{\bm{z}}
  \label{eqn:solve-bar-alpha}
\end{gather}
This defines a Gaussian distribution in $\bm{\alpha}$ with mean
$\bar{\bm{\alpha}}$ and covariance
$\left(\bar{\bm{A}}\tilde{\bm{F}}_\rho\bar{\bm{A}}\right)^{-1}\!\!\!\!$, along
with the constant term $L_{g}[\bar{\bm{\alpha}}]$.

\subsection{Multifit Calibration Marginalization}

To marginalize in the
multi-exposure case, we can simply reinterpret all of the above
results in that sense; our pixel indexes will run over all exposures,
and the full calibration parameter vectors $\bm{\beta}$ and $\bm{\rho}$
will be formed from the direct sum of their single-exposure
counterparts.  Because calibration parameters from one exposure only
affect the model on that exposure, the full matrix
$\bm{B}$ and 3-tensor $\bm{C}$ are block-diagonal:
\begin{align}
    \bm{\beta} &\rightarrow \left[ \begin{array}{ c }
      \bm{\beta}_1 \\ \bm{\beta}_2 \\ \vdots \\ \bm{\beta}_{N_{\mathrm{exp}}}
    \end{array}
    \right] &
  \bm{\rho} &\rightarrow \left[ \begin{array}{ c }
      \bm{\rho}_1 \\ \bm{\rho}_2 \\ \vdots \\ \bm{\rho}_{_{\mathrm{exp}}}
    \end{array}
    \right] &
  \bm{B} &\rightarrow \left[ \begin{array}{ c c c c }
      \bm{B}_{1,1} & 0 & 0 & 0 \\
      0 & \bm{B}_{2,2} & 0 & 0 \\
      0 & 0 & \ddots & 0 \\
      0 & 0 & 0 & \bm{B}_{N_{\mathrm{exp}},N_{\mathrm{exp}}} \\
    \end{array}
    \right]
\end{align}
\begin{equation}
  i \ne k \longrightarrow \bm{C}_{i,j,k} = 0  
\end{equation}
This structure ensures that the multi-exposure versions of
$\bm{H}_\beta$, $\bm{F}_\beta$, $\bm{H}_\rho$ and $\bm{F}_\rho$ are
all block-diagonal, with blocks corresponding to different exposures,
as long as there are no covariance terms between exposures in the
calibration parameter priors $\bm{\Sigma}_\beta$ and
$\bm{\Sigma}_\rho$.  As a result, the marginalization integrals for
each exposure are entirely separable.  Instead of constructing the
full multi-exposure matrices and vectors, we can simply construct both
the left-hand sides and the right-hand sides of
\eqnrefand{eqn:solve-tilde-alpha}{eqn:solve-bar-alpha} as the
sum of their single-exposure blocks:
\begin{align}
  \bar{\bm{A}}^T\!\bm{F}_\beta\bar{\bm{A}} &=
  \sum_{i=1}^{N_{\mathrm{exp}}}
  \bar{\bm{A}}_{i,:}^T\!\left(\bm{F}_\beta\right)_{i,i}\bar{\bm{A}}_{i,:}
  &
  \bar{\bm{A}}^T\!\bm{F}_\beta\bar{\bm{z}}
  &= 
  \sum_{i=1}^{N_{\mathrm{exp}}}
  \bar{\bm{A}}_{i,:}^T\!\left(\bm{F}_\beta\right)_{i,i}\bar{\bm{z}}_{i}
  \label{eqn:solve-tilde-alpha-sum}
  \\
  \bar{\bm{A}}^T\!\tilde{\bm{F}}_\rho\bar{\bm{A}} &=
  \sum_{i=1}^{N_{\mathrm{exp}}}
  \bar{\bm{A}}_{i,:}^T\!\left(\tilde{\bm{F}}_\rho\right)_{i,i}\bar{\bm{A}}_{i,:}
  &
  \bar{\bm{A}}^T\!\tilde{\bm{F}}_\rho\bar{\bm{z}}
  &= 
  \sum_{i=1}^{N_{\mathrm{exp}}}
  \bar{\bm{A}}_{i,:}^T\!\left(\tilde{\bm{F}}_\rho\right)_{i,i}\bar{\bm{z}}_{i}
  \label{eqn:solve-bar-alpha-sum}
\end{align}
We can similarly construct $L_g[\bar{\bm{\alpha}}]$ as the sum of the
constant terms of the individual exposures, and solve for $\tilde{\bm{\alpha}}$ and
$\bar{\bm{\alpha}}$ after computing \eqnrefand{eqn:solve-tilde-alpha-sum}{eqn:solve-bar-alpha-sum}, respectively.  This will require
at least two loops over the exposures, of course -- we cannot compute
the latter pair of sums until we have solved for $\tilde{\bm{\alpha}}$
using the results from the first pair of sums.  In practice, we will
actually use one loop in advance to compute $\bm{F}_\beta$ (which does
not depend on $\bm{\phi}$ or $\bm{\tau}$), and then perform two loops
over the exposures for each $(\bm{\phi},\bm{\tau})$ pair.

\section{Inner Sampling and Regularization}
\label{sec:modeling:inner-sampling}

With an analytic expression for
$P(\bm{z}|\bm{\phi},\bm{\alpha},\bm{\tau})$ given by \eqnrefand{eqn:Lc}{eqn:Lg}, we can turn our attention to the inner-sampling integral
\begin{align}
  P(\bm{\phi},\bm{\tau}|\bm{z}) &=
  \int\!d\bm{\alpha}\,P(\bm{\phi},\bm{\tau},\bm{\alpha}|\bm{z}) \\
  &= \frac{1}{P(\bm{z})}\int\!d\bm{\alpha}\,P(\bm{z}|\bm{\phi},\bm{\tau},\bm{\alpha})\,
  P(\bm{\phi},\bm{\tau},\bm{\alpha}) \\
  &= \frac{P(\bm{\phi})\,P(\bm{\tau})}{P(\bm{z})}\int\!d\bm{\alpha}\,
  P(\bm{z}|\bm{\phi},\bm{\tau},\bm{\alpha})\,
  P(\bm{\alpha}|\bm{\phi}) \\
  &= \frac{P(\bm{\phi})\,P(\bm{\tau})}{P(\bm{z})} P(\bm{z}|\bm{\phi},\bm{\tau})
\end{align}
Ignoring for now the terms that don't involve $\bm{\alpha}$, we will
concentrate on the likelihood marginalization:
\begin{align}
  P(\bm{z}|\bm{\phi},\bm{\tau}) &=
  \int\!d\bm{\alpha}\,P(\bm{z}|\bm{\phi},\bm{\tau},\bm{\alpha})\,
  P(\bm{\alpha}|\bm{\phi})
  \label{eqn:inner-likelihood-integral}
\end{align}
We do not wish to just compute the integral itself, however; unlike
the nuisance parameters $\bm{\beta}$ and $\bm{\rho}$, the set of
random vectors and weights used to estimate it are important pieces
of the full nested sample we consider the output of the algorithm.

Our primary goal in this section is to derive an analytic importance
distribution that will allow us to efficiently sample from the
integrand of \eqnref{eqn:inner-likelihood-integral}.  In the previous
section, we wrote the log likelihood as the sum of a term that is
quadratic $\bm{\alpha}$ and a small higher-order term; this is
equivalent to writing the likelihood as the product of an unnormalized
Gaussian and a non-Gaussian function that never differs greatly from
unity:
\begin{align}
  P(\bm{z}|\bm{\phi},\bm{\tau},\bm{\alpha}) =
  e^{-L_g[\alpha]}e^{-L_c[\alpha]} \approx e^{-L_g[\alpha]}
\end{align}
This immediately suggests a normalized Gaussian proportional to
$e^{-L_g[\bm{\alpha}]}$ as an importance function:
\begin{gather}
  h[\bm{\alpha}] = \left|\frac{\bm{H}}{2\pi}\right|^{\frac{1}{2}}
  e^{-\frac{1}{2}(\bm{\alpha}-\bar{\bm{\alpha}})^T\bm{H}(\bm{\alpha}-\bar{\bm{\alpha}})}
  \label{eqn:likelihood-only-importance}
  \\
  \bm{H} \equiv \bar{\bm{A}}^T\tilde{\bm{F}}_\rho\bar{\bm{A}}
\end{gather}
The situation here is exactly the opposite of that for the calibration
parameters discussed at the beginning of the previous section  
-- here, we generally expect the likelihood to provide a much
tighter constraint than the prior, so an importance distribution
proportional to the likelihood will be slightly broader than the
target distribution while still approximating it closely.

Unfortunately, there are common cases in which the likelihood does not
provide a tighter constraint than the prior.  Worse, we cannot
even guarantee that the matrix $\bm{H}$ is
positive definite.  When the ellipse defined by $\bm{\phi}$
is much smaller than the PSF, $\bm{H}$ will often have only one nonzero
eigenvalue, making \eqnref{eqn:likelihood-only-importance} completely
invalid.  This can happen even if the object is
well-resolved -- these conditions occur when we test the 
\emph{hypothesis} that the object is poorly-resolved, regardless of
whether or not it actually is.  A singular $\bm{H}$ 
does not stand in the way of solving for $\bar{\bm{a}}$, as we can use
a full eigendecomposition of $\bm{H}$ to find the minimum-norm
solution:
\begin{gather}
  \bm{H} = \bm{Q} \bm{S} \bm{Q}^T = 
   \left[\begin{array}{ c c }
      \bm{Q}_1 & \bm{Q}_2
    \end{array}
    \right]
  \left[\begin{array}{ c c }
      \bm{S}_1 & 0 \\
      0 & 0
    \end{array}
    \right]
  \left[\begin{array}{ c c }
      \bm{Q}_1 & \bm{Q}_2
    \end{array}
    \right]^T \\
  \bar{\bm{\alpha}} = \bm{Q}_1 \bm{S}_1^{-1}
  \bm{Q}_1^T\!
  \bar{\bm{A}}\tilde{\bm{F}}_\rho\bar{\bm{z}}
  \label{eqn:robust-solve}
\end{gather}
While the minimum-norm solution is not the only solution, it is at
least as good as any other, in that the unconstrained directions do
not contribute to the solution.
We cannot escape the possibility that $|\bm{H}|$ may be zero,
however, making \eqnref{eqn:likelihood-only-importance} an unsuitable
choice for $h[\bm{\alpha}]$.  The singularity of $\bm{H}$ here
is simply a restatement of the classic deconvolution regularization
problem; we simply do not have enough information to constrain all the
parameters of the model when the target object is poorly resolved.

One possible solution lies in the prior term
$P(\bm{\alpha}|\bm{\phi})$.  The prior must be normalized, so its
product with the likelihood is clearly normalizable, and
\eqnref{eqn:inner-likelihood-integral} hence must be finite.  We do
not wish to use the prior alone as our importance distribution,
however -- as we have noted previously, usually the likelihood
provides a much tighter constraint, so the prior would make a very
inefficient importance distribution.  Even when $\bm{H}$ is
singular, the likelihood will generally constrain some coefficient
directions much better than the prior.  With a Gaussian prior, the product
of the likelihood and prior would also be Gaussian, and we could draw
directly from that, but we cannot assume Gaussianity for the
coefficient prior (see below for more discussion on the expected form
of the prior). We could consider a Gaussian approximation to the
prior at $\bar{\bm{\alpha}}$, using a 2nd-order Taylor expansion (the
Fisher matrix approach), but it is quite likely that the prior has
very little curvature at $\bar{\bm{\alpha}}$, and may even be flat.
As a result, the product of the likelihood and a Gaussian
approximation to the prior would still result in a Gaussian defined by
a singular matrix.

Instead, we will focus on constructing our importance distribution using
the likelihood alone, combining a Gaussian distribution with a uniform
distribution we can tune to include most (and hopefully all) of the
support of the prior.  Returning to the diagonalization introduced in
\eqnref{eqn:robust-solve}, we perform an orthogonal change of
variables to an equivalent parameter vector $\bm{\eta}$ that
allows us to separate the nonzero eigenvalues from the zero eigenvalues:
\begin{gather}
  \bm{\eta} \equiv \bm{Q}^T\!\bm{\alpha} = 
  \left[\begin{array}{ c }
      \bm{Q}_1^T\bm{\alpha} \\
      \bm{Q}_2^T\bm{\alpha}
      \end{array}
    \right]
  = 
  \left[\begin{array}{ c }
      \bm{\eta}_1 \\
      \bm{\eta}_2
      \end{array}
    \right]
  \\
  \bm{\alpha} = \bm{Q}\bm{\eta} = \bm{Q}_1\bm{\eta}_1 + \bm{Q}_2\bm{\eta}_2
\end{gather}
In this parameterization, $e^{-L_g[\bm{Q}\bm{\eta}]}$ is
proportional to a Gaussian in $\bm{\eta}_1$, but it has no dependence
on $\bm{\eta}_2$:
\begin{gather}
  e^{-L_g[\bm{Q}^T\!\bm{\eta}]} \propto
  \left|\frac{\bm{S}_1}{2\pi}\right|^{\frac{1}{2}}
  e^{-\frac{1}{2}(\bm{\eta}_1-\bar{\bm{\eta}}_1)^T\bm{S}_1(\bm{\eta}_1-\bar{\bm{\eta}}_1)}
  \label{eqn:h-inner-eta1}
\end{gather}
Clearly, we would like to draw $\bm{\eta}_1$ from this Gaussian, and
draw $\bm{\eta}_2$ from a uniform distribution; the only question is
how to set the limits of the uniform distribution.  Our approach will
be to bound the norm of the unconstrained parameters
using the norm of the constrained parameters:
\begin{gather}
  \lVert\bm{\eta}_2\rVert_\infty \le r\lVert\bm{\eta}_1\rVert_\infty
  \label{eqn:h-inner-eta2}
\end{gather}
where $r$ is a tunable parameter.  For the maximum norm used above,
this defines a hypercube region for 
the uniform distribution on $\bm{\eta}_2$, with volume
$(2r\lVert\bm{\eta}_1\rVert_\infty)^{N_2}$, where $N_2$ is the dimension of
$\bm{\eta}_2$.   We can thus define the
importance function $h_\alpha[\bm{\alpha}]$ as a hybrid Gaussian and uniform
distribution:
\begin{gather}
  h_\alpha[\bm{\alpha}] = \begin{cases}
    (2r\lVert\bm{Q}^T_1\bm{\alpha}\rVert_\infty)^{-N_2}\left|\frac{\bm{S_1}}{2\pi}\right|^{\frac{1}{2}}
    e^{-\frac{1}{2}(\bm{\alpha}-\bar{\bm{\alpha}})^T\bm{S}(\bm{\alpha}-\bar{\bm{\alpha}})}
    & \lVert\bm{Q}_2^T\bm{\alpha}\rVert_\infty \le r\lVert\bm{Q}_1^T\bm{\alpha}\rVert_\infty \\
    0 & \text{otherwise}
  \end{cases}
  \label{eqn:h-inner-full}
\end{gather}
This is Gaussian in $\bm{\eta}_1$ and piecewise uniform in
$\bm{\eta}_2$, with the size of the nonzero piecewise region
set by $\bm{\eta}_1$.  We can draw from $h_\alpha[\bm{\alpha}]$ with
the following procedure:
\begin{enumerate}
\item draw $\bm{\eta}_1$ from the multivariate Gaussian distribution
  defined by \eqnref{eqn:h-inner-eta1};
\item draw $\bm{\eta}_2$ from a uniform distribution on the hypercube
  defined by \eqnref{eqn:h-inner-eta2} (using the random vector
  $\bm{\eta}_1$ on the left-hand side);
\item set $\bm{\alpha}=\bm{Q}_1\bm{\eta}_1 + \bm{Q}_2\bm{\eta}_2$.
\end{enumerate}
In theory, we can set $r$ large enough such that the support of the
product of the likelihood and prior is within the support of
$h_\alpha[\bm{\alpha}]$.  An excessively large $r$ has a large drawback,
however, as most of the random vectors we draw from $h_\alpha[\bm{\alpha}]$
would have negligible probability, increasing the Monte Carlo
variance.  The optimal $r$ is thus tuned to produce an importance
function with a uniform component only slightly larger than the
support of the target distribution.  The crucial question then becomes
whether the size of the uniform part of $h_\alpha[\bm{\alpha}]$ scales correctly.

A good prior for modeling galaxies has to account for two important
aspects of the population we are attempting to model: the distribution
with respect to observed flux must be steep, reflecting the
exponentially larger numbers of faint objects, while the distribution
of morphologies should be largely independent of observed flux.  Both
of these conditions rule out a Gaussian prior.  A change in observed
flux that does not change the morphology of the object is equivalent
to scaling the entire parameter vector; this means any norm of the
coefficient vector can be used as a rough proxy for flux.  Even for
completely unresolved objects, the flux should be well-constrained by
the likelihood, so we do not expect the norm of any ``probable'' full
coefficient vector $\bm{\eta}$ to be significantly different from than
the norm of the likelihood-constrained vector $\bm{\eta}_1$.  This is
exactly the constraint enforced by \eqnref{eqn:h-inner-eta2}.  While 
this ensures that the overall scaling is roughly correct, we also
require that the coefficients typically have the same order of
magnitude -- but this can be easily enforced by normalizing the basis
functions.  The choice of maximum norm is mostly one of convenience;
it is efficient to compute, and it defines a uniform distribution on a
hypercube, which is extremely easy to draw from.  We can easily
generalize \eqnref{eqn:h-inner-eta2} to other norms, as long as they define
a geometry we can easily sample from (for instance, another option
would be the Euclidean norm, which produces a hypersphere).

Using \eqnref{eqn:h-inner-full} as our inner importance distribution, the weights
$v_{[n,m]}$ that estimate \eqnref{eqn:inner-likelihood-integral} with
$\bm{\phi}$ and $\bm{\tau}$ fixed at $\bm{\phi}_{[n]}$ and
$\bm{\tau}_{[n]}$ are
\begin{align}
  v_{[n,m]} &=
  \frac{P(\bm{z}|\bm{\phi}_{[n]},\bm{\tau}_{[n]},\bm{\alpha}_{[n,m]})\,
    P(\bm{\alpha}_{[n,m]}|\bm{\phi}_{[n]})}{h_\alpha[\bm{\alpha}_{[n,m]}]}\\
  &= (2r\lVert\bm{Q}^T_1\bm{\alpha}_{[m]}\rVert_\infty)^{N_2}
  \left|\frac{\bm{S}_1}{2\pi}\right|^{-\frac{1}{2}}
  e^{-L_g[\bar{\bm{\alpha}}]-L_c[\bm{\alpha}_{[n,m]}]} P(\bm{\alpha}_{[n,m]}|\bm{\phi}_{[n]})
\end{align}
These likelihood weights are related to the posterior weights
$w_{[n,m]}$ by
\begin{align}
  w_{[n,m]} = \frac{P(\bm{\phi}_{[n]})\,v_{[n,m]}}{P(\bm{z})\,h_\phi[\bm{\phi}_{[n]}]}
\end{align}
where $h_\phi$ is the outer-stage importance distribution, from which
we have drawn $\bm{\phi}_{[n]}$.  The importance function for
$\bm{\tau}$ is the prior $P(\bm{\tau})$, which would otherwise appear in the
numerator as well and hence cancels out.  The final piece is the evidence
$P(\bm{z})$, whose Monte Carlo estimator is just the normalization constant:
\begin{align}
  P(\bm{z}) \approx \frac{1}{N_nN_m}\sum_{n=1}^{N_n}\sum_{m=1}^{N_m}
  \frac{P(\bm{\phi}_{[n]})\,v_{[n,m]}}{h_\phi[\bm{\phi}_{[n]}]}
\end{align}

\section{Adaptive Importance Sampling}

\label{sec:modeling:outer-sampling}

The target distribution in the outer stage is the
marginalized posterior
\begin{align}
  P(\bm{\phi}|\bm{z}) &=
  \frac{P(\bm{\phi})\,P(\bm{z}|\bm{\phi})}{P(\bm{z})}
\end{align}
Our goal is to find a suitable importance distribution
$h_\phi[\bm{\alpha}]$ that approximates $P(\bm{\phi}|\bm{z})$ and
meets the support requirements.  Unlike the inner stage, we do not
have an analytic form for the target distribution.  While the prior
$P(\bm{\phi})$ may be analytic, it will generally be too broad to make
an efficient importance distribution.  For the likelihood, all we have
is an estimate based on the inner-stage likelihood weights:
\begin{align}
  P(\bm{z}|\bm{\phi}_{[n]}) &\approx
  \frac{1}{N_m}\sum_{m=1}^{N_m} v_{[n,m]}
\end{align}
The outer stage problem is essentially one of ``black box'' sampling:
we wish to draw samples proportional to a function we can only
evaluate at discrete points.  This is much harder than the
nearly-Gaussian inner stage problem, but it is also a much more
widely-studied problem.

The most common solution is Markov-Chain Monte Carlo (MCMC),
which iteratively generates a random walk through parameter space that
is asymptotically distributed from an arbitrary target distribution.
MCMC can take thousands of steps to converge, however, and must be
preceded by a burn-in stage of sampling that should be discarded
entirely.  An alternative
that has recently become popular in 
cosmological modeling is Population Monte Carlo, or adaptive/iterative
importance sampling (see, e.g. \citealt{CosmoPMC}.  As with the
traditional importance sampling we employed in the previous section,
we draw random vectors from an analytic importance distribution and
compute the weight of each random vector as the ratio of the value of the
target distribution at that point to the value of the importance
distribution at that point.  Unlike traditional importance sampling,
we then use this weighted sample to 
update the importance distribution to make it closer to the target
distribution, and repeat.  

For many problems, MCMC and adaptive importance sampling are
comparably efficient.  While MCMC typically requires a long burn-in
phase to ``forget'' its starting position and produced unbiased
results, adaptive importance sampling can require several iterations,
each with large sample sizes, to adapt the importance function to the
target.  Adaptive importance sampling is much better suited to
parallel computing, but this is not a particularly important feature
for our purposes, because we can easily parallelize over other axes
(such as pixels or objects).  While one of the primary advantages of
MCMC in some contexts is the fact that it does not need much knowledge
of the target distribution, it also cannot make use of such knowledge if
available.  In contrast, with adaptive importance sampling we can
construct a low-variance estimator with relatively small sample sizes
if we have a good initial importance distribution.  Likewise, a good
choice for the analytic form of the importance distribution will allow
us to adapt it in fewer iterations and smaller sample sizes.

Applying iterative importance sampling to our outer-stage problem is
not significantly different from other applications of iterative
importance sampling, so we refer the reader to \citet{CosmoPMC} and
references therein for more information on the generic algorithm, and
focus here on the details of applying it in our specific context.
The number of samples required for a 
decent Monte Carlo estimate depends crucially on how close the
importance distribution is to the target distribution.  While a simple
Gaussian or multivariate Student distribution (or mixture thereof) may
be sufficient, it may also be possible to design a custom analytic importance
distribution based on the actual posterior distributions of a sample
of training galaxies.  It should also be noted that our goal need only
be to ensure that the additional variance in our measurements introduced
by the use of Monte Carlo procedures is sufficiently smaller than the
intrinsic variance of those measurements.  We can thus scale the
number of Monte Carlo samples based on the S/N ratio of the
astronomical object we are measuring; for bright, well-resolved
objects we draw many random parameter vectors to accurately
characterize the posterior, while for the vastly more common faint,
poorly-resolved objects we can draw relatively few random vectors
to save time.

However, the real power of iterative importance sampling in this
context lies in the fact that the draw produced by each iteration is
formally independent from the previous ones.  This allows us to
perform some or all of the earlier adaptive steps on an
\emph{approximation} to the target distribution, which may be much
faster to evaluate than the true distribution.  In particular, we can
optimize the importance function using the model posterior on a
coadd, potentially decreasing the time it takes to evaluate the model
by many orders of magnitude.  At the end, we need only perform one
iteration of regular importance sampling on the true, multi-exposure,
calibration-marginalized posterior to obtain a result that is not biased
by any of the peculiarities of the approximate coadd-based posterior.
Of course, we may still have a large variance if this approximate target
distribution is in fact a poor approximation to the true target, so it
may be most efficient to perform more than one iteration on the
multi-exposure posterior, but the ability to perform most iterations
on the coadd remains a huge advantage.

\section{Algorithm}
\label{sec:modeling:algorithm}

\algnewcommand\algorithmicto{\textbf{to}}
\algrenewtext{For}[3]{\algorithmicfor\ \ensuremath{#1=#2}\ \algorithmicto\ \ensuremath{#3}\ \algorithmicdo}

\subsection*{Main Function}
\noindent
Purpose: given pixel data, priors, and initial
importance function, draw weighted random parameter vectors from the posterior
distribution.
In detail, the inputs are:
\begin{itemize}\addtolength{\itemsep}{-0.5\baselineskip}
\item pixel data vector $\bm{z}$ and diagonal covariance matrix
  $\bm{\Sigma}_z$, possibly from multiple exposures;
\item Gaussian prior distributions
  $\mathcal{N}(\bar{\bm{\rho}},\bm{\Sigma}_\rho)$ and 
  $\mathcal{N}(\bar{\bm{\beta}},\bm{\Sigma}_\beta)$ that 
  characterize the uncertainty in the PSF and background models,
  respectively (we will assume we have computed the inverses 
  and determinants of the block-diagonal covariance matrices in
  advance);
\item an analytic prior distribution $P(\bm{\tau})$ that characterizes
  the uncertainty in the astrometric solution, that we can draw
  from directly;
\item the tunable parameter $r$ that sets the size of the uniform part
  of the inner importance function $h_\alpha$;
\item the model parameter prior $P(\bm{\phi},\bm{\alpha})=P(\bm{\phi})\,P(\bm{\alpha}|\bm{\phi})$;
\item initial outer-stage importance distribution $h_\phi[\bm{\phi}]$.
\end{itemize}
The outputs are the nested random vectors
$\{\bm{\phi}_{[n]},\bm{\alpha}_{[n,m]}\}$ and weights $w_{[n,m]}$, as
defined by \eqnrefand{eqn:weights-def-1}{eqn:weights-def-2},
along with an estimate of the evidence $P(\bm{z})$.

This algorithm makes use of a few auxiliary functions, which are
described and listed below the main function.
\vspace{3mm}
\begin{algorithmic}[1]
  \ssp
  \Function{SamplePosterior}{$\bm{z},\bm{\Sigma}_z,\bar{\bm{\beta}},\bm{\Sigma}_\beta,
      \bar{\bm{\rho}},\bm{\Sigma}_\rho,P(\bm{\tau}),P(\bm{\phi},\bm{\alpha}),r,h_\phi[\bm{\phi}]$}
  \State set $\bar{\bm{z}},\bm{F}_\beta,q_\beta=$
  \Call{SetupBackground}{$\bm{z},\bm{\Sigma}_z,\bar{\bm{\beta}},\bm{\Sigma}_\beta$}
  \State set $k=\frac{1}{2}\ln|2\pi\bm{\Sigma}_z| + q_\beta$
  \Repeat
  \State set $t=0$ \Comment{$t$ is an estimate of the evidence $P(\bm{z})$}
  \For{n}{1}{N_n}
    \State draw random vector $\bm{\phi}_{[n]}$ from $h_\phi[\bm{\phi}]$
    \State draw random vector $\bm{\tau}_{[n]}$ from $P(\bm{\tau})$
    \State set $q,\bar{\bm{\eta}}_1,\bm{Q},\bm{S}=$
    \Call{MarginalizeCalibrations}{$\bm{\phi}_{[n]},\bm{\tau}_{[n]},k,\bm{F}_\beta,\bar{\bm{z}},
      \bm{\Sigma}_z,\bar{\bm{\rho}},\bm{\Sigma}_\rho,r$}
    \State set $w_{[n]}=0$
    \For{m}{1}{N_m}
      \State set
      $\bm{\alpha}_{[n,m]},w_{[n,m]}=$\Call{InnerImportance}{$q,\bm{\eta}_1,\bm{Q},\bm{S},r$}
      \State set $w_{[n,m]}=w_{[n,m]} \times
      P(\bm{\phi}_{[n]},\bm{\alpha}_{[n,m]})$
      \State set $w_{[n]} = w_{[n]} + w_{[n,m]}$
    \EndFor
    \State set $t=t+w_{[n]}$
  \EndFor
  \For{n}{1}{N_n}
    \State set $w_{[n]} = w_{[n]} / t$
    \For{m}{1}{N_m}
      \State set $w_{[n,m]} = w_{[n,m]} / t$
    \EndFor
  \EndFor
  \State adapt $h_\phi[\bm{\phi}]$ to $P(\bm{\phi}|\bm{z})$ using $w_{[n]}$ and $\bm{\phi}_{[n]}$
  \Until{$h_\phi[\bm{\phi}]\sim P(\bm{\phi}|\bm{z})$}
  \Comment{i.e. until outer importance function is ``well-adapted''}
  \State return $\{\bm{\phi}_{[n]},w_{[n]}\},\{\bm{\alpha}_{[n,m]},w_{[n,m]}\},t$
  \EndFunction
\end{algorithmic}

\subsection*{Auxiliary Functions}
\noindent
{\sc SetupBackground} computes the background-subtracted pixel vector
$\bar{\bm{z}}$, marginalized pixel Fisher matrix
$\bm{F}_\beta$, and determinant term
$q_\beta=\frac{1}{2}\ln|\bm{H}_\beta||\bm{\Sigma}_\beta|$.
These operations do not depend on the model
parameters, so they can be done once at the beginning of the main algorithm.
\vspace{3mm}
\begin{algorithmic}[1]
  \ssp
  \Function{SetupBackground}{$\bm{z},\bm{\Sigma}_z,\bar{\bm{\beta}},\bm{\Sigma}_\beta$}
  \State set $q=0$
  \For{i}{1}{N_{\mathrm{exp}}}
  \State compute background matrix $\bm{B}_i$ for exposure $i$
  \State set $\bm{H} = \left(\bm{\Sigma}_{\beta}\right)_i^{-1}
      + \bm{B}_i^T\left(\bm{\Sigma}_z\right)^{-1}_i\bm{B}_i$
  \State factor $\bm{H}\rightarrow\bm{R}^T\bm{R}$
  \Comment{$\bm{R}$ upper triangular}
  \State set $q = q + \ln|\bm{R}| + \frac{1}{2}\ln|\left(\bm{\Sigma}_\beta\right)_i|$
  \State set $\bm{W}=\bm{R}^{-1}\bm{B}_i^T\left(\bm{\Sigma}_z\right)^{-1}_i$
  \Comment{back-substitution on $\bm{R}$}
  \State set $\bm{F}_i=\left(\bm{\Sigma}_z\right)^{-1}_i-\bm{W}^T\bm{W}$
  \State set $\bar{\bm{z}}_i = \bm{z}_i - \bm{B}_i\bar{\bm{\beta}}_i$
  \EndFor
  \State return $\bar{\bm{z}},\bm{F}, q$
  \EndFunction
\end{algorithmic}
\vspace{3mm}
\noindent
{\sc MarginalizeCalibrations} generates the parameters corresponding
to a second-order Taylor expansion in $\bm{\alpha}$ of the log
likelihood, marginalized over the background and PSF models: the mean
vector $\bar{\bm{\alpha}}$, the eigenvectors $\bm{Q}$ and eigenvalues
$\bm{S}$ of the Fisher matrix $\bm{H}$, and the zeroth-order scalar
term $q$.  We ignore the small non-Gaussian term $L_c$ in the PSF
marginalization.
\vspace{3mm}
\begin{algorithmic}[1]
  \ssp
  \Function{MarginalizeCalibrations}{$\bm{\phi},\bm{\tau},k,\bm{F}_\beta,\bar{\bm{z}},
    \bm{\Sigma}_z,\bar{\bm{\rho}},\bm{\Sigma}_\rho$}
  \State set $\bm{H}=0$ \Comment{dimension $N_\alpha\times N_\alpha$}
  \State set $\bm{c}=0$ \Comment{dimension $N_\alpha$}
  \State set $q=k$
  \For{i}{1}{N_{\mathrm{exp}}}
    \State compute model tensor $\bm{C}_{i,:,i}$ at $\bm{\phi},\bm{\tau}$\;
    \State set $\bm{A}_{i,:}=\bm{C}_{i,:,i}\!\times\!\{
    \begin{smallmatrix}
      \circ&\circ&\bar{\bm{\rho}}
    \end{smallmatrix}\}$
    \State set $\bm{H}=\bm{H}+\bm{A}_{i,:}^T\left(\bm{F}_\beta\right)_{i,i}\bm{A}_{i,:}$
    \State set $\bm{c}=\bm{c}+\bm{A}_{i,:}^T\left(\bm{F}_\beta\right)_{i,i}\bar{\bm{z}}_i$
    \State set $q=q+\frac{1}{2}\bar{\bm{z}}_i\left(\bm{F}_\beta\right)_{i,i}\bar{\bm{z}}_i$
  \EndFor
  \State diagonalize $\bm{H} \rightarrow
  \bm{Q}\bm{S}\bm{Q}^T=\left[\begin{array}{ c c }
      \bm{Q}_1 & \bm{Q}_2
    \end{array}
    \right]
  \left[\begin{array}{ c c }
      \bm{S}_1 & 0 \\
      0 & 0
    \end{array}
    \right]
  \left[\begin{array}{ c c }
      \bm{Q}_1 & \bm{Q}_2
    \end{array}
    \right]^T$
  \State solve for
  $\tilde{\bm{\alpha}}=\bm{Q}_1\bm{S}_1^{-1}\bm{Q}_1^T\bm{c}$
  \Comment{$\bm{S}_1$ contains nonzero elements of $\bm{S}$}
  \For{i}{1}{N_{\mathrm{exp}}}
    \State set $\bm{J}_i=\bm{C}_{i,:,i}\!\times\!\{
    \begin{smallmatrix}
      \circ&\tilde{\bm{\alpha}}&\circ
    \end{smallmatrix}\}$ 
    \State set $\bm{H}_\rho=\left(\bm{\Sigma}_\rho\right)_i^{-1}+\bm{J}_i^T\bm{J}_i$
    \State factor $\bm{H}_\rho\rightarrow\bm{R}^T\bm{R}$
    \Comment{$\bm{R}$ upper triangular}
    \State set $q=q + \ln|\bm{R}| + \frac{1}{2}\ln|\left(\bm{\Sigma}_\rho\right)_i|$
    \State set $\bm{W}=\bm{R}^{-1}\bm{J}_i^T(\bm{\Sigma}_z)^{-1}_i$
    \Comment{back-substitution on $\bm{R}$}
    \State set $\bm{H}=\bm{H}-\bm{A}_{i,:}^T\bm{W}^T\bm{W}\bm{A}_{i,:}$
    \State set $\bm{c}=\bm{c}-\bm{A}_{i,:}^T\bm{W}^T\bm{W}\bar{\bm{z}}_i$
    \State set $q=q-\frac{1}{2}\bar{\bm{z}}_i\bm{W}^T\bm{W}\bar{\bm{z}}_i$
  \EndFor
  \State diagonalize 
  $\bm{H} \rightarrow \bm{Q}\bm{S}\bm{Q}^T=\left[\begin{array}{ c c }
      \bm{Q}_1 & \bm{Q}_2
    \end{array}
    \right]
  \left[\begin{array}{ c c }
      \bm{S}_1 & 0 \\
      0 & 0
    \end{array}
    \right]
  \left[\begin{array}{ c c }
      \bm{Q}_1 & \bm{Q}_2
    \end{array}
    \right]^T$
  \State solve for $\bar{\bm{\eta_1}}=\bm{S}_1^{-1}\bm{Q}_1^T\bm{c}$
  \State set
  $q=q-\frac{1}{2}\bm{c}^T\bm{Q}_1\bar{\bm{\eta}}_1$
  \State return $q,\bar{\bm{\eta}}_1,\bm{Q},\bm{S}$
  \EndFunction
\end{algorithmic}
\vspace{3mm}
\noindent
{\sc InnerImportance} draws a random vector $\bm{\alpha}$ using the
importance distribution $h_\alpha$ defined by
\eqnref{eqn:h-inner-full}, returning that vector along with the
and the initial (likelihood-only) weights.
\vspace{3mm}
\begin{algorithmic}[1]
  \ssp
  \Function{InnerImportance}{$q,\bar{\bm{\eta}}_1,\bm{Q},\bm{S},r$}
    \State draw $\bm{\eta}_1$ from
    $\mathcal{N}(\bar{\bm{\eta}}_1,\bm{S}_1^{-1})$
    \State set $x = \max[\bm{\eta}_1]$
    \State draw $\bm{\eta}_2$ from a uniform hypercube with corners at
    $\pm r x$
    \State set $\bm{\alpha} = \bm{Q}_1\bm{\eta}_1 +
    \bm{Q}_2\bm{\eta}_2$
    \State set
    $w=(2rx)^{N_2}\left|\frac{\bm{S}_1}{2\pi}\right|^{-\frac{1}{2}}e^{-q}$
    \Comment{$N_2$ is the dimension of $\bm{\eta}_2$}
    \State return $\bm{\alpha}, w$
  \EndFunction
\end{algorithmic}

\chapter[Multi-Scale Elliptical Shapelets]{Multi-Scale Elliptical Shapelets\footnote{Much of this
  chapter was previously published separately in \citet{Bosch2010}.}
}
\label{sec:shapelets}
\defcitealias{NB07}{NB07}
\defcitealias{BJ02}{BJ02}
\section{Introduction}
\label{sec:shapelets-intro}

One of the primary challenges in implementing the algorithm developed
in the previous section is finding a galaxy model ``basis'' that we
can efficiently convolve.  While we can define a model tensor $\bm{C}$
using \eqnref{eqn:model-convolution-tensor} for any
linear PSF model and linear galaxy model, in general we have to
convolve a PSF basis function with a galaxy basis function to compute
each element of the tensor, and repeat this process for every new
nonlinear parameter vector.  This makes a fast convolution
algorithm a virtual necessity.  Models based on a linear combination of
Gauss-Hermite or Gauss-Laguerre functions (also known as Cartesian or
polar ``shapelets'', respectively) have become
particularly important in weak-lensing ellipticity measurement, and
have also been used for morphological classification and in building
realistic mock image data (see, for instance,
\citet{BJ02,R03,MR05,Massey2004,KellyMcKay2004,Andrae2010}).  These
have exactly the sort of convolution relation we require: we can
compute the tensor $\bm{C}$ using fast recurrence relations.

Because the standard shapelet bases consist of polynomials weighted
by a circular Gaussian, they are not a terrible approximation to typical
galaxy and PSF morphologies.  Recent work has
highlighted the drawbacks of standard shapelet-based galaxy 
modeling, however, and demonstrated that even high-order shapelet
expansions are often poor representations of real galaxies.
\citet{Melchior2010} have shown that these deficiencies can introduce
serious systematic ellipticity biases in shapelet-based weak lensing
measurements.  In particular, shapelets cannot reproduce the sharp
core and broad wings of galaxies with high S\'{e}rsic indices, and become
increasingly distorted at high ellipticities.

We propose here a shapelet-based modeling technique that
can much more compactly represent real galaxies, while preserving the
lossless analytic convolution and other useful properties of the
standard shapelet expansion.  By combining multiple low-order shapelet
expansions with different scales, our technique can simultaneously
represent the extended wings and cuspy cores of real galaxies.  We
also present a new convolution relation for
Gauss-Hermite functions that allows the convolution tensor to be
computed exactly when the PSF and galaxy have different
ellipticities.  This eliminates the distortion of the models at high
ellipticity, eliminating one source of shear bias
present in shapelet-based weak lensing methods.

In section~\ref{sec:shapelets:limitations}, we summarize the limitations of
standard shapelet modeling techniques.  In
section~\ref{sec:shapelets:convolution}, we derive an exact convolution relation
for elliptical shapelets, and discuss the combination of multiple
shapelet expansions into a single compound expansion in
section~\ref{sec:shapelets:compound}.  We provide a simple demonstration of
the compound shapelet technique using the \citet{Frei1996} sample of
nearby galaxies in section~\ref{sec:shapelets:demos}.

\section{Limitations of standard shapelet techniques}
\label{sec:shapelets:limitations}

\subsection{Ellipticity}
\label{sec:shapelets:limitations:ellipticity}

The standard Cartesian shapelet basis functions are formed from the
product of two one-dimensional Gauss--Hermite functions:
\begin{align}
  \Phi_{\bm{n}}[\bm{\theta}] &= 
  \frac{H_{n_1}[x_1] H_{n_2}[x_2] e^{-\frac{1}{2}(x_1^2 + x_2^2)}} 
       {\sqrt{2^{n_1+n_2}\,\pi\,n_1!\,n_2!}} .
  \label{eqn:2d-basis}
\end{align}
This is clearly an expansion about a circular Gaussian, and is
naturally poor at representing functions with high ellipticity.
However, we can create an elliptical expansion by transforming the
coordinate grid, just as in \eqnref{eqn:model-ideal}:
\begin{equation}
  \tilde{\Phi}_{\bm{n}}\left[\bm{\theta},\bm{\phi}\right]
  = \Phi_{\bm{n}}[\bm{T}[\bm{\phi}]^{-1}\bm{\theta}]
  \label{eqn:elliptical-shapelets}
\end{equation}
This transformation is exact, but the standard shapelet convolution
formula \citep[see][]{R03} applies only to shapelet expansions with
identical ellipticities; this limits analytic convolution to the
case where both the galaxy and the PSF are approximately circular.

\begin{figure*}
  \begin{center}
    \includegraphics[width=\textwidth]{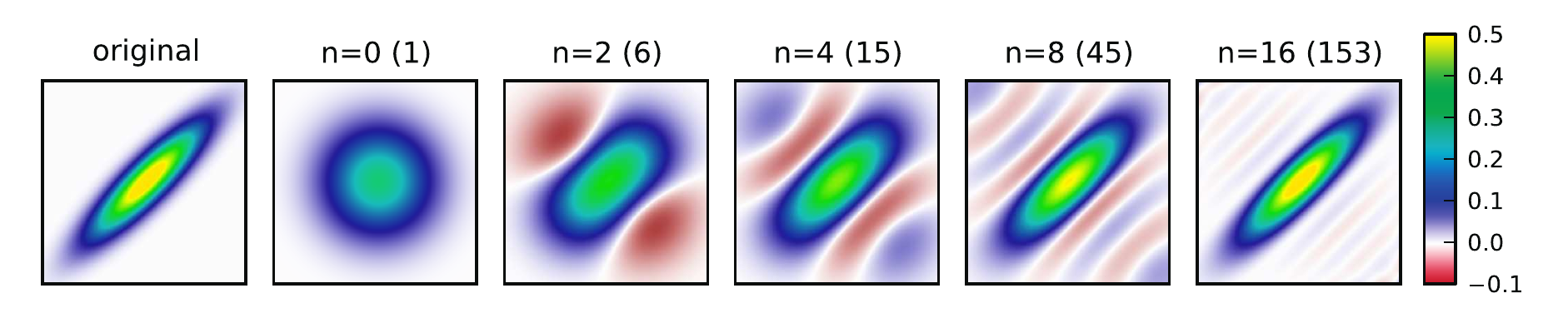}
  \end{center}
  \caption{Circular shapelet approximations of increasing order $n$ to
    an elliptical Gaussian with an axis ratio of $1/4$.  The number of
    basis functions ($\frac{1}{2}(n+1)(n+2)$) used in each approximation
    is given in parenthesis.  Because the number of basis functions
    increases so dramatically at high order, the order of the
    expansion must be kept low for computationally feasible
    techniques, but low-order expansions produce distorted
    representations of even simple high-ellipticity
    morphologies.
    \vspace{2cm}
    \label{fig:ellipticity}}
\end{figure*}

Instead, most shapelet techniques make use of the
shapelet-space shear operator, which can be represented by a matrix
multiplication on the basis vector \citep[see][]{R03,MR05}:
\begin{equation}
  \tilde{\Phi}_{\bm{n}}[\bm{\theta},\bm{\phi}] \approx
   \sum_{\bm{m}}^{\infty}
   \hat{\bm{T}}[\bm{\phi}]_{\bm{n},\bm{m}}
   \Phi_{\bm{m}}(\bm{\theta})
\end{equation}
This relation must be truncated at finite $\bm{m}$ in practice,
however, making the shear operation lossy.  A simple elliptical
Gaussian cannot be represented exactly by a finite circular shapelet
expansion, and the approximation introduces artifacts that become
increasingly severe as the ellipticity increases
(Figure~\ref{fig:ellipticity}).  Unless the average galaxy morphology
mimics these distortion patterns (highly unlikely, given that they
involve regions with negative flux), this necessarily makes the
goodness of fit of shapelet models worse, on average, as ellipticity
increases.  While this is most noticeable at high ellipticities, it is
important even at low ellipticities, as these approximation-induced
artifacts systematically bias shapelet-based lensing techniques
\citep{Melchior2010}.

A better solution, proposed by \citet{NB07} (hereafter
\citetalias{NB07}), is to use elliptical basis functions to model 
the galaxy.  To convolve with the PSF, one applies the inverse
shapelet-space shear operator to the PSF to transform it into the
coordinate system in which the galaxy is round.
However, as the ellipticity of the galaxy model increases, the magnitude of the
inverse shear transform that must be applied to the PSF model also
increases, introducing approximation artifacts in the PSF model.

This trade-off is advantageous for two reasons.  First, one can often
afford to use a higher-order expansion for the PSF, because the PSF
coefficients are not fully free parameters when fitting an
individual galaxy; the PSF model is mostly determined separately using images
of stars in the same field.  In addition, when the galaxy radius is
larger than the PSF, the undistorted galaxy model plays a
larger role in determining the ellipticity of the convolved model than
the now-distorted PSF model.  Still, this technique does not eliminate
the ``shear artifact'' ellipticity bias; it merely decreases it by
substituting a distorted, approximate PSF model for a distorted,
approximate galaxy model.

\subsection{Radial Profiles}
\label{sec:shapelets:limitations:profiles}

\begin{figure}
  \begin{center}
  \includegraphics[width=0.5\textwidth]{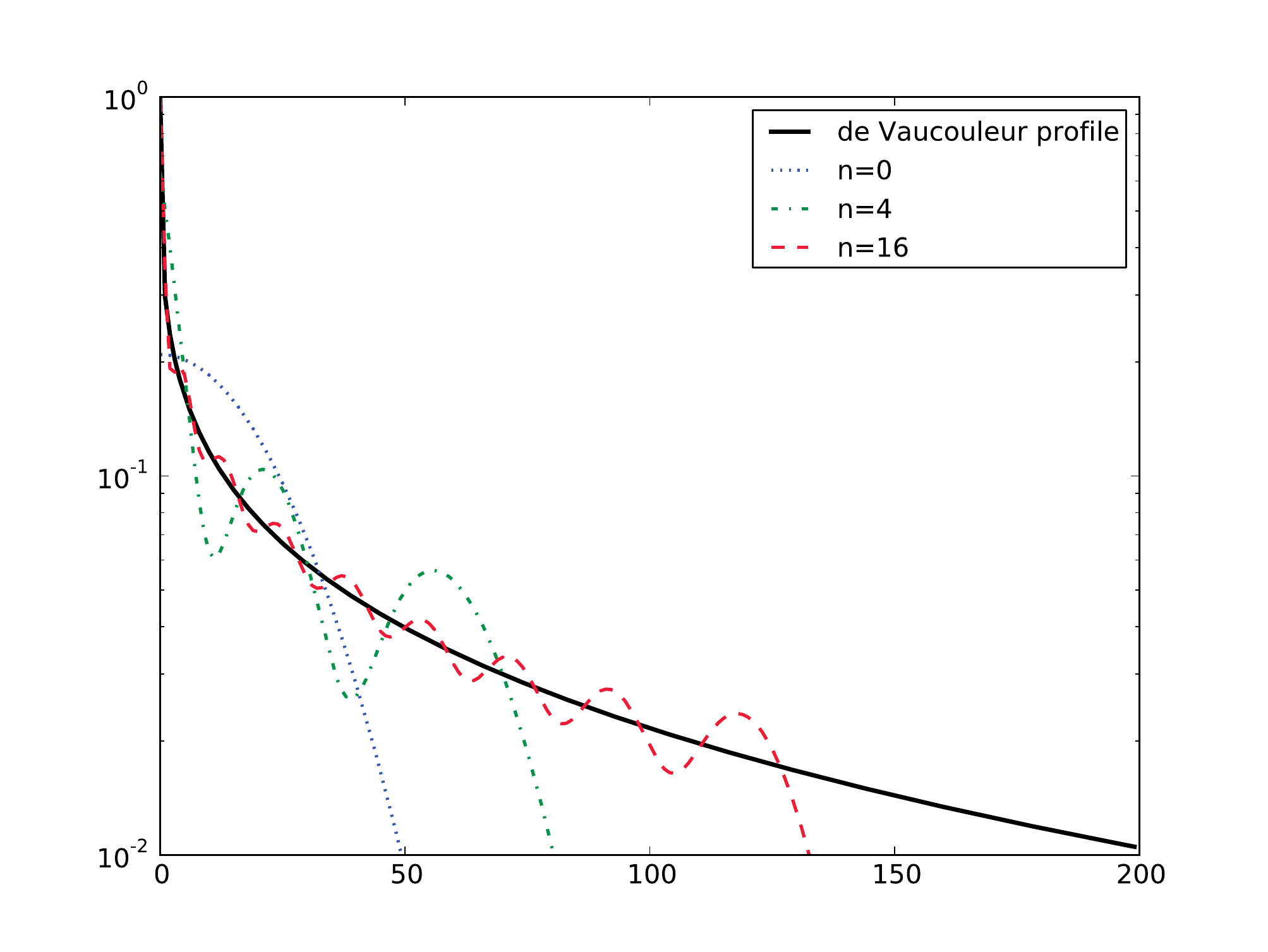}
  \caption{Shapelet approximations to a de Vaucouleurs (S\'{e}rsic $n=4$)
    profile.  The radius and flux units are arbitrary, and the fit can
    be tuned to perform well at either small or moderate radii, but not
    both.  Regardless of the scale of shapelet expansion, it will
    always fall off much faster at large radii compared to a pure de
    Vaucouleurs profile.  This artificial truncation of the radial
    profile strongly affects flux and size estimates made using the
    model, and can bias the measured ellipticity as
    well, even when the truncation occurs at low surface brightness. 
    \label{fig:profiles}} 
  \end{center}
\end{figure}

Shapelet expansions also have difficulty reproducing realistic galaxy
radial profiles.  The azimuthally averaged radial profile of a galaxy often
follows a S\'{e}rsic law:
\begin{equation}
  \mathrm{flux} \propto e^{-r^{1/n}}
\end{equation}
with the \emph{S\'{e}rsic index} $n$ generally greater than one.  Disk
galaxies typically have $n \approx 1$ (an 
exponential profile), while spheroidal galaxies often have $n \approx 4$
(the de Vaucouleurs profile) or greater.  The shapelet expansion is
based on the Gaussian function ($n=1/2$), and hence has a much softer
core and sharper cutoff at large radii than a S\'{e}rsic profile with 
$n\ge 1$.  In theory, the shapelet basis is complete,
and can absorb these differences in higher-order terms, but in
practice a finite shapelet expansion converges to a S\'{e}rsic model with
high $n$ extremely slowly, producing a clear ``ringing'' pattern in
the approximation.  Figure~\ref{fig:profiles} shows a typical case.

This poses a clear problem for the use of shapelets as a tool for
morphological classification: the slope of the radial profile,
arguably the clearest and most obvious distinction between galaxy
types, is a nonlinear function of high-order terms in shapelet space.
In contrast, it can be easily be estimated with a simple S\'{e}rsic fit
or concentration measurement.

The inability to accurately reproduce realistic radial profiles also
has implications for shear measurement.  Ellipticity estimators based
on model fitting have been shown to produce a multiplicative bias when
the model is a systematically poor fit to the data 
\citep{VB09,Bernstein2010}.  Using mock galaxies with S\'{e}rsic profiles,
\citet{Melchior2010} have shown that this ``underfitting bias'' exists for
shapelets even at relatively high orders.  Even at low surface
brightness, modeling the wings of galaxies properly can be very
important in shear estimation, as the wings contain
the low-spatial-frequency shape information that is corrupted least by
convolution with the PSF \citep{Bernstein2010}.

A shapelet-like expansion based on general S\'{e}rsic profiles rather
than Gaussians has been proposed as a possible solution to this
problem \citep{Ngan2009,Andrae2011b}, but these ``S\'{e}rsiclets'' lack many of the
advantages of the shapelet basis, including analytic convolution and
fast numerical evaluation, and thus far have not been used to
construct a practical shear or morphological estimator.

\section{Convolution of elliptical shapelets}
\label{sec:shapelets:convolution}

In this section, we derive an exact analytic convolution formula for
Gauss-Hermite functions.  This provides the ``missing ingredient'' for
making use of the elliptical shapelet basis of
\eqnref{eqn:elliptical-shapelets}
to address the limitations of the standard shapelet basis at high
ellipticity.\footnote{
  An equivalent convolution
  relation for Gauss-Laguerre functions was derived independently in
  \citet{HirataSeljak2003}, but their form is numerically
  unstable at high order, and has been largely
  ignored by subsequent shapelet-based methods.
}  This allows the PSF model and unconvolved galaxy
model to be represented by elliptical shapelets with different basis
ellipses, and determines an optimal basis ellipse for the convolved
shapelet expansion that makes the convolution exact.  

Consider a pair of functions and their convolution, each represented
by a finite elliptical shapelet expansion in image coordinates:

\begin{align}
  f[\bm{x}] &= \sum_{\bm{n}}^{|n|\le N_f} f_{\bm{n}}
  \Phi_{\bm{n}}\!\left[\bm{U}^{-1}\bm{x}\right] \\
  g[\bm{x}] &= \sum_{\bm{n}}^{|n|\le N_g} g_{\bm{n}}
  \Phi_{\bm{n}}\!\left[\bm{V}^{-1}\bm{x}\right] \\
  (f * g)[\bm{x}] &= \sum_{\bm{n}}^{|n|\le N_h} h_{\bm{n}}
  \Phi_{\bm{n}}\!\left[\bm{W}^{-1}\bm{x}\right]
\end{align}

The boldface two-dimensional indices $\bm{n}$ each run over pairs
of one-dimensional indices 
$\{n_1,n_2\}$ as in \eqnref{eqn:2d-basis}, but for conciseness we will
define the ``magnitude'' of a vector index as $|n| \equiv n_1 + n_2$, as
this combination will appear often.

The Fourier transforms of the circular
shapelet basis functions are proportional to the basis functions
themselves \citep{R03}, and the Fourier-space ellipse transform is
simply the transpose of the inverse of the real-space ellipse
transform, so

\begin{align}
  \tilde{f}[\bm{k}] &= \left|\bm{U}\right|\!\sum_{\bm{n}}^{|n|\le N_f} f_{\bm{n}}
  i^{|n|} \Phi_{\bm{n}}\!\left[\bm{U}^T\bm{k}\right] \\
  \tilde{g}[\bm{k}] &= \left|V\right|\!\sum_{\bm{n}}^{|n|\le N_g} g_{\bm{n}}
  i^{|n|}\Phi_{\bm{n}}\!\left[\bm{V}^T\bm{k}\right]\\
  2\pi \tilde{f}[\bm{k}]\,\tilde{g}[\bm{k}] &=
  \left|\bm{W}\right|\!\sum_{\bm{n}}^{|n|\le N_h} h_{\bm{n}} 
  i^{|n|}\Phi_{\bm{n}}\!\left[\bm{W}^T\bm{k}\right] \\ 
  &= 2\pi \left|\bm{U}\right|\left|\bm{V}\right|\!\sum_{\bm{p}}^{|p|\le
    N_f} \sum_{\bm{q}}^{|q|\le N_g}
  i^{|p|+|q|} \Phi_{\bm{p}}\!\left[\bm{U}^T\bm{k}\right]
  \Phi_{\bm{q}}\!\left[\bm{V}^T\bm{k}\right]
\end{align}
We can use the orthogonality of the basis functions to isolate the
coefficients of the convolved expansion by multiplying by
$\Phi_{\bm{m}}\!\left[\bm{W}^T\bm{k}\right]$ and
integrating
\begin{align}
  h_{\bm{n}} &= 2\pi
  \left|\bm{U}\right|\left|\bm{V}\right|\!\sum_{\bm{p}}^{|p|\le N_f}
  \sum_{\bm{q}}^{|q|\le N_g}
  i^{|p|+|q|-|n|} \int \!d^2\bm{k}\,
  \Phi_{\bm{p}}\!\left[\bm{U}^T\bm{k}\right]
  \Phi_{\bm{q}}\!\left[\bm{V}^T\bm{k}\right]
  \Phi_{\bm{n}}\!\left[\bm{W}^T\bm{k}\right]
  \label{eqn:conv-unsimplified}
  \\
  &= 2\pi \left|U\right|\left|V\right|\!\sum_{\bm{p}}^{|p|\le N_f}
  \sum_{\bm{q}}^{|q|\le N_g}
  i^{|p|+|q|-|n|}\;
  I_{\bm{p},\bm{q},\bm{n}}\!\left[\bm{U},\bm{V},\bm{W}\right]
\end{align}
To obtain the model convolution tensor $\bm{C}$ at pixel coordinates
$\{\bm{x}_j\}$, we simply remove the sums over $\bm{p}$ and $\bm{q}$,
and evaluate the expansion
\begin{align}
  C_{j,\bm{p},\bm{q}} =
  2\pi\left|\bm{U}\right|\left|\bm{V}\right|\sum_{\bm{n}}^{|n|\le N_h}
  \Phi_{\bm{n}}[\bm{W}^T\bm{x}_j]\; i^{|p|+|q|-|n|}\;
  I_{\bm{p},\bm{q},\bm{n}}\!\left[\bm{U},\bm{V},\bm{W}\right]
\end{align}
To compute the integral, it is useful to separate the exponential and
polynomial terms
\begin{align}
  I_{\bm{p},\bm{q},\bm{n}}\!\left[\bm{U},\bm{V},\bm{W}\right]
  &=
  \int \!d^2\bm{k}\,
  \Phi_{\bm{p}}\!\left[\bm{U}^T\bm{k}\right]
  \Phi_{\bm{q}}\!\left[\bm{V}^T\bm{k}\right]
  \Phi_{\bm{n}}\!\left[\bm{W}^T\bm{k}\right]\\
  &=
  \int \!d^2\bm{k}\, e^{-\frac{1}{2}\bm{k}^T\left(\bm{U}\bm{U}^T +
    \bm{V}\bm{V}^T +
    \bm{W}\bm{W}^T\right)\bm{k}}\;
  Z_{\bm{p}}\!\left[\bm{U}^T\bm{k}\right]
  Z_{\bm{q}}\!\left[\bm{V}^T\bm{k}\right]
  Z_{\bm{n}}\!\left[\bm{W}^T\bm{k}\right]
  \label{eqn:conv-integral-split}
  \\
  \intertext{with}
  Z_{\bm{p}}\!\left[\bm{U}^T\bm{k}\right] &\equiv
  \Ht_{p_1}\!\left[ U_{11}x_1 + U_{21}x_2 \right]\,
  \Ht_{p_2}\!\left[ U_{12}x_1 + U_{22}x_2 \right] \\
  \Ht_n\!\left[x\right] &\equiv
  \left(\sqrt{\pi}\,2^n\,n!\right)^{-1/2} H_n\!\left[x\right]
  \label{eqn:normalized-hermite-def}
\end{align}
We prefer the ``normalized'' Hermite polynomials $\Ht_n$ here
over the standard Hermite polynomials $H_n$ both because they provide
a more concise notation and because their recurrence relations (see
Appendix~\ref{sec:appendix:normalized-hermite-polynomials}) are more
numerically stable \citep{NumRecipes}.

Returning to \eqnref{eqn:conv-integral-split}, we can simplify the
exponential factor by requiring
\begin{equation}
  \bm{W}\bm{W}^T = \bm{U}\bm{U}^T +
  \bm{V}\bm{V}^T
  \label{eqn:conv-transform}
\end{equation}
This reduces to the familiar formula for the convolution of elliptical
Gaussians at zeroth order; for an elliptical shapelet expansion with
ellipse transform $\bm{S}$, $\bm{S}\bm{S}^T$ is the
covariance matrix of the Gaussian.\footnote{Note that this relation
  between the basis ellipses does not
  assume or require that the ellipticity of the functions we are
  convolving obey the Gaussian convolution rule; while the basis ellipse
  uniquely determines the ellipticity for the zeroth-order term in a
  shapelet expansion, the ellipticity of a higher-order expansion is
  also a function of the coefficients.}  We then make the
substitution $\bm{W}^T\bm{k} \rightarrow \bm{y}$:
\begin{align}
  I_{\bm{p},\bm{q},\bm{n}}\!\left[\bm{U},\bm{V},\bm{W}\right]
  &=
  \int \!d^2\bm{k}\, e^{-\bm{k}^T\bm{W}\bm{W}^T\bm{k}}\;
  Z_{\bm{p}}\!\left[\bm{U}^T\bm{k}\right]
  Z_{\bm{q}}\!\left[\bm{V}^T\bm{k}\right]
  Z_{\bm{n}}\!\left[\bm{W}^T\bm{k}\right] \\
  &= \frac{1}{|W|}\int \!d^2\bm{y}\, e^{-\bm{y}^T\bm{y}}\;
  Z_{\bm{p}}\!\left[\bm{U}^T\bm{W}^{-T}\bm{y}\right]
  Z_{\bm{q}}\!\left[\bm{V}^T\bm{W}^{-T}\bm{y}\right]
  Z_{\bm{n}}\!\left[\bm{y}\right]
  \label{eqn:conv-integral-substituted}
\end{align}

Each normalized Hermite polynomial can be represented as a linear
combination of monomials:
\begin{equation}
  \Ht_n[x] = \sum_m^N M_{n,m} x^m\;,
\end{equation}
where $M$ is the lower-triangular matrix of normalized
Hermite polynomial coefficients.  We can also write a monomial as a
linear combination of normalized Hermite polynomials using the inverse
matrix:
\begin{equation}
  x^m = \sum_n^N M^{-1}_{m,n} \Ht_n[x]\;.
\end{equation}
Along with the binomial theorem, these can be used to factor the
ellipse-transform elements out of the polynomials:
\begin{align}
  Z_{\bm{n}}\!\left[\bm{T}\bm{x}\right]
  &=
  \Ht_{n_1}\!\left[ T_{11}x_1 + T_{12}x_2 \right]\,
  \Ht_{n_2}\!\left[ T_{21}x_1 + T_{22}x_2 \right] \\
  &=
  \sum_{m1}^N \sum_{m2}^{N-m_1} M_{n_1,m_1} M_{n_2,m_2} 
  \left( T_{11}x_1 + T_{12}x_2 \right)^{m_1}\,
  \left( T_{21}x_1 + T_{22}x_2 \right)^{m_2} \\
  &=
  \sum_{m1}^N \sum_{m2}^{N-m_1} \sum_{k_1}^{m_1} \sum_{k_2}^{m_2} \biggl(
  M_{n_1,m_1}M_{n_2,m_2}\,T_{11}^{m_1-k_1}\,T_{12}^{k_1}\,T_{21}^{m_2-k_2}T_{22}^{k_2}\,
  \biggr.\nonumber\\
  & \quad\quad\quad \biggl. \times \;
  \binom{m_1}{k_1}\binom{m_2}{k_2}\,
  x_1^{m_1+m_2-k_1-k_2}\,x_2^{k_1+k_2} \biggr)\\
  &=
  \sum_{m1}^N \sum_{m2}^{N-m_1} \sum_{k_1}^{m_1} \sum_{k_2}^{m_2}
  \sum_{j_1}^N \sum_{j_2}^{N-j_1} \biggl(
  M_{n_1,m_1} M_{n_2,m_2}
  M^{-1}_{m_1+m_2-k_1-k_2,j_1} M^{-1}_{k_1+k_2,j_2}
  \biggr.\nonumber\\
  & \quad\quad\quad \biggl. \times \;
   \binom{m_1}{k_1}\binom{m_2}{k_2} 
  T_{11}^{m_1-k_1}\,T_{12}^{k_1}\,T_{21}^{m_2-k_2}\,T_{22}^{k_2}\,
   \biggr)\Ht_{j_1}\![x_1]\,\Ht_{j_2}\![x_2]\\
  &=
  \sum_{\bm{j}}^{|j|\le N}
  Q_{\bm{n},\bm{j}}\!\left(\bm{T}\right)
  Z_{\bm{j}}\!\left(\bm{x}\right)
  \label{eqn:hermite-transform-matrix}
\end{align}
Plugging \eqnref{eqn:hermite-transform-matrix} into
\eqnref{eqn:conv-integral-substituted}, we have
\begin{align}
  I_{\bm{p},\bm{q},\bm{n}}\!\left[\bm{U},\bm{V},\bm{W}\right]
  &= \frac{1}{|W|}\sum_{\bm{l}}^{|l|\le N_f} \sum_{\bm{m}}^{|m|\le N_g} 
  Q_{\bm{p},\bm{l}}\!\left[\sqrt{2}\bm{U}^T\bm{W}^{-T}\right]
  Q_{\bm{q},\bm{m}}\!\left[\sqrt{2}\bm{V}^T\bm{W}^{-T}\right]\nonumber\\
  &\quad\quad \times \quad \int \!d^2\bm{y}\, e^{-\bm{y}^T\bm{y}}\;
  Z_{\bm{l}}\!\left[\bm{y}/\sqrt{2}\right]
  Z_{\bm{m}}\!\left[\bm{y}/\sqrt{2}\right]
  Z_{\bm{n}}\!\left[\bm{y}\right]\\
  &= \frac{1}{|W|}\sum_{\bm{l}}^{|l|\le N_f} \sum_{\bm{m}}^{|m|\le N_g} 
  Q_{\bm{p},\bm{l}}\!\left[\sqrt{2}\bm{U}^T\bm{W}^{-T}\right]
  Q_{\bm{q},\bm{m}}\!\left[\sqrt{2}\bm{V}^T\bm{W}^{-T}\right]\nonumber\\
  &\quad\quad \times \quad \int \!d^2\bm{y}\, 
  \Phi_{\bm{l}}\!\left[\bm{y}/\sqrt{2}\right]
  \Phi_{\bm{m}}\!\left[\bm{y}/\sqrt{2}\right]
  \Phi_{\bm{n}}\!\left[\bm{y}\right]\\
  &= \frac{2}{|W|}\sum_{\bm{l}}^{|l|\le N_f} \sum_{\bm{m}}^{|m|\le N_g} 
  Q_{\bm{p},\bm{l}}\!\left[\sqrt{2}\bm{U}^T\bm{W}^{-T}\right]
  Q_{\bm{q},\bm{m}}\!\left[\sqrt{2}\bm{V}^T\bm{W}^{-T}\right]\nonumber\\
  &\quad\quad \times \quad 
  B_{l_1,m_1,n_1}\!\left[\sqrt{2},\sqrt{2},1\right]\;
  B_{l_2,m_2,n_2}\!\left[\sqrt{2},\sqrt{2},1\right]
  \label{eqn:convolution-complete}
\end{align}
where $B_{l,m,n}$ is the one-dimensional triple product integral defined by
\citet{RB03}; it can be computed directly with recurrence relations (see
Appendix~\ref{sec:appendix:triple-product-integral}).
The complete formula for convolution of elliptical shapelets is thus
\begin{align}
  h_{\bm{n}} &= 
  \frac{4\pi \left|U\right|\left|V\right|}{\left|W\right|}\!
  \sum_{\bm{p}}^{|p|\le N_f} \sum_{\bm{q}}^{|q|\le N_g}
  \sum_{\bm{l}}^{|l|\le N_f} \sum_{\bm{m}}^{|m|\le N_g} 
  i^{|p|+|q|-|n|}\;
  Q_{\bm{p},\bm{l}}\!\left[\sqrt{2}\bm{U}^T\bm{W}^{-T}\right]
  Q_{\bm{q},\bm{m}}\!\left[\sqrt{2}\bm{V}^T\bm{W}^{-T}\right]\nonumber\\
  &\quad\quad\times\quad B_{l_1,m_1,n_1}\!\left[\sqrt{2},\sqrt{2},1\right]\;
  B_{l_2,m_2,n_2}\!\left[\sqrt{2},\sqrt{2},1\right]
\end{align}
and the model convolution tensor is
\begin{align}
  C_{j,\bm{p},\bm{q}} &= 
  \frac{4\pi \left|U\right|\left|V\right|}{\left|W\right|}\!
  \sum_{\bm{n}}^{|n|\le N_h}
  \sum_{\bm{l}}^{|l|\le N_f} \sum_{\bm{m}}^{|m|\le N_g} 
  i^{|p|+|q|-|n|}\;
  Q_{\bm{p},\bm{l}}\!\left[\sqrt{2}\bm{U}^T\bm{W}^{-T}\right]
  Q_{\bm{q},\bm{m}}\!\left[\sqrt{2}\bm{V}^T\bm{W}^{-T}\right]\nonumber\\
  &\quad\quad\times\quad B_{l_1,m_1,n_1}\!\left[\sqrt{2},\sqrt{2},1\right]\;
  B_{l_2,m_2,n_2}\!\left[\sqrt{2},\sqrt{2},1\right]\;\Phi_{\bm{n}}[\bm{W}^T\bm{x}_j]
\end{align}

Because $B_{l,m,n}\!\left[\sqrt{2},\sqrt{2},1\right]$ is zero for
$n>l+m$, these relations are exact if $N_h \ge N_f + N_g$ and $W$ is
chosen as the solution to equation \eqnref{eqn:conv-transform},
allowing an elliptical shapelet galaxy model to be convolved with an
elliptical shapelet PSF model without approximation. 

\section{Modeling with Compound Shapelets}
\label{sec:shapelets:compound}

In this section we propose a simple solution to
the radial profile problem: instead of increasing the order of a
shapelet expansion in order to fit S\'{e}rsic-like profiles, we create a
compound shapelet expansion that combines multiple low-order shapelet
expansions with different radii.  By adding a low-order expansion
with small radius, a compound basis can represent a realistically cuspy
core.  Likewise, additional expansions with large radii should be able
to compactly represent extended wings without introducing
oscillatory artifacts.

In particular, we will consider compound basis functions of the form
\begin{equation}
  \Psi_n\!\left[\bm{x}\right] = \sum_j^{j \le N_{\beta}} \sum_k^{k \le N^{(j)}_{\Phi}} M^{(j)}_{k,n}\,
  \Phi_k\!\left[\bm{x}/s_j\right]
  \label{eqn:compound-basis}
\end{equation}
The compound basis is thus composed of several
shapelet bases, each with a different relative scale $s_j$.  Each
compound basis function may be a linear combination of several
shapelet basis functions, with weights given by the matrices
$\bm{M}^{(j)}$.  When an ellipse transform is applied, the same
transform is applied to all of the shapelet basis functions that
comprise the compound basis.  A compound basis is thus defined by a
sequence of weight matrices $\bm{M}^{(j)}$ and relative scales $s_j$
(note that all the weight matrices will have the same number of
columns, but may have differing numbers of rows).

\subsection{Properties of the Compound Basis}
\label{sec:shapelets:compound:properties}

Two notable properties of the standard shapelet basis are its
orthonormality and completeness.  Both of these are in general broken
in the construction of a compound basis.
The orthonormality condition for standard shapelets,
\begin{equation}
  \int\!d\bm{x} \,\Phi_{\bm{j}}[x]\,\Phi_{\bm{k}}[x] = \delta_{\bm{j},\bm{k}}\;,
\end{equation}
is defined for a continuous integral with infinite limits.  While this
is an important property for deriving analytic formulae involving
shapelet basis functions, it is of limited practical use in
decomposing image data into shapelets, because the pixelization and
finite limits of image data do not match the idealized orthonormality
condition; even standard shapelet basis functions are not orthonormal
when projected to images \citep{Massey2004}.  As a result, most
standard shapelet decomposition techniques use linear least-squares
methods that do not require orthonormality of the basis
functions.
In deriving analytic formulae (such as the convolution relation) for
the compound basis functions, it is simpler to start with the
analogous formulae for standard shapelets and multiply by the weight matrices $\bm{M}^{(j)}$ than to
attempt to derive these from scratch on a new orthonormal basis.

However, the
inner-stage importance function derived in section
\ref{sec:modeling:inner-sampling} does assume that a reasonable coefficient
vector will have elements that are all within a few orders of
magnitude, and a normalized basis will make this much more
likely.  We can orthonormalize an arbitrary basis by computing the
actual matrix of inner products $\bm{P}$ for an arbitrary compound basis:
\begin{align}
  P_{m,n} &\equiv \int\!d\bm{x} \,\Psi_{m}[x]\,\Psi_{n}[x] \\
  &= \sum_i^{i \le N_{\beta}} \sum_j^{j \le N^{(i)}_{\Phi}}
  \sum_k^{k \le N_{\beta}} \sum_l^{l \le N^{(k)}_{\Phi}}
  M^{(i)}_{j,m}\,
  M^{(k)}_{l,n} \int\!d\bm{x}\,
  \Phi_j\!\left[\bm{x}/s_i\right]\,\Phi_l\!\left[\bm{x}/s_k\right]
  \label{eqn:inner-product-integral}
\end{align}
A diagonal $\bm{P}$ corresponds to an orthogonal basis, and
$\bm{P}=\bm{I}$ corresponds to an orthonormal basis.
The integral in \eqnref{eqn:inner-product-integral} can be computed
using recurrence relations (see
Appendix~\ref{sec:appendix:inner-product-integral}).  We can then
modify the basis by multiplying the
mapping 
matrices $\bm{M}$ by the inverse of a square root of $\bm{P}$.  
For instance, we can use the Cholesky factorization (where $\bm{L}$ is
lower-triangular):
\begin{gather}
  \bm{L}\bm{L}^T = \bm{P} \\
  \bm{M}^{(j)} \rightarrow \bm{M}^{(j)}\bm{L}^{-1}
\end{gather}
The new inner product matrix $\bm{P}$ is then the identity matrix by
construction.

The formal completeness of the standard shapelet basis is also rarely useful in
practice; while an infinite shapelet expansion can perfectly represent
any function in $\mathbb{R}^2$, in practice we must work with finite
expansions.  As we have noted, a finite-size shapelet basis can be
poor at representing the particular functions -- those that mimic
galaxy and PSF profiles -- we are most interested in.  While the
finite-size compound basis thus lacks the
completeness properties of the infinite standard shapelet basis, our
essential argument is that a compound basis can be significantly
more complete than a similarly-sized standard shapelet basis
over the domain of functions we are most interested in.

\subsection{Ellipse Degeneracies}
\label{sec:shapelets:compound:basis-ellipse}

While we have made a point of avoiding shapelet-space geometric
transformation operators because they are not exact at finite shapelet
order, the fact that they exist implies that the parameters
$\bm{\phi}$ that enter through the ellipse transform will be
highly degenerate with those linear combinations of basis coefficients
$\bm{\alpha}$ that produce shapelet-space translation, shear, and
scaling transforms.  For a compound basis composed of low-order
shapelet expansions, these degeneracies are reduced, because the basis
functions poorly approximate the ellipse transform.  However, even
first- or second-order shapelet components will generally make the
model too flexible, as first-order terms approximate a
centroid shift, and second-order terms approximate changes in
ellipticity and size. 

\citet{BJ02} propose determining their
equivalent of the ellipse 
parameters using a ``null test'' algorithm, in which
the $\bm{\phi}$ parameters are iterated until the resulting
expansion has zero centroid, zero ellipticity, and unit size in the
ellipse-transformed coordinate system.  This is equivalent to the
linear constraint 
\begin{align}
  \sum_n K_{m,n} \alpha_n &= 0
  \label{eqn:ellipse-constraint}
\end{align}
with
\begin{align}
  K_{1,n} &= \int d^2\bm{\theta}\, \theta_1\, \Phi_{n}\![\bm{\theta}]&
  K_{2,n} &= \int d^2\bm{\theta}\, \theta_2\, \Phi_{n}\![\bm{\theta}]\nonumber\\
  K_{3,n} &= \int d^2\bm{\theta}\, (\theta_1^2 - \theta_2^2)\, \Phi_{n}\![\bm{\theta}]&
  K_{4,n} &= \int d^2\bm{\theta}\, \theta_1 \theta_2\, \Phi_{n}\![\bm{\theta}]\nonumber\\
  K_{5,n} &= \int d^2\bm{\theta}\, (\theta_1^2+ \theta_2^2-1)\, \Phi_{n}\![\bm{\theta}]
\end{align}
These can generally be computed using recurrence
relations.\footnote{\citet{BJ02} actually use weighted moments, which are
  necessary when operating directly on images, but should be less
  important when operating on a model (which imposes its own effective
  weight function).  We will return to the question of weight
  functions in section~\ref{sec:applications:shear-estimation}.}

We can construct the analogous constraint matrix for a compound shapelet basis
simply by replacing $\Phi$ with $\Psi$ above.  Rather than applying the
constraint when fitting individual galaxies, however, we can instead
``circularize'' the basis itself by modifying the weight
matrices $\bm{M}$ in equation \eqnref{eqn:compound-basis}.  Given a set of
input basis functions $\Psi_n(\bm{\theta})$ 
and its constraint matrix $\bm{K}$, we first compute the singular
value decomposition of $\bm{K}^T$:
\begin{equation}
  \bm{U} \bm{D} \bm{V}^T = \bm{K}^T\;.
\end{equation}
The diagonal matrix $\bm{D}$ will have at most five non-zero
elements, corresponding to the first five columns of 
$\bm{U}$.  The remaining columns of $\bm{U}$ give the mapping from
the original basis to the circularized basis:
\begin{equation}
  \Psi^c_n(\bm{\theta}) = \sum_m \Phi_m(\bm{\theta}) \; U_{m,n+5} 
\end{equation}
or
\begin{equation}
  M^c_{k,n} = \sum_m M_{k,m} \; U_{m,n+5}\;.
\end{equation}
The resulting basis is naturally five elements smaller than the input
basis, and its constraint matrix is the zero matrix, so
\eqnref{eqn:ellipse-constraint} is always true.

While a ``circularized'' basis may be useful when using a greedy optimizer to
maximize the likelihood or posterior, ellipse degeneracies in a basis are
actually desirable in our Monte Carlo algorithm.  Because we can
change the centroid, radius, or ellipticity of the model without
changing the ellipse transform parameters, when we evaluate the model
at a single ellipse point $\bm{\phi}$, we can actually
represent a range of ``effective'' ellipses with variation in
$\bm{\alpha}$.  This allows us to 
more sparsely sample $\bm{\phi}$-space, as $\bm{\phi}$ points that are
not sampled can be approximated by different combinations of
$\bm{\alpha}$ parameters at nearby sample points.  We can make this
even more effective by ensuring that the ellipse derivatives of any
particularly important basis functions can be represented as a linear
combination of other basis functions -- this allows the basis to
function as a Taylor series in the ellipse parameters, and better
approximate the ``in-between'' $\bm{\phi}$ points.

Allowing the ellipse degeneracies to remain in the basis clearly means
any measurement of an ellipse quantity depends on the coefficients
$\bm{\alpha}$, as well as the ellipse parameters $\bm{\phi}$.  Unlike
the flux measurement operator, these measurement operators are not
linear in the coefficients.  We can measure the effective ellipse by
measuring the first and second moments of the model; the first moment
vector is the centroid, and the second moment matrix can be related to
the ellipticity and radius by the relations given in
Appendix~\ref{sec:appendix:ellipse-parameterizations}.  These moments can be
related to a set of linear operation vectors $\bm{e}_i$ that can be
computed directly from the basis functions (similar to the rows of the
constraint matrix $\bm{K}$), and are transformed by the 
inverse of the ellipse transform matrix $\bm{T}[\bm{\phi}]$.  The
first and second moment operators are:
\begin{align}
  \left[\begin{array}{c}
    I_x \\
    I_y
    \end{array}
    \right]
  &= \frac{1}{\bm{e}_0^T\bm{\alpha}}\bm{T}[\bm{\phi}]
  \left[
    \begin{array}{c}
      \bm{e}^T_x\bm{\alpha} \\
      \bm{e}^T_y\bm{\alpha}
    \end{array}
    \right]
  + \hat{\bm{\theta}}[\bm{\phi}]
  \\
  \left[\begin{array}{c c}
      I_{xx} & I_{xy}\\
      I_{xy} & I_{yy}
    \end{array}
    \right]
  &= \frac{1}{\bm{e}_0^T\bm{\alpha}}\bm{T}[\bm{\phi}]
  \left[
    \begin{array}{c c }
      \bm{e}^T_{xx}\bm{\alpha} - (\bm{e}^T_{x}\bm{\alpha})^2 &
      \bm{e}^T_{xy}\bm{\alpha} - (\bm{e}^T_{x}\bm{\alpha})(\bm{e}^T_{y}\bm{\alpha})
      \\
      \bm{e}^T_{xy}\bm{\alpha} -
      (\bm{e}^T_{x}\bm{\alpha})(\bm{e}^T_{y}\bm{\alpha}) &
      \bm{e}^T_{yy}\bm{\alpha} - (\bm{e}^T_{y}\bm{\alpha})^2
    \end{array}
    \right]
  \bm{T}[\bm{\phi}]^{T}
\end{align}
with
\begin{align}
  (e_0)_n &= \int\!d^2\bm{\theta}\, \Psi_{n}\![\bm{\theta}] &
  (e_x)_n &= \int\!d^2\bm{\theta}\,\theta_1\,\Psi_{n}\![\bm{\theta}]\nonumber\\
  (e_y)_n &=
  \int\!d^2\bm{\theta}\,\theta_2\,\Psi_{n}\![\bm{\theta}] &
  (e_{xx})_n &= \int\!d^2\bm{\theta}\,\theta_1^2\,\Psi_{n}\![\bm{\theta}]\nonumber\\
  (e_{yy})_n &=
  \int\!d^2\bm{\theta}\,\theta_2^2\,\Psi_{n}\![\bm{\theta}] &
  (e_{xy})_n &= \int\!d^2\bm{\theta}\,\theta_1\theta_2\,\Psi_{n}\![\bm{\theta}]
\end{align}

We will return to the question of measuring ellipse parameters when we
discuss shear estimation applications in
section~\ref{sec:applications:shear-estimation}; it may be necessary
to take additional steps to avoid systematic biases at extremely low
levels.

\subsection{Radii and Shapelet Order}
\label{sec:shapelets:compound:radii-and-order}

The primary disadvantage of the compound basis technique is its
flexibility.  The user must select in advance the scales $s_j$ and
shapelet orders $N^{(j)}_{\Phi}$ to be used in fitting an ensemble of
galaxy images.  This will typically require some experimentation on a
smaller training sample of galaxies.

It is not immediately clear what metric to optimize in selecting the
scales and shapelet orders, even for an ideal training set; one does
not want to simply ensure the compound basis fits the mean morphology
well.  A better choice would be to optimize the full posterior of
of all galaxies in the training sample, treating the
individual galaxy parameters as nuisance parameters.  This is a huge
global nonlinear optimization problem, however, requiring each galaxy
in the sample to be fit for a new ellipticity and set of coefficients
for every iteration in the basis parameters.  Furthermore, this
metric imposes a signal-to-noise weighting on the members of the
training sample, which may or may not be appropriate for different
training sets.  Most importantly, it is clear that a basis with a
large total number of basis functions will generally outperform a
basis with fewer, and thus part of the challenge is in determining
the correct number of basis functions to use.  There are also
computational advantages to using fewer shapelet expansions, just
as there are advantages in decreasing the order of the shapelet
components, and a useful metric must assign weights to each of these
opposing priorities. 

\begin{figure*}
    \includegraphics[width=\textwidth]{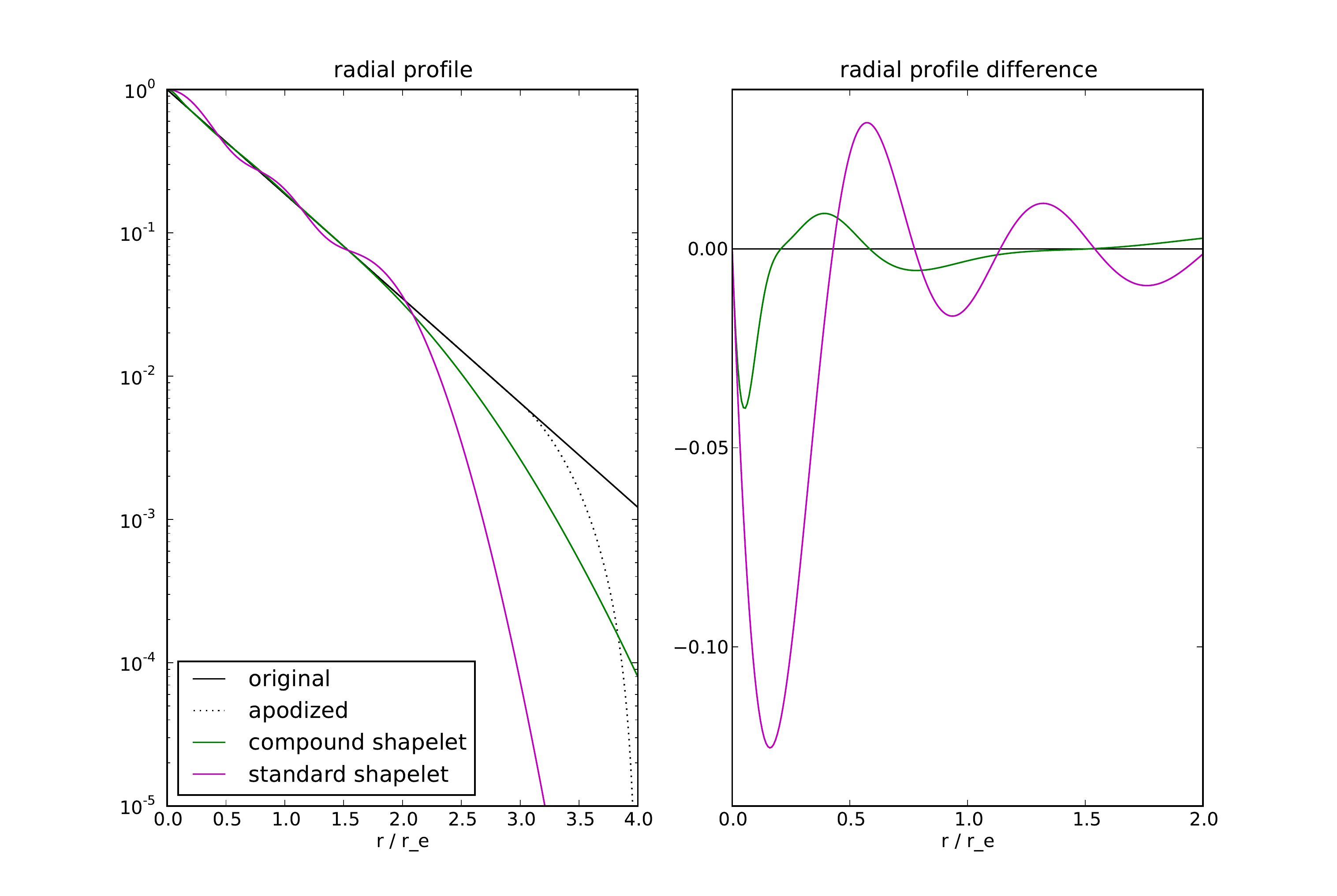}
    \caption{
      Standard and compound shapelet fits to a modified exponential
      (S\'{e}rsic $n=1$) profile, where the modification is a
      polynomial truncation at $r_e=4$ (the same prescription
      as used in the SDSS galaxy modeling).  The standard shapelet approximation
      is 8th order, while the compound shapelet fit uses only 4
      zeroth-order (Gaussian) profiles.  The panel on the right shows
      the difference between the original S\'{e}rsic profile and the
      shapelet approximation at small radius.
      \label{fig:expdisk}
    }
\end{figure*}

\begin{figure*}
    \includegraphics[width=\textwidth]{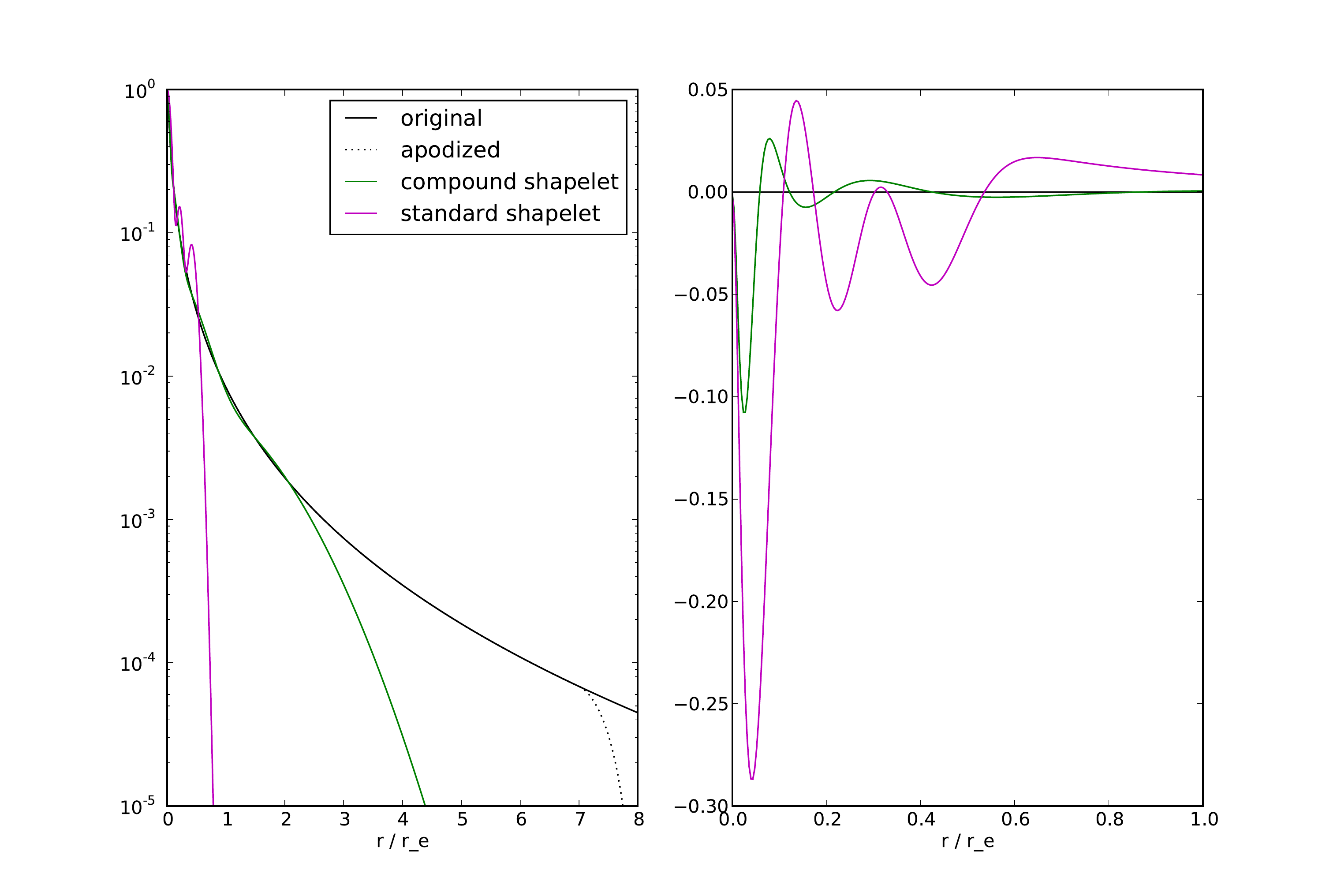}
    \caption{
      Standard and compound shapelet fits to a modified de Vaucouleurs
      (S\'{e}rsic $n=4$) profile, where the modification is a
      polynomial truncation at $r_e=8$ (the same prescription
      as used in the SDSS galaxy modeling).  As in Figure~\ref{fig:expdisk},
      the standard shapelet approximation
      is 8th order, while the compound shapelet fit uses only 4
      zeroth-order (Gaussian) profiles (but the relative scales of the
      Gaussians is different from Figure~\ref{fig:expdisk}).  Note
      that the difference panel on the right uses a different radius
      range than that of Figure~\ref{fig:expdisk}.  While even the
      compound shapelet approximation truncates much earlier than the
      original profile, it does a good job in the region we are likely
      to have data.
      \label{fig:devauc}
    }
\end{figure*}

One approach is to set the scales and shapelet orders to match a
particular analytic function, such as an exponential or other
fixed-index S\'{e}rsic profile.  However, to do so, one must choose a
metric that defines similarity between the analytic profile and its
shapelet approximation.  While a compound basis can approximate a
S\'{e}rsic function over a wider range of scales than a standard
shapelet basis, it still cannot reproduce a general S\'{e}rsic profile
over the full infinite range of the function, and a metric that puts
significant weight at large or very small radii typically produces a
poor approximation over the range of radii where galaxy profiles can
realistically be observed.  These problems can be greatly mitigated by
matching the basis to an apodized S\'{e}rsic function, which is
modified to decrease rapidly to zero beyond some multiple of the
half-light radius.  Such apodized functions are common even when
fitting pure S\'{e}rsic functions, reflecting the fact that we
generally do not have any useful data beyond 4-5 half-light radii.
Compound basis fits to apodized exponential ($n=1$) and de
Vaucouleurs ($n=4$) profiles are shown in Figures~\ref{fig:expdisk} and
\ref{fig:devauc}.  We can
combine a few such profile-trained basis functions to form a basis
with similar flexibility to a full S\'{e}rsic fit with variable $n$;
we can then add terms with higher-order shapelet functions to support
more complex morphologies.

\subsection{Dimensionality Reduction}
\label{sec:shapelets:compound:dimensionality-reduction}

Because the radii and shapelet orders must be chosen in advance, a
compound basis will typically have a fixed number of elements, and
unlike the standard shapelet basis, there is no 
consistent way to increase or decrease the size of the basis to match
the signal-to-noise ratio (S/N) and resolution of the image data being modeled.

Given a training sample, one can attempt to mitigate some of
these problems by changing the weight matrices $\bm{M}$, eliminating
those linear combinations of basis functions which do not play an important
role in fitting the training sample.  This can be used to produce a
``reduced'' compound basis, in which the number of basis functions can
be much smaller than the number of shapelet basis functions used to
build it.  Quantifying the ``importance''
of a basis function is difficult, however, and may vary between
different modeling applications and datasets.

In spite of this, we expect some sort of dimensionality reduction to
be a crucial part of any computationally feasible weak-lensing or
morphological analysis technique based on compound shapelets, and we
will explore this more fully in a future paper.  It may also be an
important step in using the training sample to determine the optimal
scales and shapelet orders of the component expansions; while we have
discussed the two problems separately here, constructing a basis
from a training sample involves optimizing $s_j$, $N^{(j)}_\Phi$, and
$\bm{M}^{(j)}$ together.

\subsection{Priors and Linear Constraints}
\label{sec:shapelets:compound:priors}

The problem of building a basis from a training sample is also
closely linked to the problem of building an empirical coefficient
prior to go with it.  In an ideal sense, the problem is to
characterize the intrinsic distribution of galaxy morphologies, using
an arbitrary linear basis; we then wish to project to a new basis that
somehow simplifies that distribution.  The standard method for basis
selection, principal components analysis (PCA), would treat this
distribution as Gaussian, and identify the most important basis
functions as the directions with the largest variance.  We can
imagine extending this to a mixture-of-Gaussians clustering approach
that could handle multimodal distributions.  As we have noted, a
Gaussian or mixture of Gaussians is not appropriate as a coefficient
prior on its own, as it does not account for the steep slope of the
prior as a function of flux.  We may be able to use a Gaussian or
Gaussian mixture distribution in the normalized coefficients
(i.e. $\tfrac{\alpha}{\lVert\alpha\rVert}$), however, and this would
naturally work well with the importance distribution of
\eqnref{eqn:h-inner-full}.

The first roadblock to such efforts is the need to ``bootstrap'' the
initial distribution using some other, non-optimal basis.  We have had
some success modeling simulated galaxies using a simple basis
consisting of a few S\'{e}rsic profile approximations (such as those of Figures~\ref{fig:expdisk} and
\ref{fig:devauc}), along with a single second-order shapelet expansion
to provide desirable ellipse degeneracies.  To build a distribution
from a training sample, however, we require a much more flexible
basis, such as the combinations of higher-order shapelet functions listed
in Table~\ref{table:basis-sets}.  A more flexible basis has greater
need for an informative prior, however, unless the data quality is
extremely good.  We have sidestepped this in the next section by
simply using a maximum-likelihood fit, which may also be an option in
the bootstrap phase of building a basis and corresponding prior.

Even if we must start with a flat prior, however, we can make use of
linear inequality constraints to reduce the volume of the space that prior
includes, and reject unphysical models.  We can view these constraints
as part of the prior; they essentially define a
polytope in parameter space, with the prior set to zero outside this region.
Highly flexible compound
shapelet models can suffer from severe extrapolation problems without
such constraints, as a compound basis can easily have flexibility well
beyond the maximum radius constrained by the data.  These are easily
addressed through positivity and monotonicity constraints, which
can be written as linear inequality constraints on the coefficients
(and do not depend on $\bm{\phi}$).  To avoid removing
too much flexibility, we can even define a smoothed monotonicity
constraint, in which we require a smoothed radial profile to be
monotonic.  These constraint-based flat priors should be viewed as a
temporary solution, however; the real goal is an empirical prior.

\section{Demonstration and Results}
\label{sec:shapelets:demos}

To demonstrate the improvements that compound and elliptical shapelets
offer relative to more standard shapelet techniques, we apply the
compound shapelet models to
the \citet{Frei1996} sample of nearby galaxies.  This data set has high
S/N and a very small PSF relative to the galaxy sizes, and
spans a wide range of morphologies.  Even though these conditions are
not typical of most weak-lensing observations, this is an ideal
data set for testing the suitability of our basis functions in fitting
the true, unconvolved morphologies of galaxies, even with application
to shear measurement.  While the distribution of morphologies for the
moderate- and high-redshift galaxies used in weak-lensing measurements
is not identical to the distribution of 
morphologies in our sample, the same morphological types will be
present in both, and a nearby galaxy sample affords relative image
quality that is significantly better than even the best space-based
or adaptive optics imaging of more distant galaxies.

\begin{table}
  \center
  \begin{tabular}{c|ccccc}
    \textbf{$\beta$} & 
    \textbf{0.25} & 
    \textbf{0.50} & 
    \textbf{1.00} &
    \textbf{2.00} & 
    \textbf{4.00} \\
    \hline
    \texttt{\_\_8\_\_} & - & - & 8 & - & - \\
    \texttt{\_44\_4} & - & 4 & 4 & - & 4 \\
    \texttt{\_444\_} & - & 4 & 4 & 4 & - \\
    \texttt{4\_4\_4} & 4 & - & 4 & - & 4 \\
    \texttt{4\_44\_} & 4 & - & 4 & 4 & - \\
    \texttt{206\_3} & 2 & 0 & 6 & - & 3 \\
    \texttt{2063\_} & 2 & 0 & 6 & 3 & - \\
    \texttt{026\_3} & 0 & 2 & 6 & - & 3 \\
    \texttt{0263\_} & 0 & 2 & 6 & 3 & - \\
  \end{tabular}
  \caption{Shapelet order at five different radii for the nine
    demonstration basis sets.\label{table:basis-sets}} 
\end{table}

Of the 113 galaxies in the
sample, 82 were observed in $R$ and $B_j$, and the other 31 in $g$,
$r$, and $i$.  We limit our attention to the $B_j$ and $g$ images (the
two filters are very similar).
We fit each of the 113 galaxies in the sample with both the
elliptical and circular forms of each basis described below, but
discard the 31 galaxies in which the fit did not converge in one or
more cases (these failures are mostly due to large gradients in the
background present in some of the images).

\begin{figure*}
    \includegraphics[width=\textwidth]{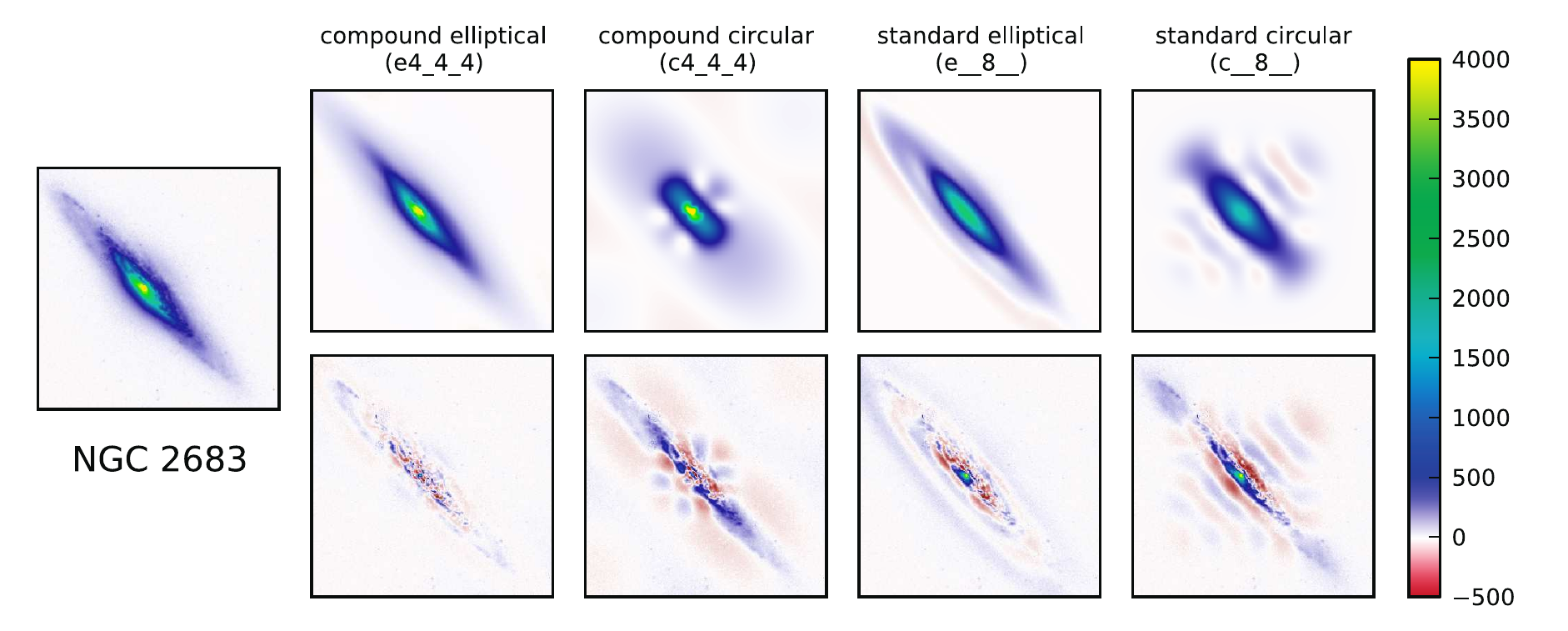}
    \caption{Image, shapelet models (upper),
      and residuals (lower) for NGC~2683, a typical edge-on spiral
      galaxy in our test sample.  Each model has 45 free parameters,
      equivalent to an 8th-order standard circular shapelet fit.
      This example clearly illustrates the artifacts
      introduced in circular shapelet fits to high-ellipticity galaxy
      images; in this case, the circular models actually have negative
      flux in certain regions.  As expected, the compound and standard shapelet fits
      perform similarly at moderate radii, but the compound
      basis is noticeably better at representing structure at small
      and large scales, modeling the core well while reducing or
      eliminating the ``ringing'' residuals present in the wings of
      the standard basis fits.  The basis labels are defined in
      Table~\ref{table:basis-sets}, and the data and fitting procedure
      are described in section~\ref{sec:shapelets:demos}.
      \label{fig:ngc2683}
    }
\end{figure*}

\begin{figure*}
    \includegraphics[width=\textwidth]{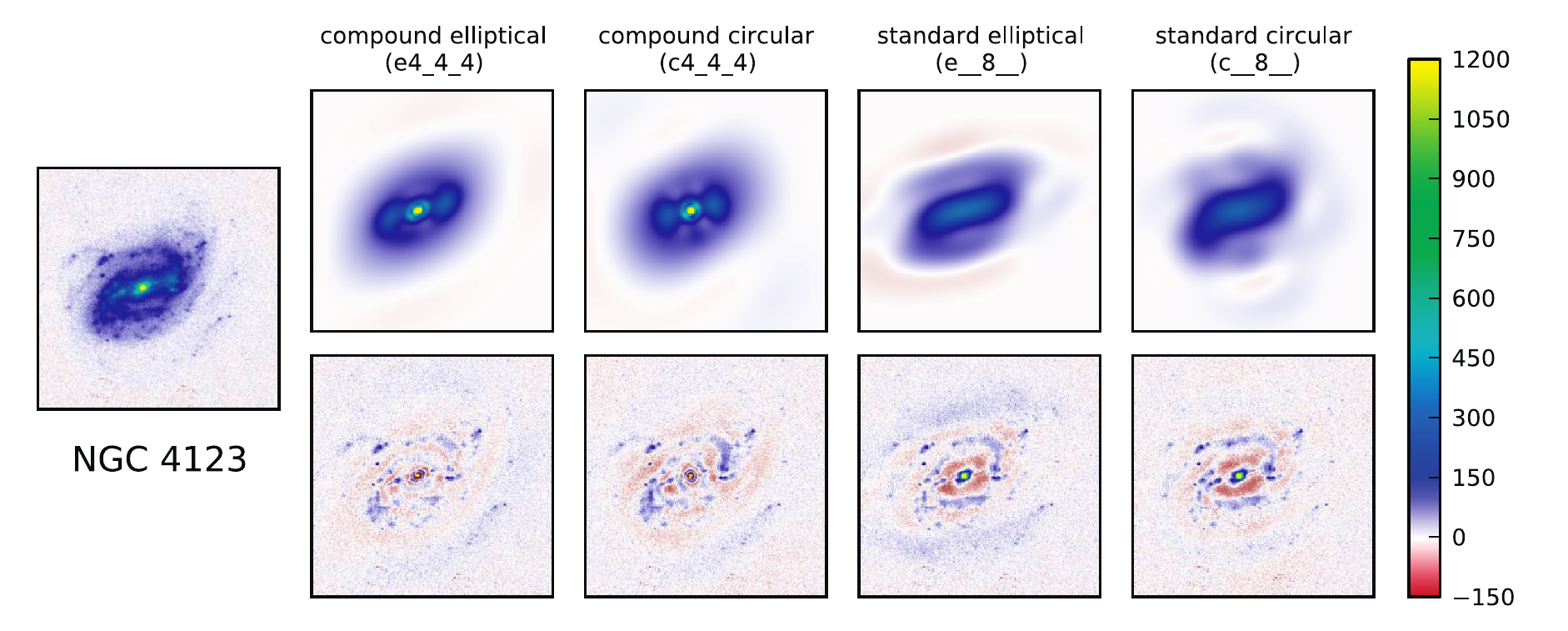}
    \caption{Image, circular shapelet models (upper),
      and residuals (lower) for NGC~4123, a typical face-on spiral
      galaxy in our test sample.  The elliptical compound basis models
      the core, bar, and ring structure reasonably well, but none of
      the models capture much of the spiral arm structure.  This is
      generally the case for the face-on spirals in our sample.
      \label{fig:ngc4123}
    }
\end{figure*}

\begin{figure*}
    \includegraphics[width=\textwidth]{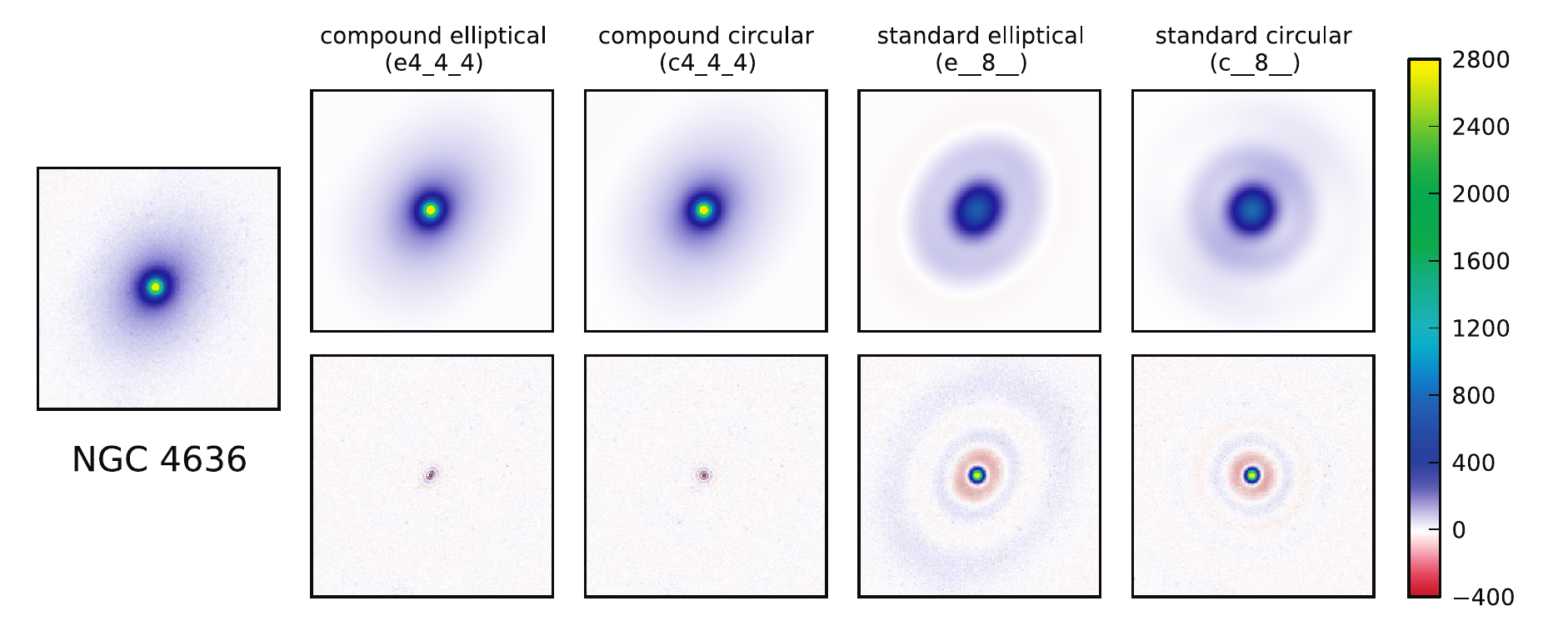}
    \caption{Image, shapelet models (upper),
      and residuals (lower) for NGC~4636, a typical massive elliptical
      galaxy in our test sample.  The difference between the compound
      basis fits and the standard basis fits is striking, particularly
      in the oscillating residuals and in the core of the galaxy,
      which is not fit at all by the standard basis.  Both circular
      fits show residuals which are clearly circular, rather than
      matched to the ellipticity of the galaxy, and the standard
      circular shapelet model is quite clearly circular at large
      radii, even though the galaxy is not; the limited basis simply
      does not have the modeling power to apply the correct shear at
      both moderate and large radii.
      \label{fig:ngc4636}
    }
\end{figure*}

We consider nine compound basis sets, including a single-scale
standard shapelet basis, each with 45 basis functions (before
circularization).  These are summarized in
Table~\ref{table:basis-sets}.  For elliptical shapelet fits
(\texttt{e} prefix), we circularize the basis and fit for the
coefficients and ellipse parameters  
simultaneously using the Levenberg--Marquardt nonlinear least-squares
optimizer.  The initial ellipse parameters are determined from the
unweighted moments of the image.  For circular shapelet fits
(\texttt{c} prefix), we fix
the centroid and size to the results from the elliptical fit, and
perform a linear least-squares fit using the uncircularized basis.  In
each case, there are 45 free parameters.  We use a simple circular
Gaussian PSF model (with FWHM as given in the image headers).

\subsection{Results}

The results for three representative galaxies are shown in
Figures~\ref{fig:ngc2683}, \ref{fig:ngc4123}, and \ref{fig:ngc4636}.
As expected, the edge-on spiral galaxy (Figure~\ref{fig:ngc2683})
demonstrates most clearly the improvement from circular to elliptical
shapelets, both for the compound basis (\texttt{4\_4\_4}) and the
standard basis (\texttt{\_\_8\_\_}).  The difference can also be seen
in the \texttt{c\_\_8\_\_} fit to the elliptical galaxy
(Figure~\ref{fig:ngc4636}).  Because the standard shapelet expansion
already has such difficulty representing the de Vaucouleurs profile, it
cannot afford to ``spend'' coefficients on matching the ellipticity of
the galaxy. The compound basis provides better fits to
both the inner radii and outer radii of all the galaxies shown, and
eliminates the ``ringing'' residuals except at very small radii; this
is particularly dramatic for NGC 4636.  None of the basis sets
provide enough flexibility to fit the spiral arms of NGC 4123
(Figure~\ref{fig:ngc4123}) well, and this is generally the case for
the face-on spirals in the sample.  The compound basis fits do provide
a reasonably good approximation to the bar and ring structure, and all
the fits do a moderately good job of fitting the overall profile of
the spiral galaxies.
\begin{figure*}
  \begin{center}
    \includegraphics[width=\textwidth]{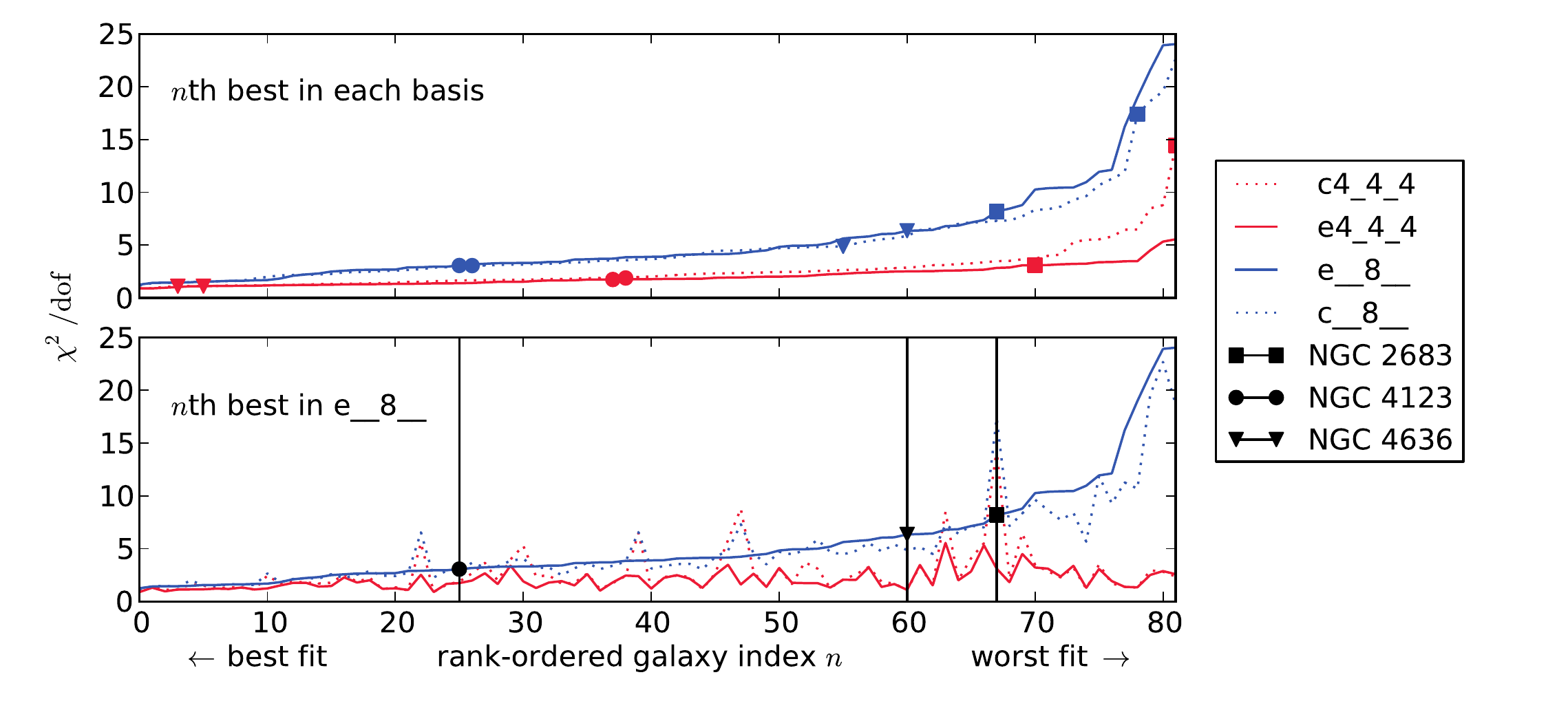}
  \end{center}
  \caption{Rank-ordered goodness of fit for 82 galaxies from the
    \citet{Frei1996} sample of nearby galaxies, for selected
    elliptical and circular shapelet basis sets.  In the upper plot,
    the galaxies are rank-ordered by reduced $\chi^2$ for each basis
    separately; each point on the $x$-axis thus corresponds to the $n$th
    best-fit galaxy for each basis.  In the lower
    plot, the reduced $\chi^2$ from the \texttt{e\_\_8\_\_} fit is
    used to set the galaxy rank index for all bases; a point on the
    $x$-axis thus corresponds uniquely to a single galaxy.  The
    goodness of fit
    for the galaxies shown in Figures~\ref{fig:ngc2683}, \ref{fig:ngc4123},
    and \ref{fig:ngc4636} are given by the square, circle, and triangle
    points, respectively.  The circular standard basis actually
    produces slightly lower average and worst-case residuals than the
    elliptical standard basis, but the compound elliptical basis is
    significantly better than either.  The high-ellipticity edge-on spirals
    (such as NGC 2683) that present a particular problem for the
    circular shapelet expansion can be clearly seen in the lower plot
    as spikes in the circular basis
    goodness-of-fit. \label{fig:residuals-ec}}  
\end{figure*}

These tendencies are quantified in Figure~\ref{fig:residuals-ec},
which shows the goodness-of-fit distribution for the 82
galaxies successfully fit with all basis sets.  The elliptical compound
basis consistently has the 
lowest residuals, and even the circular compound basis generally
outperforms both the elliptical and circular standard shapelet basis.
The elliptical standard shapelet basis is not uniformly better
than the circular standard shapelet basis, but the cases where it is --
high-ellipticity edge-on spirals, like NGC 2683 -- stand out clearly as
spikes in the lower plot, in which the galaxy index is consistent
across all the fits.  It is also worth noting that when the compound
basis is used, featureless, low-ellipticity early-type galaxies like
NGC 4636 have some of the lowest residuals in the sample.  In
contrast, these have moderately large residuals compared to the rest
of the sample for the standard shapelet basis fits, demonstrating that
using a compound basis essentially solves the problem of fitting de
Vaucouleurs profiles with shapelets, at least from a goodness-of-fit
standpoint.

\begin{figure*}
  \begin{center}
    \includegraphics[width=\textwidth]{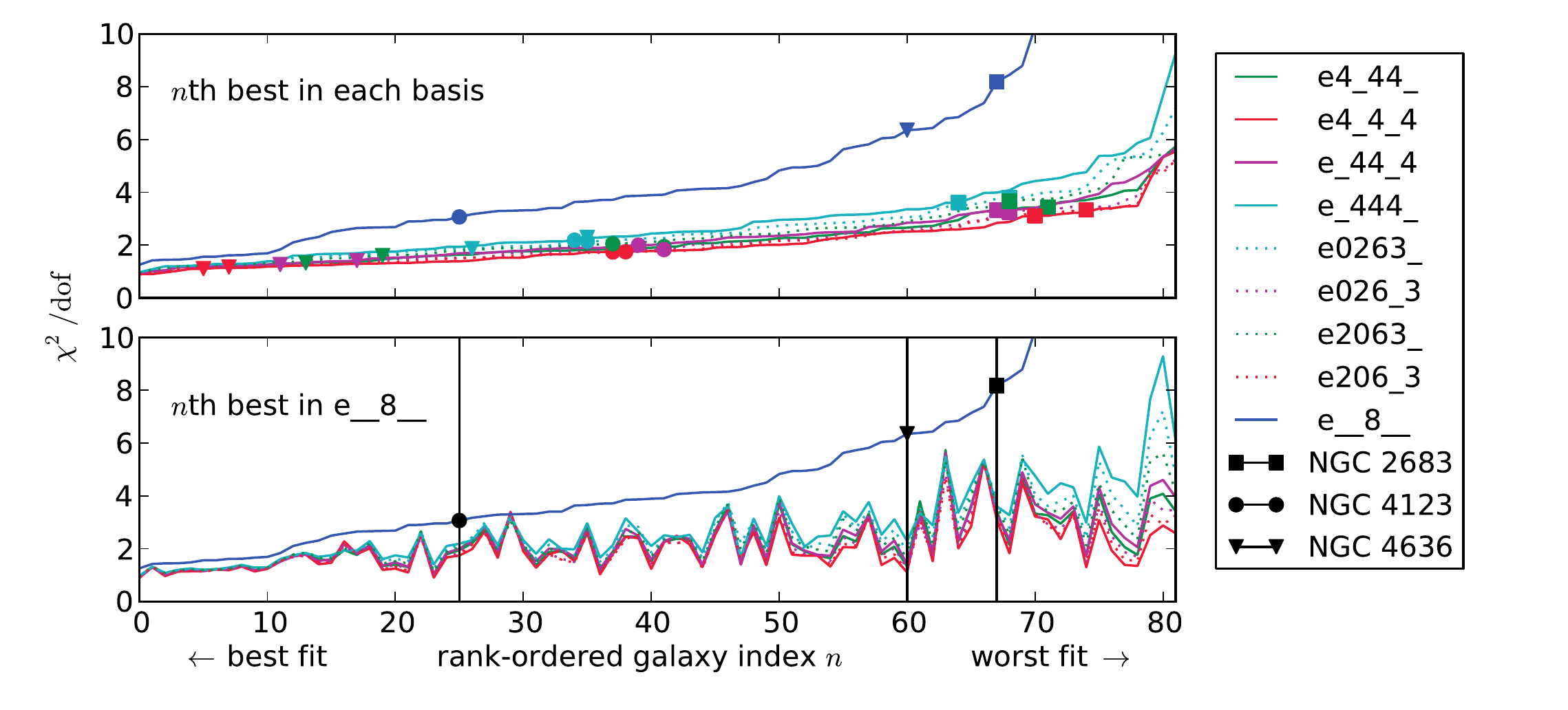}
  \end{center}
  \caption{Goodness-of-fit comparison for all elliptical
    basis sets. The rank-ordered indexing and galaxy sample are the
    same as in Figure~\ref{fig:residuals-ec}, but the $y$-axis scale
    has been changed to show more detail.  All of the compound
    shapelet basis sets outperform the standard shapelet basis for all
    galaxies but one, in which the results are comparable, and the
    compound bases show considerably better worst-case performance.
    Between compound basis sets, those with components at both the
    largest and smallest radii (\texttt{4\_4\_4} and \texttt{206\_3})
    consistently have the lowest residuals. \label{fig:residuals}}
\end{figure*}

Figure~\ref{fig:residuals} similarly shows the distribution
goodness-of-fit for all elliptical basis sets.  The most obvious
result is that all the 
compound shapelet bases tested outperform the standard shapelet basis,
both in aggregate and (with one slight exception) for every galaxy
individually.  Between the compound 
basis sets, those with shapelet expansions at both $\beta=0.25$ and
$\beta=4.0$ tend to perform slightly better, especially compared to
basis sets that contain neither, suggesting that compound bases are
most effective when constructed with shapelet orders at radii that
differ by more than a factor of two.  This provides a partial answer
to the question of how to choose the radii and orders of the component
shapelet bases that is both frustrating and somewhat comforting: while
there is no clear choice for an optimal combination of radii and
orders, any reasonable choice that samples the radial range of
significant morphology is still a marked improvement over the
standard shapelet basis.

\section{Summary and Future Work}
\label{sec:shapelets:summary}

Despite the problems discussed in section~\ref{sec:shapelets:limitations},
shapelets have become one of the most important tools in galaxy
modeling, particularly in weak lensing.  The inability of the standard
shapelet basis to 
represent morphologies with high ellipticity or steep profiles puts
limits on the accuracy of shapelet-based shear estimators, however,
and likely degrades the performance of shapelet-based morphological
analyses.  We have presented here two techniques to address these
limitations. 

The first is a
convolution relation for elliptical shapelets.  This eliminates the
need to use the lossy shapelet-space shear operator, and allows
ellipse-parameterized shapelets to be used in place of a circular basis
parameterized only by centroid and radius.  While the elliptical basis
requires five nonlinear parameters to be fit instead of three,
the two additional parameters more than pull their weight in modeling
power, particularly for high-ellipticity galaxies.  Perhaps most
importantly, an exact elliptical shapelet convolution allows
shapelet-based shear estimators to be constructed that do not suffer
from the ``shear artifact'' bias discussed in
section~\ref{sec:shapelets:limitations:ellipticity}.  
The results of section~\ref{sec:shapelets:demos} do not demonstrate the advantages
of the elliptical convolution formula;
while we used it in fitting the galaxies, the size of our test
galaxies relative to the PSF makes the difference between
our exact convolution and the approximation of \citetalias{NB07}
negligible.  However, this formula 
can be immediately put to use in existing elliptical shapelet
shear--measurement methods, such as that of \citetalias{NB07},
eliminating one source of bias for galaxies near the resolution
limit.

To address the difficulties of the standard shapelet basis in fitting
galaxies with high S\'{e}rsic indices, we have introduced the concept of a
``compound'' shapelet basis -- a basis composed of multiple low-order
shapelet expansions with different scale radii.  Even with simple, ad-hoc
choices for the radii and orders of the component shapelet expansions,
the compound basis is significantly better than a single higher-order
shapelet expansion at fitting realistic galaxy morphologies,
particularly in its worst-case performance.  By combining
low-order shapelet basis functions at multiple radii, a compound basis
can compactly represent both the sharp cores and extended wings of galaxies
with high S\'{e}rsic indices.  As we have seen, this can make a
significant difference even in spiral galaxies with relatively low
S\'{e}rsic indices, and almost completely eliminates the oscillatory
artifacts often present in standard shapelet fits. 
This should reduce the underfitting shear bias
for shapelet-modeling shear measurement methods, but whether the
bias can be reduced to acceptable levels for future surveys is a
question we reserve for a future paper.  Morphological analyses such as
those of \citet{KellyMcKay2004} and \cite{Andrae2010} should also benefit from
utilizing compound shapelet basis sets, as the more compact compound
shapelet representation essentially gives classification and
clustering algorithms a head start in reducing the dimensionality of
the problem.  

Unlike the elliptical convolution formula, further work is needed to
make full use of the compound shapelet concept in a complete shear
measurement or morphological analysis method.  While we have
demonstrated success in 
fitting nearby galaxies with unoptimized basis sets and a simple
fitting algorithm, more challenging modeling problems may require
techniques that tune the parameters and weights of a compound basis
using a training sample of galaxies.  Such a basis of
``eigenmorphologies'', built from analytically tractable shapelet
building blocks, would be an extremely powerful tool, not only for
morphological analysis, but also in further quantifying and reducing
underfitting biases in model-based shear measurement techniques.
\chapter{Applications}
\label{sec:applications}

\section{Photometry}
\label{sec:applications:photometry}
Measuring the fluxes and colors of astronomical objects using the
methods we have developed is fairly straightforward.  The total flux
of the model is a simple linear operator on the coefficients
$\bm{\alpha}$:
\begin{align}
  \text{flux} &= \int\!d\bm{\theta}
  g[\bm{\theta},\bm{\phi},\bm{\alpha}]
  \\
  &= \int\!d\bm{\theta} \sum_{i=1}^N
  \tilde{\Psi}_i[\bm{\theta},\bm{\phi}]\alpha_i
  \\
  &= \sum_{i=1}^N\alpha_i\bigl|\bm{T}[\bm{\phi}]\bigr|
  \int\!d\bm{\theta} 
  \Psi_i\!\left[\bm{T}[\bm{\phi}]^{-1}(\bm{\theta} -
    \hat{\bm{\theta}}[\bm{\phi}])\right]\\
  &= \sum_{i=1}^N\alpha_i\int\!d\bm{\theta}^\prime
  \Psi_i\!\left[\bm{\theta}^\prime\right]\\ 
  &= \bm{v}^T\bm{\alpha}\\
  v_i &\equiv \int\!d\bm{\theta}^\prime
  \Psi_i[\bm{\theta}^\prime]
\end{align}
Because the model is multiplied by the determinant of of the ellipse transform
$\bm{T}$, we can change variables to remove the dependence on
$\bm{\phi}$, and as a result the flux operator $\bm{v}$ does not
depend on the ellipse parameters.  The weighted sample
$\{\bm{v}^T\bm{\alpha}_{[n,m]},w_{[n,m]}\}$ is thus representative of
the posterior distribution of the flux.  Similarly, we can apply a
logarithm transform to our sample to obtain the posterior distribution
of the magnitude.

This has some clear advantages over the more common procedure of
estimating ``best-fit'' fluxes with symmetric variance-based error
bars.  While the magnitude system does not permit negative fluxes,
``best-fit'' fluxes can be negative (and their lower bounds based on
symmetric error bars are very often negative).  These do not transform
easily to the magnitude system, necessitating the development of
alternate definitions such as the
hyperbolic arcsin magnitudes used by the SDSS \citep{Luptitudes}, and making
the transformation of uncertainties to regular magnitudes somewhat
problematic.  While arcsinh magnitudes have other advantages, it is
worth noting that the negative flux problem is limited to non-Bayesian
techniques.  Any reasonable prior will assign zero
probability to a model with an unphysically negative flux, and this
will naturally carry over to the posterior.  This may present some
problems in terms of publishing (asymmetric) confidence limits in
public catalogs, but it is clearly a better reflection of reality.

The photometric measurements produced by our Monte Carlo sampling
approach also differ from standard photometric measurements in that
they cannot easily be related to an aperture magnitude.  Most
maximum likelihood modeling methods for measuring flux can be seen in
some sense as an aperture method by viewing the profile of the model
as a continuous aperture; while the slope and radius of the profile
model may be allowed to vary, the flux is measured with 
these fixed at their maximum-likelihood values.  With our approach,
the flux is marginalized over the other parameters of the model, so
the effective profile is the posterior-weighted average of the profile
at all the Monte Carlo sample points.  This provides a much more
complete and robust accounting of the uncertainties involved in
measuring the flux, but it may also make it difficult to compare
our measurements with historical measurements, or more importantly, to
calibrate our measurements to a standardized magnitude system.  This
should be viewed as a challenge, not a disadvantage; we should not a
reject careful and robust treatment of uncertainties to make our
results match up with expectations.

This challenge is perhaps most evident in the question of how
to measure galaxy colors.  Because of the intrinsic difficulty in
measuring the absolute magnitude of a galaxy, quality estimates of
flux ratios across different filters (colors) are generally more
important.  Historically galaxy colors have been measuring by applying
the same aperture (or equivalently, a model with all parameters but
amplitude fixed) to data from each filter.  We do not have a single
set of parameters to ``fix'' a model at in our paradigm, however.  The
closest approximation would be to run our algorithm on one canonical
band, and use the exact same Monte-Carlo points in other bands, allowing
only the amplitude of the model to vary and changing the weights
accordingly.  Real galaxies have different morphologies at different
wavelengths, however, producing color gradients that inevitably make
the choice of aperture a factor in the measured color.  There is no
``magic'' aperture or model that produces correct colors; the real
reason we use the same aperture or model in different bands is simply
because it ensures consistency.  Measuring colors with maximum
likelihood modeling techniques that allow the profile parameters to
vary across different bands lacks this needed consistency simply
because the modeling technique is not robust -- fluxes are not
marginalized over uncertainties in the profile parameters, and can
change significantly when the best-fit profile parameters are
perturbed by noise.  With our more robust modeling methods, this
consistency should be easier to achieve, but it is still crucial to
require the models to be similar in different bands, simply
because the S/N may be extremely low in some bands and large in
others due to the spectrum of the object.  This can be achieved by
fitting to all filters simultaneously, and allowing the object to have
different linear parameters $\bm{\alpha}$ on different filters but the
same nonlinear parameters $\bm{\phi}$.  We can then use the prior on
the linear parameters to ensure that the colors are reasonable and the
morphologies in different filters are similar.

We can use a similar approach when fitting variable objects, using a
single set of nonlinear parameters and a different set of linear
parameters for each exposure.  Most variable astronomical objects are
point sources, so this results in only one linear parameter per exposure.
For point sources, the nonlinear parameters will
generally model stellar motions, rather than the ellipse of an
extended object; sharing those parameters across exposures allows us
to construct models for variable stars with proper motion and
parallax.  Just as with fitting galaxies to multiple bands, we can
define a coefficient prior for the variable point source model that
will ensure the variability levels are physically reasonable.

In some cases, we may wish to use more complex models that connect
these multiple observations in place of a simple prior.  When modeling
galaxies, for instance, we could evaluate the posterior of a
photometric redshift model in place of the coefficient prior.  This
would entail adding additional nonlinear parameters to account for the 
redshift and spectrum of the object, which must be explored in the
outer stage.  In the inner stage, we would evaluate the posterior of
the photometric redshift model using the fluxes defined by the linear
coefficients as the data.  Similarly, we could fit a light curve model
directly to the pixel values when modeling variable objects, rather
than allow the fluxes to vary independently on different
exposures.  These proposals may be significantly more computationally
expensive than the methods used today, but they would provide better
measurements for the faintest objects, as well as a more robust
accounting of uncertainties.  The question of \emph{how much} better
is an open one, however, and it is likely that we could obtain similar
results just by making use of the full flux posteriors when modeling
redshifted galaxy spectra or light curves, even if we do not apply
them directly to the pixels.

Finally, we should note that almost all of these methods require the
final photometric calibration to be in place before the modeling
algorithms are run.  The photometric calibration sets the weights of
individual exposures, and these are combined with the rest of the
inputs in a complex way by the modeling algorithm.  For the case of
galaxy modeling, in which a single set of linear parameters are shared
across multiple exposures, there is no way to correct the outputs of
the algorithm for changes in the photometric calibration without
recomputing all the Monte Carlo weights (which would require
reevaluating the model at all of the nonlinear parameter points).   For
variable objects, the situation is slightly better, as we can identify
a single set of linear parameters that correspond to each exposure.
This breaks down if we make use of scientifically interesting priors
on the variability, or simultaneously fit nonvariable and variable
objects.  While we could use a different set of linear parameters for
each exposures even for nonvariable objects, this would entail a
significant increase in the degrees of freedom of the model and would
thus make any measurements based on the model much noisier, especially
in the short-exposure limit.

\section{Astrometry}

We have already mentioned the possibility of including proper motion
and parallax among the nonlinear parameters when fitting stars to
multiple exposures simultaneously.  This
approach has already been demonstrated with a non-Monte Carlo modeling
method by \citet{Lang2009}.  It is particularly valuable for stars
that are too faint to appear in individual exposures, a regime
where both coadd-based approaches and single-exposure modeling are
completely inadequate; while we may still rely on the coadd for
detection, we cannot measure motion there, and we cannot even measure
centroids on individual exposures.  This does present some problems
for our adaptive importance sampling algorithm; unlike the galaxy
models, we cannot adapt our importance distribution to an approximate
coadd-based posterior before switching to more expensive
multi-exposure modeling.  This could require more iterations of
importance sampling, or perhaps initialization with a greedy
optimizer.  Point source models are significantly faster to evaluate
than PSF-convolved galaxy models, however, so this may not be
impractical, even at large scales.  

\section{Shear Estimation}
\label{sec:applications:shear-estimation}

Weak gravitational lensing is one of the most important -- and most
demanding -- applications of large multi-epoch surveys.  Our goal here
is limited to measuring an unbiased ellipticity for each lensing
source galaxy.  A lack of bias is by far the most important
requirement; the lensing shear is a small fraction of the observed
ellipticity of a typical galaxy, so we must spatially average the
ellipticities of thousands of galaxies to obtain a measurable lensing
signal.  Any spatially correlated or constant bias in the ellipticity
measurement will thus be amplified many times, potentially dwarfing
the lensing effect, especially in the extremely weak regime of lensing
by large-scale structure.  The PSF is the main source of difficulties
in weak lensing measurements; the spatial variation in the
ellipticity of the PSF can easily mimic a physical shear signal.
Furthermore, even convolution by an isotropic PSF will modify the
observed ellipticity, usually decreasing it.

Simply using a forward-fitting modeling technique (in which we convolve the
model with the PSF) corrects for both of these problems, assuming the
PSF model is adequate.  Basing the ellipticity measurement on
an imperfect model introduces its own systematic 
effects, however, and some of these are exacerbated by the PSF
convolution.  The most obvious are underfitting biases, in which the
model is not flexible enough to match the data.  These do not fit
easily into the standard picture of ellipticity measurement as a
process that depends only on low-order morphological information, and
hence can be performed on faint and barely-resolved galaxies.  One of
the reasons modeling biases are difficult to address is that
higher-order morphological information in the model is transferred to
lower spatial frequencies by the PSF convolution, making the observable
ellipticity dependent on aspects of the model that are naturally hard
to constrain from the data.

 We have already
discussed underfitting problems with standard shapelet models in
Chapter~\ref{sec:shapelets}, and demonstrated that they can
be greatly mitigated by using multi-scale elliptical shapelets.  When a
model is flexible enough to represent realistic galaxy profiles,
however, it is likely it will often be under-constrained by the
likelihood, and the prior may have a significant effect on the
result.  This is an improvement over the underfitting problem,
however, as long as we marginalize our ellipticity measurements over
the other parameters in the model, and make use of the full resulting
ellipticity distribution.  Most ellipticity measurement
algorithms do not marginalize, however, and most weak lensing
techniques do not even make use of any uncertainty estimate for the
ellipticities of individual galaxies, let alone the full distribution.  
This likely accounts for some of the success of the shear estimation
method of \citet{Miller2007}, which does make use of the full
ellipticity distribution.  This cannot fully account for the fact that
some aspects of our model are unconstrained by the data, however, and
the prior will necessarily impact the measurement.  We can choose a
prior that reduces any biases we measure in simulations, but this is
little better than the current practice of applying an
empirically-determined ``fudge-factor'' to our measurements, as it is
still subject to differences between our simulated distribution of
morphologies and the real distribution.

\citet{Bernstein2010} approaches this problem from another
perspective.  We can add a weight function to the ellipticity measurement
operators defined in
section~\ref{sec:shapelets:compound:basis-ellipse}, and
define this measurement in Fourier space, as an operator that
acts on the Fourier transform of the model.  Because convolving the
model with the 
PSF destroys high spatial frequency features of the model, we can
define a Fourier-space weight function that ignores these high spatial
frequency features and hence does not depend on features of the model
that have been censored by the PSF.  The practical implementation
proposed by \citet{Bernstein2010} involves using a standard shapelet
decomposition to generate a Fourier-space representation of the data
that the Fourier-space weight function can be applied to.  As a
result, the method is still subject to some of the limitations of the
shapelet basis in representing real galaxies that we discuss in
section~\ref{sec:shapelets:limitations}.  The final problem with this
method is that it creates a different definition of ellipticity for
each galaxy, as the weight function depends on the relative sizes of
the galaxy and the PSF.  This complicates the procedure of
estimating the shear from a catalog of ellipticity measurements,
because the meaning of each measurement -- and hence its sensitivity
to the lensing effect -- is slightly different.

Our compound-shapelet galaxy models can also be easily
Fourier-transformed, and should represent a much better way to
generate the Fourier-space realization needed by the modified weight
function.  Moreover, by applying this ellipticity measurement to our
full Monte Carlo sample of parameter vectors, we can more correctly
account for the full uncertainty distribution of the ellipticity, and
hence address model biases that are not strictly due to the convolution;
these will generally be much less severe than those imposed by the 
convolution, but they should not be ignored entirely.  This may also
allow us to relax the truncation of the ellipticity weight function at
large spatial frequency, which would make the definition of the
ellipticity for different galaxies more similar.

With this general approach, a number of other features of our
modeling method provide additional advantages for weak lensing
measurements.  First, by marginalizing over the centroid parameters of
the model, we can eliminate the ``centroid bias'' that allows PSF
spatial variation to print through to ellipticity measurements in a
subtle way \citep{Kaiser2000}.  Essentially, an elongation in the PSF in one direction
increases the noise in the centroid in that same direction;
meanwhile, an error in the centroid produces an error in the estimated
ellipticity if the centroid is held fixed rather than marginalized.

By fitting multiple objects simultaneously, we can also reduce the
effects of neighboring objects on shear measurements.  While we may
not be able to obtain usable measurements from severely blended
objects, we should be able to reduce any small biases that
result from slight overlaps.  This should be particularly important
for cluster and galaxy-galaxy lensing, in which source galaxies close
to the line of sight of the center of the mass distribution are of
course more likely to be at least partially obscured 
by foreground galaxies associated with that mass distribution.

Finally, we should note that making full use of many of the ideas in
this section will require a fundamental change in how we view
``higher-level'' weak lensing measurements.  Current methods typically
make use of catalogs of ellipticities; future methods should be
prepared to deal with catalogs of ellipticity \emph{distributions},
with the added complication that these ellipticities may each have a
slightly different definition.  Some weak lensing studies already make
use of Bayesian methods, and may be easy to adapt to these different
ellipticity measurements.  Others -- the 2-point functions (and
higher-order statistics) commonly used in cosmic shear studies in
particular -- may be harder to adapt.

\section{Morphology and Classification}
\label{sec:applications:morphology}

The flexibility of our galaxy model is also important for studies of
galaxy morphologies.  While morphological studies of based on
poorly-resolved or low S/N galaxies are always difficult, a generative
modeling approach at least has the potential to avoid strong
systematic effects due to the PSF and low surface brightness.  Many
popular automated morphological estimators
\citep{Abraham1994,Bershady2000,Lotz2004} make use of direct 
statistics on the observed image, and such estimates are inherently
biased by the image quality (as are morphological classifications
based on visual inspection).  Traditional methods thus almost always tend
to find more complicated morphologies as data improves, and correcting
for this known bias is a difficult problem.  This may be true in the modeling case as
well -- we still lose high spatial frequency information due
to the PSF, and our best-fit models may be systematically smoother
and less cuspy as a result.  Similarly, our model may lack low surface
brightness features when these are buried in the noise  -- but by
characterizing the full posterior and using a sufficiently flexible
model we can at least ``know how much we don't know'' in a robust
sense; when the data is poor, we may not choose more complicated
morphologies as the best fit, but we \emph{can} establish when they
cannot be ruled out.  And as we have discussed, if we can train an
informative prior on a small sample of higher-quality data, we can
essentially make the right guesses when we are forced to consider the
parts of the model we cannot constrain with the data.

A galaxy model with many linear parameters is a natural fit to a
machine learning approach to morphological classification, and studies
based on standard shapelets have already made some progress in this
area \citep{KellyMcKay2004,Andrae2010}.  We believe these were limited by
the use of a non-elliptical shapelet basis, as much of the variance in
the measured shapelet coefficients was simply due to variations in
ellipticity.  As a result, both studies were able to
easily distinguish edge-on and face-on spiral galaxies, and showed
some ability to distinguish both from ellipticals, but could not go
much further.  Improving upon these efforts is central to the
question of building an optimal basis and training an empirical prior
on the basis coefficients.  As we have discussed, such
a basis and accompanying prior would have great potential for
improving the models, but it is extremely challenging.  The work so far on
automated morphological classification using linear galaxy models is
still in a ``proof of concept'' phase, limited to convenient,
magnitude-limited datasets and the most basic properties of galaxies,
but the stage is set for much more advanced work that makes use of
larger, more representative samples, more flexible models, and a more
robust accounting of modeling errors.

Robust modeling also has an important role to play in the much simpler
question of how to distinguish stars from
galaxies.  Most Bayesian model selection methods make use of the
evidence ratio of the two models (see, for example, chapter~4 of
\citealt{Sivia2006}), which we naturally compute as part 
of our modeling algorithm.  We can construct other estimators using
other properties of the model as well, such as the half-light radius
of the galaxy model, or by comparing the measured flux in both
models.  An important feature for any of these is our robust
accounting of the model uncertainties; any naive classifier based on the
PSF model that does not take into account its uncertainty will fail
when that uncertainty is large, as will any classifier based on
estimated galaxy size when we lack the ability to put rigorous
confidence limits on the measured radii.

Finally, Bayesian modeling methods can play an important role in
classifying objects that are actually ``garbage'', or have significant
data quality issues on one or more exposures.  This is an important
part of automated data quality assessment -- we should not expect a
perfect set of masks as input, especially when the evidence also
provides a powerful way to indicate the goodness-of-fit and hence
discover bad data.  Unlike a
high $\chi^2$, a low evidence does not merely indicate large
residuals; it can also flag good fits with ``unlikely'' combinations
of parameter values, indicated by a low overlap of the prior with the
likelihood.  This is also indicative of a problem with the data (or
possibly a new class of astronomical object).  While the evidence for
a single object is difficult to interpret, we will fit many objects
with similar brightness and size, and we can flag outliers in the
evidence for additional inspection after binning along these axes.

\section{Deblending}
\label{sec:applications:deblending}

We have emphasized throughout that our modeling algorithm can be
applied to multiple astronomical objects simultaneously.  This is an
important part of analyzing crowded fields, but it is not a complete
solution by itself; we also require a way of generating a deblending
hypothesis for how many objects are present, roughly where they are,
and what types they may be.  While our modeling technique may allow us
to rigorously compare these hypotheses, it is easy to imagine even
simple blends producing an unwieldy number of hypotheses.  With just
three objects and two types of models (variable
point sources with proper motions and static galaxies, for instance),
we would need to fit the system $2^3$ times to account for all
possibilities.  In addition, complex deblending models will have
many more parameters than single-object models, making each system
significantly more expensive.  Happily, much of the additional expense
will be in additional adaptive importance sampling iterations that
should be performed on the coadd.  Still, as survey depth increases,
more objects will overlap, and it may be that most objects will be fit in
pairs or triplets.  It will become increasingly important to have a
deblender that can generate good hypotheses and reject in advance
highly unlikely permutations of object types, to reduce the number of
deblending models that must be fit.

Additional computational expense aside, simultaneous modeling provides
some significant scientific advantages for dealing with blended
objects.  The primary advantage is that our rigorous Monte Carlo
characterization of the uncertainties allows us to marginalize over
neighboring objects we do not care about, correcting measurements for
the presence of undesirable foreground or background objects.  When
multiple objects in a blended set are of interest, the Monte Carlo
samples correctly account for the correlations between their measured
properties.

When fitting multiple exposures simultaneously, our models also allow
us to take advantage of additional time domain information when
assigning flux to different sources.  The classic example in this case
is strong gravitational lensing of quasars by galaxies.  By fitting a static
galaxy model simultaneously with a set of variable point-source
models, we can better measure both the light curves of the quasars and
the profile of the galaxy.  If we marginalize over uncertainties in the
PSF model rather than hold it fixed, we can even use this
additional information to make up for deficiencies in the PSF model.
We can also use this technique to deblend supernovae from their host
galaxies, particularly if we put an informative prior on the light
curve.  Finally, if we fit multiple filters simultaneously, we can use
a strong prior on the colors of objects to aid in deblending as
well.  While this may make it more difficult to find rare objects with
extreme colors, it would provide an improvement in the photometry of
most blended objects, and may represent the only way to attempt to
separate severely blended objects.

The current state-of-the-art in deblenders, represented by the SDSS
deblender \citep{Lupton2005}, operates entirely
separately from the measurement process, assigning a fraction of
each pixel's flux to each object in such a way that all of the flux in
each pixel is accounted for.  Measurement algorithms, including modeling
methods, then operate on these fractional pixel values, making it
unnecessary to fit multiple objects simultaneously.  Accounting
for all the flux is clearly an attractive feature, and the
pixelized, symmetry-based models used by the SDSS deblender
are much more flexible than we expect our shapelet models to be.
Unfortunately, generalizing these models to the multi-exposure case is
difficult, and while they can still be used to generate a deblending
hypothesis on the coadd, they cannot be used directly to divide
per-pixel flux on individual exposures.  While we may be able to
develop a procedure to transform these sort of pixelized models to the
individual exposures (by convolving them with the difference kernel
between the coadd and each exposure, for instance), we argue that it
will be better \emph{not} to divide per-pixel fluxes before the
modeling stage.  Clearly simultaneous modeling provides 
much better control over the uncertainty; dividing the pixel fluxes
essentially ignores uncertainty in deblending that may in fact be the
dominant source of error in the measurements.  The only way avoiding
simultaneous modeling will produce better results -- aside from
performance questions, which admittedly may be a factor -- is if the
models used to divide pixel fluxes are a significantly better fit to
the data than the measurement models, to the extent that the
measurement models cannot be used to reconstruct most of the data.
This is certainly possible if the measurement models are too simple,
but this situation would also produce large modeling biases in our
measurements even when deblending is not a factor, so we view improving
the measurement models as the best solution to this problem.

\section{Public Catalogs}
\label{sec:applications:catalogs}

All of our approaches to measurement have focused on characterizing the
full posterior distribution of quantities by applying measurement
operators to our Monte Carlo sample of parameter vectors.  This is
only possible while we actually \emph{have} the Monte Carlo sample
available, and thus requires that we know the desired measurements in
advance (and can reduce their posterior distributions to a few
descriptive numbers), or that we store the Monte Carlo sample and make
it available as one of the data products of the survey.

Ignoring for a moment the question of whether the latter proposal is
practical, consider the advantages:
\begin{itemize}
\item We do not need to realize nearly as many concrete measurements,
  making catalog database tables much ``thinner''.  Because we can
  efficiently compute most measurements on the fly, we do not need
  to anticipate all catalog-like measurements in advance.
\item Users will be able to make use of the full posterior
  distribution of any measurement, including
  much higher-level measurements that are usually based on catalog
  entries for many objects, with full
  control over how to marginalize over other model parameters
  and neighboring objects.
\item New operations that would have otherwise required access to the
  image data will be able to make of the model samples instead,
  potentially reducing the costs of image access and public compute
  resources.
\item New data from outside sources can be correctly combined with the
  measurements, by using the same models and multiplying the
  posteriors.
\end{itemize}
Because the priors are shared and easily
stored, this also allows users to use a slightly different prior,
simply by multiplying the weights by the ratio of the new prior to the
original one.  As discussed by \citet{Hogg2011}, a probabilistic
approach involving generative models, like this one, is simply the
correct way to view the outputs of an astronomical survey; catalogs
are at best convenient approximations.

Unfortunately, this approach may also be totally impractical -- while we
do not have good estimates of either the size of the Monte Carlo samples
needed or the number of linear parameters, the uncompressed size of
the samples could easily approach the size of the imaging data.
Whether the samples could be compressed efficiently is an open
question, however, and we can always reduce the sample size by
extracting a random subsample.  If we choose such a subsample
by resampling with the weights, we will draw more points from the regions with higher
probability, and possibly decrease the sample size without
increasing the variance.  And if making the samples available to the
user does reduce direct image access, this change might be viewed as a
trade-off between compute resources and storage resources.

Regardless, while providing direct access to the full modeling results
is a challenging technical problem, and it represents a paradigm shift
in how public surveys publish their data, it is almost certainly a
positive paradigm shift, and the challenge should be viewed as one
worth some technical effort.

\chapter{Implementation}
\label{sec:implementation}
\section{Pipeline Overview}

\begin{figure*}
  \begin{center}
    \includegraphics[width=\textwidth]{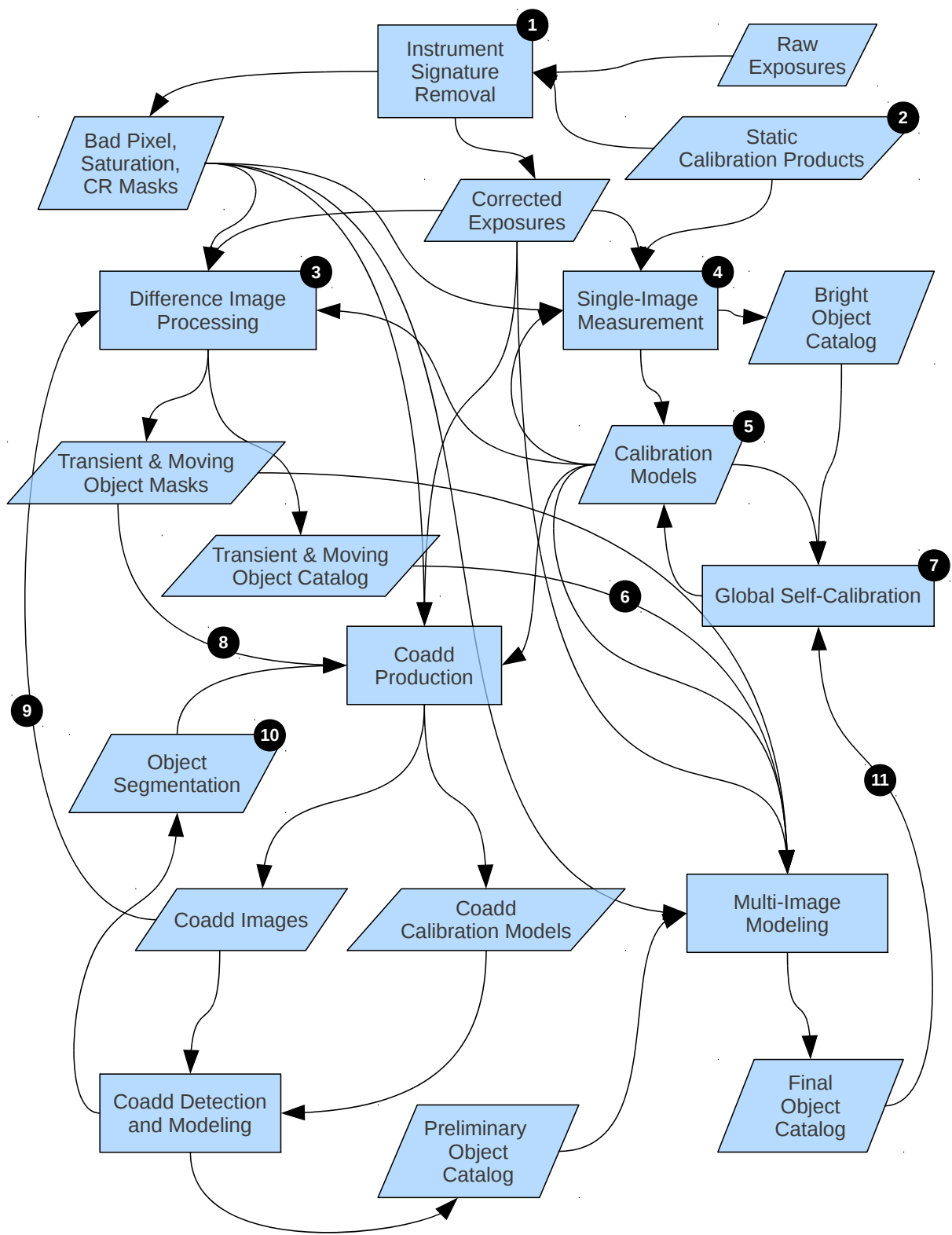}
  \end{center}
  \caption{
    Idealized data reduction pipeline data flow and dependencies.
    Most processes and relationships are self-explanatory; numbered
    items are described in the text.
    \vspace{2cm}
    \label{fig:pipeline}}
\end{figure*}

A complete discussion of the data flow and dependencies for a full
survey reduction pipeline is beyond the scope of this paper, but it is
worth spending some time to discuss how our measurement methods will
interact with the rest of the pipeline.  Figure~\ref{fig:pipeline} is
a high-level diagram of the \emph{idealized} data flow and
dependencies.  It includes several circular dependencies, reflecting
the fact that reduction is an iterative process; measurements feed
into self-calibration procedures that are used to improve the
measurements.  A practical pipeline diagram would involve
decisions regarding which relationships to ignore, and which
iterative procedures should loop until a condition is met and which
should simply be performed a fixed number of times.

A few aspects of Figure~\ref{fig:pipeline} require some additional
discussion and/or explanation.  The numbers below refer to the numbers
in the figure.

\begin{enumerate}
\item ``Instrument Signature Removal'' (a name taken from the LSST
  pipelines) refers simply to the basic image reduction operations
  that operate on entire images: de-bias, flat-field, fringe removal,
  etc.  Here we also assume it includes cosmic ray and saturation
  detection.
\item ``Static Calibration Products'' are inputs to calibration models that do
  not depend on the pipeline outputs (or depend so weakly that they
  are updated on an entirely different schedule).  These include
  detector correction images like bias frames, as well as global
  astrometric and photometric standard catalogs.
\item ``Difference Image Processing'' is a stage in which a coadd is
  convolved to match an individual exposure and subtracted from it,
  allowing transient, moving, and variable objects to be detected.  This
  generates a catalog of objects (after the detections from each exposure
  are matched and reconciled).  We can also build a mask to avoid
  including pixels affected by moving objects and transients in the
  coadd and during multifit.
\item In ``Single-Image Measurement'', we detect and measure the
  properties of objects that are bright enough to appear in individual
  exposures, using only one exposure at a time.  We also
  generate the first version of the calibration models for the PSF,
  background, astrometric solution and photometric solution.
\item By ``Calibration Models'', we mean exactly those quantities we
  have included in our modeling algorithm as nuisance parameters: the
  PSF and background models for each exposure, as well as the
  relative astrometric and photometric calibrations.  It is only the
  relative calibrations that participate in the dependency cycles;
  changes to the absolution calibration can be easily applied to the
  final catalogs with no need for iteration.
\item We do not limit multifit to just those objects that are detected
  on the coadd; we also want to include transient objects that are
  detected only in the difference images (and often
  deblend them from the static and slow-moving objects found on the
  coadd).  Extremely fast-moving objects like asteroids can be fit
  in a multi-epoch sense as well, though this may present a challenge
  for organizing the image data efficiently.
\item In ``Global Self-Calibration'', the relative astrometric and
  photometric models are improved by requiring measurement consistency
  across all overlapping sets of exposures (e.g. \citealt{Ubercal}).  We
  also envision 
  a global PSF modeling step such as the PCA approach
  developed by \citet{JarvisJain2004}.  All of these operations are
  formally global, as the entire survey must be tied together, but an
  implementation that operates on large subsets of the data may be
  sufficient for some iterations, as long as these are in agreement at
  the end.  Unfortunately, global calibration steps may also introduce
  small inter-exposure covariance terms in the uncertainties of the
  calibrations, which would complicate slightly the marginalization
  algorithm in section~\ref{sec:modeling:marginalizing-calibrations}.
\item When building a coadd, there are several options on how to
  handle time-domain information.  One option is to use masks from the
  difference image processing to try to leave such objects out of the
  coadd entirely.  We can also use an outlier rejection method to
  attempt to remove the objects without using masks.  On the other
  hand, it may be useful to define the coadd as a strict mean that
  includes time-domain objects.  Our
  tentative recommendation would be to build coadds that include
  variable and slow-moving objects (i.e. those that move less than a
  typical PSF over the course of the survey) in the coadd, while
  removing true transients and fast-moving objects.  The best
  measurements for all faint time-domain objects will have to come
  from a multi-epoch modeling procedure, but a coadd defined this way
  will maximize the usefulness of the coadd in setting up those
  models.  Using outlier rejection is an attractive possibility due to
  its lack of dependency on the difference imaging, but it makes the
  construction of a PSF model for the coadd much  more difficult, as
  the effective weights of different exposures differ from pixel to
  pixel in a way that depends on the noise.
\item A coadd is a necessary ingredient in the sort of difference
  imaging we imagine (though it may be necessary to start by
  differencing pairs of individual exposures).  The choices we
  discussed above in how to handle time-domain objects in the coadd
  have important implications for what difference image detections and
  measurements mean, and play a role in how deciding how best to
  resolve the circular relationship between coadds and difference
  images.
\item As we have discussed, an optimal coadd cannot have masked pixels,
  image edges, or a variable PSF within any pixel region that contains
  an astronomical object of interest.  This means we should segment
  the sky based on the locations of objects, and ``round-off'' all
  defects and edges on individual exposures to the nearest
  segmentation boundary.  This segmentation can 
  only be built by detecting on a full-depth coadd,
  however, making at least two phases of coadd production necessary if
  coadd measurements need to be high-quality.
\item If we expect to use faint objects in our self-calibration
  procedures, the final self-calibration stage must occur after the
  multifit procedure.  However, we cannot fully make use of
  improved calibration models without performing the multi-epoch
  modeling after the final self-calibration.  This is one of the most
  difficult cycles to resolve.  Ideally, we would be able to meet our
  self-calibration goals using only bright objects that are measured
  on individual exposures.  If not, all the options are poor: we can
  either perform a second multifit step (faster than the first, since
  we only need to recompute the weights, but still expensive), or we
  can correct the measurements for the changes in the calibration at
  the end.  This would require fitting different measurements on all
  exposures even for galaxy models, dramatically reducing the S/N of
  those measurements while dramatically increasing the size of the
  catalogs.
\end{enumerate}

\section{Scaling and Parallelization}

Implementations of the algorithm and modeling procedures we have
described are not yet mature enough for concrete performance measurements
to be anything but misleading.  However, it is useful to note briefly
how the algorithm scales along certain axes, and discuss some of the
options for parallelization.

First, it is worth noting that in the many-exposure limit, the number
of pixels will be far larger than the number of linear or nonlinear
parameters, even for small objects.  It is thus very important that
the worst-case complexity of the model in the number of pixels (and
similarly, the number of exposures) is linear, except in the case of
variable objects (in which the number of coefficients scales with the
number of exposures).  Happily, we expect to fit variable objects with
simple point source models, so the computational expense involved in
evaluating those models on each exposure is also considerably less.
Since each set of coefficients only affects one exposure, the 
model matrix is also sparse in a block sense, and has exactly as many
nonzero elements as the matrix for a nonvariable model.
For galaxies, we expect the dominant computational expense to be
evaluating the models on each exposure, even though we perform many
operations with worse than linear complexity in the number of
linear parameters for each model evaluation.

To evaluate a (compound) convolved shapelet model on a single exposure, we must
evaluate the ellipse-transformed shapelet basis functions on each pixel; the order
of the convolved basis is equal to the sum of the PSF order and the
intrinsic galaxy model order, and the number of elements in the basis
is proportional to the \emph{square} of the convolved basis order.  Reducing
the order of the shapelet bases involved in modeling galaxies and PSFs
is thus an important avenue for optimization; many low-order bases
with different orders can easily outperform fewer high-order
bases when convolution is involved.  We expect the total number of
basis functions to be between 10 and 50, so the size of these basis
matrices also virtually ensures that the algorithm will be limited by
memory or (most likely) CPU constraints, rather than I/O constraints.
For each segmented region containing a single galaxy, we will need to
evaluate perhaps 20-30 outer-stage Monte Carlo points; for each of
these, we must evaluate a matrix with 10-50 times more elements than
the number of data pixels.

Any modeling procedure for individual astronomical objects can easily
be parallelized along the catalog axis; we simply assign each
segmentation region to a different processor.  While multi-object
deblending segmentations will take longer, this parallelization
requires virtually no communication between nodes.  Our nested Monte
Carlo procedure suggests two more fine-grained natural parallelization
axes that may also prove useful.  For a more traditional (e.g. multicore CPU) computing
environment, parallelizing along outer-stage Monte Carlo points is an
attractive option, requiring synchronization only when updating the
outer-stage importance function.  Each core would then be responsible
for a single model matrix evaluation, as well as sampling from the
corresponding inner-stage importance function.  In a manycore
(i.e. GPU) environment, we should parallelize along individual pixels
when computing the model matrices; each exposure maps easily to a
block of lightweight threads, and the shapelet recurrence relations
are exactly the sort of math-heavy, single-instruction
multiple-data operations that GPU processors excel at.  The remainder of
the operations are essentially dense linear algebra operations, and
would also naturally translate extremely well to a manycore
implementation.

\chapter{Related Work}
\label{sec:related-work}

We have already touched on several modeling methods that have features
in common with the one we have developed here.  In this chapter, we
will discuss a few of these in detail, emphasizing the advantages and
disadvantages of their approach relative to ours.

\section{Shapelets and S\'{e}rsiclets}

Gauss-Laguerre and Gauss-Hermite functions first started to see
extensive use modeling galaxies following
\citet{BJ02} and \citet{R03}.  The latter coined the term
``shapelets'' to refer to both sets of basis functions, which have
since been used in a variety of applications.  Our approach is most
similar to \citet{NB07}, who also transform the basis functions by
transforming the input coordinates rather than using shapelet-space
operators, and also convolve the model with the PSF before comparing
it with data.  Unlike a full generative model approach, however, their
``forward fitting'' algorithm iterates until a constraint similar to
\eqnref{eqn:ellipse-constraint} is met, rather than maximizing the
full likelihood or posterior.  They also use an approximate
convolution using shapelet shear operators, but the exact convolution
relation we have derived in section~\ref{sec:shapelets:convolution} could
easily be used instead in their algorithm and would likely improve the
results. 

The failure of shapelet models to fit realistic galaxy profiles has
led to the proposal of similar basis sets based on general S\'{e}rsic
profiles by \citet{Ngan2009} and \citet{Andrae2011b}, which they have
called S\'{e}rsiclets.  While they do provide much better fits to
typical galaxies, these basis functions do not have many of the
convenient properties of shapelets.  In particular, they do
not have an analytic convolution relation, making them inconvenient to
use in generative models, which must be convolved efficiently with a
PSF model.  While building an approximate S\'{e}rsiclet basis using
compound shapelet basis techniques may provide a way around that
problem, the S\'{e}rsiclet basis functions do not provide any benefits
beyond an empirically reasonable profile at zeroth order and an
orthogonal set of higher-order functions; there is no reason to
believe the first or second-order S\'{e}rsiclet basis functions are
more useful than higher-order terms.  We can much more easily obtain a
compound shapelet basis with these properties simply by using the
Cholesky orthogonalization method introduced in
section~\ref{sec:shapelets:compound:properties}.

\section{Lensfit}

The \textsc{Lensfit} code of \citet{Miller2007} and \citet{Kitching2008}
was the first to apply Bayesian methods to the shear estimation
problem.  The success of this method, both with the GREAT08
simulations \citep{GREAT08} and the CFHTLS lensing survey, is one of
the primary reasons we believe Bayesian methodology and a robust
accounting of uncertainties based on generative models have an
important role to play in shear estimation.

The \textsc{Lensfit} algorithm is different from ours in several respects,
however.  First, they sample the likelihood on a grid, rather than
using Monte Carlo methods.  This provides a more deterministic
accounting of errors, but does not scale well to higher dimensions
(i.e. large numbers of parameters).
This has two important implications.  First, the \textsc{Lensfit} models
are relatively simple; they use a fixed S\'{e}rsic profile, and allow
only the ellipse parameters to vary.  These models are probably
sufficient for most lensing source galaxies in terms of information
content, but they may impose a small systematic modeling bias in the
shear estimate; \citet{VB09} have shown that S\'{e}rsic model biases
are not negligible when fit with maximum-likelihood methods, and it is
not clear whether this persists with the more sophisticated Bayesian
modeling approach.  \textsc{Lensfit} also makes use of a clever
Fourier-space method to marginalize over the centroid automatically,
which reduces the dimensionality of the grid. 

This automatic marginalization may be problematic in the
many-short-exposures limit, however.  \textsc{Lensfit} combines data from
multiple exposures by simply multiplying the likelihoods at each grid
point.  This means the centroid marginalization is carried out
\emph{before} combining the likelihoods, effectively considering the
model to have a different centroid on each exposure.  When each
exposure has sufficient depth to effectively measure a centroid using
data from that exposure alone, this increase in the number of degrees
of freedom is relatively unimportant, and it may even help account for
errors in the astrometry.  When we have many short exposures, however,
the centroid likelihood on an individual exposure is broad, and
marginalizing over these broad distributions individually (instead of
marginalizing over their much narrower product once) will produce a
significant degradation in the ellipticity measurement.

\section{Coadd-Based Shear Estimation}

Prospects for effectively measuring ellipticities for weak-lensing on
image coadds are not as bleak as they were once thought to be, thanks
largely to a practical procedure for building a PSF model for the
coadd due to \citet{JeeStackfit2011}.  Combined with a
practical implementation of the Kaiser optimal coadd algorithm
mentioned earlier, under certain conditions a coadd-based modeling
method should be equivalent to a multi-exposure modeling method.  None
of these conditions will be exactly met in realistic conditions, but
most will be nearly met most of the time:
\begin{itemize}
\item The PSF must not be spatially varying; in practice, this is
  almost always a safe approximation on the spatial scales of a single
  source galaxy image.
\item No pixels must be masked out, and there must be no edges to
  images in the modeling region.  In practice, this means we must discard
  images with bad pixels, cosmic ray hits, saturation, or image
  boundaries near an object of interest; this would then require a
  separate coadd for each object (or blended group of objects) to
  avoid throwing away too much data.  We would suffer a small loss in
  signal-to-noise, and building these multiple coadds will be more
  expensive than a single coadd (because some overlap regions may be
  coadded many times).  In contrast, with multi-exposure modeling we
  discard bad data pixel-by-pixel, instead of
  exposure-by-exposure.
\item Image noise must be stationary.  In practice, this means we must
  be in the sky-noise-dominated limit, which will generally be the
  case for most lensing source galaxies, but will not be true for
  brighter foreground galaxies and stars that may be nearby.
\item Calibration models must have negligible uncertainties.  While we
  cannot say that correctly marginalizing over calibration
  uncertainties is impossible with a coadd-based approach, no method
  for doing so exists, so at present we must require that calibration
  models be well-determined enough that their uncertainties do not contribute
  significantly to the total error budget.
\end{itemize}
Because coadd-based methods cannot be used at all for variable or moving
objects, multi-exposure methods should also be considered necessary
when deblending galaxies from possibly variable or moving stars.

These stringent requirements, along with the fact that we \emph{must}
use multi-exposure methods when fitting stars, may suggest that we
should simply avoid coadd-based procedures.  But fitting on a coadd --
even the per-object coadds discussed above -- can be orders of
magnitude faster than fitting to multiple exposures, especially if
most of the computation time is spent evaluating and convolving the
model.  We have already discussed the importance of using a coadd for
most iterations of our adaptive importance sampling algorithm, in
which we do as little work as possible in ``multifit mode''.  In the
same vein, we should consider the possibility that for some objects
all of the above conditions will be met satisfactorily, and we may not
need a final multifit step.  We should thus view the coadd generation
process as an important one, and our goal should be the best possible
coadd-based measurement (rather than simply a ``quick and dirty''
measurement needed to bootstrap a lengthy multifit procedure).  Even
if we can only rarely skip the multifit step, doing so is a powerful
optimization.  If we cannot skip the multifit step, better
characterization of the importance distribution on the coadd will
still allow us to save time in the multifit stage.

\chapter{Conclusion and Future Work}
\label{sec:conclusion}

The methodology we have described in this paper is essentially an
ambitious proposal to change how we build astronomical catalogs from
photometric survey data.  In the ideal case, it is even a proposal to
change what we consider ``catalog'' to mean.  It is also, necessarily,
an incomplete proposal.  The core of this work -- the Monte Carlo
algorithm described in Chapter~\ref{sec:modeling}, and the
shapelet-based approach to build and convolve galaxy models in
Chapter~\ref{sec:shapelets} -- constitutes a full prescription for a
method to robustly model galaxies.  But testing and iterative
improvement is a necessary part of making any method practical, and
we have proposed many other ideas that require significant further
development.

Perhaps the most important piece is simply the idea of a Bayesian
approach to survey data reduction.  We are not the first to advocate
this approach, but very little work has been done to make the idea
practical.  There are two important aspects of a Bayesian approach.
First, we should rigorously characterize the actual distribution of
uncertainties on measurements, including correlations between
measurements.  This does not mean we must avoid, for instance, Fisher
matrix methods -- but we must validate such approximations.  Second,
we should consider uncertainties in calibration models, and include them
in our measurements, ideally through marginalization.  This is the
crucial step in moving from generative models for individual
astronomical objects to a single self-consistent generative model for an
entire survey.  In both cases, the crucial point is not that
approximations are to be avoided -- the concern is that
when we do not take these steps, we tend to systematically
underestimate uncertainties.  Most real-world distributions have
broader tails than Gaussians, and marginalization of a nuisance
parameter rarely tightens a constraint.

When we cannot approximate some distributions with Gaussians, or
ignore the impact of calibration parameters on our measurements, data
reduction is almost inevitably more computationally expensive.  We
cannot rely on Moore's Law and the rapid improvement in hardware
capabilities, however -- these same forces drive the increase in astronomical data
volumes, and we must instead look to algorithmic improvements.
Moreover, these complex algorithms are \emph{more} important from a
science standpoint as data volumes grow, as the increase in sample
sizes leads to a decrease in stochastic sources of error, which makes
understanding and addressing systematic errors more important.

The nested Monte Carlo importance sampling algorithm we have developed
provides an example of both the challenge and the sort of algorithmic
research needed.  Even with a highly-optimized implementation, it will
almost certainly be slower than an optimized version of most modeling
methods commonly used today, but it does much more, marginalizing over
calibration parameters and supporting an extremely flexible and
general class of models, all while making very few assumptions about
the behavior of the likelihood or prior distributions.  By using a
linear model and taking advantage of the near-Gaussianity of the
marginal likelihood, we can draw many Monte Carlo points
for each model evaluation, decreasing the number of expensive
multi-exposure model evaluations needed to produce reasonable
estimates.

It remains to be seen whether a compound shapelet approach will be the
most effective way to provide convolvable models for this algorithm
and others like it.  Shapelet methods seem well positioned with regard
to certain trends in hardware architectures (namely GPU-based
computing), in which a simple, highly-parallelizable brute-force method
may be more efficient than a more complex and memory-intensive
lookup-table approach, but such trends are difficult to predict, and
of course actual performance tests will have to decide.  The shapelet
convolution relation and ``compound basis'' techniques we have
developed provide simple solutions to the biggest drawbacks of
current shapelet methods, but the compound shapelet approach in
particular could benefit greatly from further study.  While we can
construct usable compound shapelet models from combinations of
S\'{e}rsic-profile components, and fit these with the flat,
linear-constraint priors we have described, the method has a great
deal more potential with an empirically-trained basis and informative
prior.

This is where perhaps the most scientifically interesting future work
is needed.  Current efforts to characterize galaxy morphologies using
linear basis functions have focused almost exclusively on shapelets
and shapelet-inspired bases so far, and have been limited by the
flexibility of those models and, frankly, the extreme difficulty of
the problem.
PCA and machine-learning methods have been powerful in
similar classification and dimensionality reduction problems in other
areas of astronomy and cosmology, however, and our ellipse-transformed
galaxy models and robust handling of the PSF may provide the necessary
boost to make these approaches more useful in quantifying galaxy
morphologies.  This is an interesting topic in its own right, but it
also provides a way to take advantage of both narrow 
space-based surveys and wider ground-based surveys.  By training a
galaxy basis and a prior on higher-resolution space-based data, we can
better infer the properties of much larger samples of galaxies
observed from the ground.

Tying these galaxy modeling methods into a larger pipeline must also
be the subject of a great deal of future work.  We cannot make use of
algorithms that handle calibration uncertainties unless those
uncertainties are themselves well-characterized.  This is in some
respects a more difficult challenge than modeling galaxy morphologies;
calibration models must essentially be able to model all the myriad
ways the optical system can (mis)behave, or at least provide a way to
reject data from modes that cannot be modeled.  We must also deal with
the fact that many of the steps in a survey pipeline have
intrinsically circular dependencies, including the measurement and
calibration modeling stages; these must be resolved by a carefully
study of the implications of changes in ordering and the number of
iterations.  Finally, we should put serious effort into addressing the
challenge of making a full
characterization of the generative models for astronomical objects
available to the public.  We have proposed doing this by making the
actual Monte Carlo samples and weights our algorithm produces the
actual ``catalog'' users see (along with indices for searching and
filtering some common measurements, of course).  This may not be
necessary, and it may not be the most efficient way to accomplish this
goal -- but the information we do make available in catalog form
should be equivalent in terms of information content, or at least
represent a conservative rather than optimistic characterization of
the uncertainties.  While this may be more expensive than
standard catalog access, it allows us to provide support
for Bayesian analyses without requiring users to access the raw pixel
data.

The topics addressed in these thesis have often been at the boundary
of astronomy, and have considerable overlap with more
``methodological'' fields such as computer science, applied
mathematics, and statistics.  These topics are extremely important for
astronomy, however, and must be addressed at least in part by
astronomers and physicists.  They also must be addressed in advance;
many of the ideas in this thesis are most applicable to the LSST,
which is nearly a decade away from operation, though they will also be
applied to smaller surveys in the interim.  Yet this is likely
scarcely enough time for some of these ideas to be fully realized,
requiring us to work on simulation data or precursor surveys that have
already been mined for their most important results.
Efforts to improve data reduction algorithms -- and build practical software
implementations of them -- are an integral part of building a public
survey facility, and while they may sometimes seem removed from the 
high-level science questions that drive the fields of astronomy and
astrophysics as a whole, they is no less important to the march of
scientific progress.

%
%

\pagestyle{plain}

\ssp   
\bibliography{references}
\bibliographystyle{custom2}

%
\part*{\addcontentsline{toc}{part}{Appendices}Appendices}
\appendix

\pagestyle{fancyplain}

\chapter{Ellipse Parameterizations}
\label{sec:appendix:ellipse-parameterizations}

Ellipses are widely used in astronomy to represent the shapes of
extended objects, and can be parameterized in a number of ways,
usually involving exactly five numbers.
Almost all of these use a center coordinate point to indicate the
position, leaving three numbers to define the ellipticity, radius, and
orientation of the ellipse.  The most familiar form uses the semimajor
axis $a$, the semiminor axis $b$, and the position angle $\varphi$ (we
will measure $\varphi$ as the angle from the $x$-axis to the major
axis).

Another common form, which we will call the ``quadrupole'' or
``moments'' representation with parameters $\{q_{xx},q_{yy},q_{xy}\}$,
is more easily related to the moments of a function $f[x,y]$:
\begin{align}
  q_0 &\equiv \int\!dx\,dy\,f[x,y] \\
  q_x &\equiv  \frac{1}{q_0}\int\!dx\,dy\,x\,f[x,y] \\
  q_y &\equiv  \frac{1}{q_0}\int\!dx\,dy\,y\,f[x,y] \\
  q_{xx} &\equiv
  \frac{1}{q_0}\int\!dx\,dy\,(x-q_x)^2\,f[x,y] \\
  q_{yy} &\equiv
  \frac{1}{q_0}\int\!dx\,dy\,(y-q_y)^2\,f[x,y] \\
  q_{xy} &\equiv \frac{1}{q_0}\int\!dx\,dy\,(x-q_x)(y-q_y)\,f[x,y]
\end{align}
When $f[x,y]$ is a noisy image, it is necessary to include an
apodizing weight function in the integrals.  As we discuss
in sections~\ref{sec:shapelets:compound:basis-ellipse} and
\ref{sec:applications:shear-estimation}, this is less important and
takes on a different role when $f[x,y]$ is a model.  Not all weight
functions define an ellipse that transforms properly under changes of
coordinates, however; we must choose a weight
function for which the ellipse measurement operation commutes with the
operation of applying a linear coordinate transform.  We refer the
reader to \citet{BJ02} and \citet{NB07} for further discussion of
which weight functions are appropriate.

Viewing $f[x,y]$ as a probability distribution, we can see that these
are simply the elements of the covariance matrix:
\begin{align}
  \bm{\Sigma} = \left[\begin{array}{c c}
      q_{xx} & q_{xy} \\
      q_{xy} & q_{yy}
      \end{array}
    \right]
\end{align}
It is natural
then to relate the moments definition of an ellipse to the ``axes''
definition by defining $a$, $b$, and $\varphi$ by the 1-$\sigma$
contour of a Gaussian distribution with this covariance matrix.  The
direct conversions are:
\begin{align}
  q_{xx} &= a^2\cos^2\varphi + b^2\sin^2\varphi \\
  q_{yy} &= a^2\sin^2\varphi + b^2\cos^2\varphi \\
  q_{xy} &= (a^2-b^2)\cos\varphi\sin\varphi
\end{align}
\begin{align}
  a &=
  \sqrt{\frac{1}{2}\left(q_{xx}+q_{yy} +
    \sqrt{(q_{xx}-q_{yy})^2+4q_{xy}^2}\right)}
  \\
  b &=
  \sqrt{\frac{1}{2}\left(q_{xx}+q_{yy} -
    \sqrt{(q_{xx}-q_{yy})^2+4q_{xy}^2}\right)}
  \\
  \varphi &=
  \frac{1}{2}\arctan\!\left[\frac{2q_{xy}}{q_{xx}-q_{yy}}\right]
\end{align}
The ellipse transform matrix defined in
\eqnref{eqn:ellipse-transform-matrix} is a square root of the covariance
matrix $\bm{\Sigma}$:
\begin{align}
  \bm{\Sigma} &= \bm{T}\bm{T}^T = \left[
    \begin{array}{c c}
      \cos{\varphi} & -\sin{\varphi} \\
      \sin{\varphi} & \cos{\varphi}
    \end{array}
    \right]
  \left[
    \begin{array}{ c c }
      a^2 & 0 \\
      0 & b^2
    \end{array}
    \right]
  \left[
    \begin{array}{c c}
      \cos{\varphi} & \sin{\varphi} \\
      -\sin{\varphi} & \cos{\varphi}
    \end{array}
    \right]
\end{align}
The axes representation of the ellipse is thus simply a
diagonalization of the covariance matrix.
Note that we can define many other ellipse transform matrices other
than the one we have chosen, by adding another orthogonal matrix on
the right side of \eqnref{eqn:ellipse-transform-matrix}.  The shear
transform used by \citet{BJ02} does just this, adding a rotation by
$-\varphi$ on the right.  Unlike our form, this does not consistently
identify one axes of an ellipse-transformed basis function with the
major axes of the ellipse.

In weak gravitational lensing, it is common to use a complex
ellipticity and radius parameterization.  There are many definitions
of ellipticity, and two primary definitions of radius (other than the
semimajor and semiminor axes).  The radius definitions are based on
the two most important scalar functions of a matrix, the trace and
determinant, applied to $\bm{\Sigma}$:
\begin{align}
  r_{\text{tr}} &\equiv \frac{1}{2}\sqrt{\tr[\bm{\Sigma}]}
  = \frac{1}{2}\sqrt{q_{xx} + q_{yy}}
  = \frac{1}{2}\sqrt{a^2+b^2} \\
  r_{\text{det}} &\equiv \left(\det[\bm{\Sigma}]\right)^{1/4}
  = \left(q_{xx}q_{yy}-q_{xy}^2\right)^{1/4}
  = \sqrt{ab}
\end{align}
Both of these radii reduce to the conventional radius for a circle.  The two
radii are most different in the opposite limit; when the ellipse
approaches a line segment, the determinant radius approaches zero
while the trace radius approaches half the length of the line segment.
The square of the determinant radius is proportional to the area,
regardless of the ellipticity.

All definitions of complex ellipticity use the same definition of the
phase, but define the magnitude of the ellipticity differently.  Any
of these can be combined with any definition of radius and a center to
form a complete parameterization of an ellipse.  Using the terminology
of \citet{BJ02}, these are the reduced shear $g$, the distortion
$\delta$, and the conformal shear $\eta$.  The relations between these
and the axes and quadrupole parameterizations are:
\begin{gather}
  g = \frac{a-b}{a+b} = \tanh{\frac{\eta}{2}} =
  \frac{1-\sqrt{1-\delta^2}}{\delta} 
  = \sqrt{g_+^2 + g_\times^2}\\
  g_+ = g\cos{2\varphi} =
  \frac{q_{xx}-q_{yy}}{q_{xx}+q_{yy}+2\sqrt{q_{xx}q_{yy}-q_{xy}^2}} \\
  g_\times = g\sin{2\varphi} =
  \frac{2q_{xy}}{q_{xx}+q_{yy}+2\sqrt{q_{xx}q_{yy}-q_{xy}^2}}
\end{gather}
\begin{gather}
  \delta = \frac{a^2-b^2}{a^2+b^2} = \tanh{\eta} =
  \frac{2g}{1+g^2} = \sqrt{\delta_+^2 + \delta_\times^2} \\
  \delta_+ = \delta\cos{2\varphi} =
  \frac{q_{xx}-q_{yy}}{q_{xx}+q_{yy}} \\
  \delta_\times = \delta\sin{2\varphi} =
  \frac{2q_{xy}}{q_{xx}+q_{yy}}
\end{gather}
\begin{gather}
  \eta = \ln\frac{a}{b} = 2\arctanh{g} = \arctanh\delta
  = \sqrt{\eta_+^2 + \eta_\times^2} \\
  \eta_+ = \eta\cos{2\varphi} \\
  \eta_\times = \eta\sin{2\varphi}
\end{gather}
Both reduced shear and distortion have a maximum magnitude of one
(corresponding to a line segment); a line segment corresponds to an
infinite reduced shear.  While the former are easier to compute from
image moments, the infinite domain of the latter makes it generally a
better choice for approximating a probability distribution with an
analytic function such as a Gaussian or Student distribution.

\chapter{Gauss-Hermite Functions}
\section{Normalized Hermite Polynomials}
\label{sec:appendix:normalized-hermite-polynomials}
The normalized Hermite polynomials are simply the product of the
standard Hermite polynomials with the Gauss--Hermite normalization factor
\begin{equation}
  \Ht_n\!\left[x\right] \equiv
  \left(\sqrt{\pi}\,2^n\,n!\right)^{-1/2} H_n\!\left[x\right].
  \tag{\ref{eqn:normalized-hermite-def}}
\end{equation}

Like the standard Hermite polynomials, these can be efficiently
evaluated using recurrence relations
\begin{equation}
  \Ht_0[x] = \pi^{-1/4}
\end{equation}
\begin{eqnarray}
  \Ht_{n}[x] &=& x\sqrt{\frac{2}{n}}\Ht_{n-1}[x]
  -\sqrt{\frac{n-1}{n}}\Ht_{n-2}[x] \\
  &=& x\sqrt{\frac{2}{n}}\Ht_{n-1}[x]-\frac{1}{\sqrt{2n}}\D{\Ht_{n-1}[x]}{x}
\end{eqnarray}
\begin{equation}
  \D{\Ht_{n}[x]}{x} = \sqrt{2n}H_{n-1}[x]
\end{equation}
These recurrence relations are more numerically stable than those
for the standard Hermite polynomials, which suffer from round-off
error at high order.

\section{Triple Product Integral}
\label{sec:appendix:triple-product-integral}

In section~\ref{sec:shapelets:convolution}, we defined the elliptical shapelet
convolution formula \eqnref{eqn:convolution-complete} in terms of
the \emph{triple product integral} 
\begin{equation}
  B_{l,m,n}[\alpha,\beta,\gamma]
  = \frac{1}{\sqrt{\alpha\beta\gamma}}\int^{\infty}_{-\infty}
  \Ht_l\!\bigl[\tfrac{x}{\alpha}\bigr]\,
  \Ht_m\!\bigl[\tfrac{x}{\beta}\bigr]\,
  \Ht_n\!\bigl[\tfrac{x}{\gamma}\bigr]\,
  e^{-\frac{1}{2}\left(\frac{x^2}{\alpha^2}+\frac{x^2}{\beta^2}+\frac{x^2}{\gamma^2}\right)}dx
\end{equation}
with $\alpha=\beta=\sqrt{2}$ and $\gamma=1$.  The recurrence
relation given by \citet{RB03} to compute this integral suffers from
the same round-off error problems as the standard Hermite
polynomial recurrence relations.  Using the normalized Hermite
polynomials, we present here a more stable relation.  Let
\begin{equation}
  B_{l,m,n}[\alpha,\beta,\gamma]
  = \frac{\sqrt{2}v}{\sqrt{\alpha\beta\gamma}}\Lt_{l,m,n}[a,b,c],
\end{equation}
with
\begin{equation}
  v^{-2}=\alpha^{-2}+\beta^{-2}+\gamma^{-2},\;
  a=\sqrt{2}\frac{v}{\alpha},\:
  b=\sqrt{2}\frac{v}{\beta},\:
  c=\sqrt{2}\frac{v}{\gamma}
\end{equation}

\begin{equation}
  \Lt_{0,0,0} = \pi^{-1/4}
\end{equation}
\begin{eqnarray}
  \Lt_{l,m,n}
  &=& \sqrt{\frac{l-1}{l}}(a^2-1)\Lt_{l-2,m,n}
  + \sqrt{\frac{m}{l}}\,ab\,\Lt_{l-1,m-1,n} 
  + \sqrt{\frac{n}{l}}\,ab\,\Lt_{l-1,m,n-1} \\
  &=& \sqrt{\frac{l}{m}}\,ab\,\Lt_{l-1,m-1,n} 
  + \sqrt{\frac{m-1}{m}}(b^2-1)\Lt_{l,m-2,n}
  + \sqrt{\frac{n}{m}}\,bc\,\Lt_{l,m-1,n-1} \\
  &=& \sqrt{\frac{l}{n}}\,ac\,\Lt_{l-1,m,n-1}
  + \sqrt{\frac{m}{n}}\,bc\,\Lt_{l,m-1,n-1} 
  + \sqrt{\frac{n-1}{n}}(c^2-1)\Lt_{l,m,n-2}
\end{eqnarray}

Note that while our $B$ is the same as that defined by
\cite{RB03}, our $\Lt$ is not the same as their $L$:
\begin{equation}
  \Lt_{l,m,n}[a,b,c] = \left(2^{l+m+n}\,l!\,m!\,n!\,\pi^{3/2}\right)^{-1/2}\,
  L_{l,m,n}[a,b,c]
\end{equation}

\section{Inner Product Integral}
\label{sec:appendix:inner-product-integral}

We define the 1-d Gauss-Hermite inner product integral as
\begin{align}
  \Pt_{m,n}[a,b] \equiv \int^{\infty}_{-\infty}\!
  \Ht_m\!\left[ax\right]\,\Ht_n\!\left[bx\right]\,e^{-\frac{1}{2}(a^2+b^2)x^2}\,dx
\end{align}
The orthogonality of the Gauss-Hermite functions requires
$\Pt_{m,n}[a,a]=\frac{\delta_{m,n}}{a}$, but we need to be able to
evaluate this integral for $a\ne b$ when computing the compound basis inner
product matrix defined by \eqnref{eqn:inner-product-integral}.  This
can be computed using the recurrence relation below.
\begin{gather}
  \Pt_{0,0}[a,b] = \sqrt{\frac{2}{a^2+b^2}}
\end{gather}
\begin{align}
  \Pt_{m,n}[a,b] &=
  \frac{2ab}{a^2+b^2}\sqrt{\frac{m}{n}}\Pt_{m-1,n-1}[a,b] +
  \frac{b^2-a^2}{b^2+a^2}\sqrt{\frac{n-1}{n}}V_{m,n-2}[a,b] \\
  &=\frac{2ab}{a^2+b^2}\sqrt{\frac{n}{m}}\Pt_{m-1,n-1}[a,b] +
  \frac{a^2-b^2}{a^2+b^2}\sqrt{\frac{m-1}{m}}V_{m-2,n}[a,b]
\end{align}
Because the Hermite polynomials are all strictly odd or strictly even,
we also have $\Pt_{m,n}=0$ whenever $m+n$ is odd.

\end{document}